\documentclass[11pt]{article}

\usepackage{pstricks}
\usepackage{pst-plot}
\usepackage{pst-node}
\usepackage{amsmath}
\usepackage{pst-grad}
\usepackage{amssymb}
\usepackage{xspace}
\usepackage{pst-math}
\usepackage{pst-xkey}
\usepackage{pst-coil}
\usepackage{pst-3dplot}
\usepackage[T1]{fontenc}
\usepackage{ae}
\usepackage[ansinew]{inputenc}

\newrgbcolor{darkred}{0.8 0 0}
\newrgbcolor{darkerred}{0.7 0 0}
\newrgbcolor{darkestred}{0.5 0 0}
\newrgbcolor{darkgreen}{0 0.5 0.5}
\newrgbcolor{darkergreen}{0 0.6 0}
\newrgbcolor{darkestgreen}{0 0.5 0}
\newrgbcolor{darkblue}{0 0 0.6}
\newrgbcolor{darkestblue}{0 0 0.5}
\newrgbcolor{redgreen}{0.8 0.8 0}
\newrgbcolor{bluegreen}{0 0.4 0.7}
\newrgbcolor{bluewhite}{0 0.9 1}
\newrgbcolor{lightgreen}{0.6 1 0.6}
\newrgbcolor{lightred}{1 0.8 0.8}
\newrgbcolor{lightyred}{1 0.4 0.4}
\newrgbcolor{lightyblue}{0.5 0.5 1}
\newrgbcolor{lightblue}{0.8 1 1}
\newrgbcolor{lilas}{0.5 0.3 0.7}
\newrgbcolor{lilaswhite}{1 0.8 1}
\newrgbcolor{redy}{1 0.6 0.6}
\newrgbcolor{bluish}{0.6 0.6 1}
\newrgbcolor{sand}{0.9 0.9 0.7}
\newrgbcolor{brown}{0.4 0.4 0}
\newrgbcolor{blue1}{0 0.35 0.65}
\newrgbcolor{blue2}{0 0.4 0.6}
\newrgbcolor{blue3}{0 0.45 0.55}
\newrgbcolor{blue4}{0 0.5 0.5}
\newrgbcolor{blue5}{0 0.55 0.45}
\newrgbcolor{blue6}{0 0.6 0.4}
\newrgbcolor{blue7}{0 0.65 0.35}

\newcommand{\R }{\mathbb R}
\newcommand{\N }{\mathbb N}
\newcommand{\Z }{\mathbb Z}
\newcommand{\Q }{\mathbb Q}
\newcommand{\C }{\mathbb C}
\newcommand{\sen }{\ \! {\rm sen}\ \! }
\newcommand{\tg }{\ \! {\rm tg}\ \! }
\newcommand{\cosec }{\ \! {\rm cosec}\ \! }
\newcommand{\cotg }{\ \! {\rm cotg}\ \! }
\newcommand{\dx }{\ \! dx}
\newcommand{\e }{\ \! e}
\newcommand{\senh }{\ \! {\rm senh}\ \! }
\newcommand{\tgh }{\ \! {\rm tgh}\ \! }
\newcommand{\cotgh }{\ \! {\rm cotgh}\ \! }
\newcommand{\sech }{\ \! {\rm sech}\ \! }
\newcommand{\cosech }{\ \! {\rm cosech}\ \! }
\newcommand{\arcsen }{\ \! {\rm arcsen}\ \! }
\newcommand{\arctg }{\ \! {\rm arctg}\ \! }
\newcommand{\arccotg }{\ \! {\rm arccotg}\ \! }
\newcommand{\arcsenh }{\ \! {\rm arcsenh}\ \! }
\newcommand{\arccosh }{\ \! {\rm arccosh}\ \! }
\newcommand{\arctgh }{\ \! {\rm arctgh}\ \! }
\newcommand{\arccosech }{\ \! {\rm arccosech}\ \! }
\newcommand{\arcsech }{\ \! {\rm arcsech}\ \! }
\newcommand{\arcsec }{\ \! {\rm arcsec}\ \! }
\newcommand{\arccosec }{\ \! {\rm arccosec}\ \! }
\renewcommand{\arctgh }{\ \! {\rm arctgh}\ \! }
\newcommand{\arccotgh }{\ \! {\rm arccotgh}\ \! }
\newcommand{\nega }{\neg \ }
\newcommand{\eq }{\Leftrightarrow }

\begin{document}

\title{Survivability and centrality measures for networks of financial market indices}

\author{Leonidas Sandoval Junior \\ \\ Insper, Instituto de Ensino e Pesquisa}

\maketitle

\begin{abstract}
Using data from 92 indices of stock exchanges worldwide, I analize the cluster formation and evolution from 2007 to 2010, which includes the Subprime Mortgage Crisis of 2008, using asset graphs based on distance thresholds. I also study the survivability of connections and of clusters through time and the influence of noise in centrality measures applied to the networks of financial indices.
\end{abstract}

\section{Introduction}

The study of the dynamics of clusters has been occupying the minds of many researchers in the past years. In particular, one would like to study how clusters form and evolve in time. In the present work, I build and analize networks based on 92 indices of stock exchanges around the globe. The indices cover the five habitable continents, spanning a variety of markets and economies. In a companion article \cite{clusters1}, I analyzed data from some of the major international financial crises of the past decades using asset graphs in order to visualize the formation and evolution of clusters. There, I also analyzed how the second highest eigenvalue for the correlation matrix of indices may show an internal structure peculiar to financial market indices which is a consequence of them operating at different times. Here, I again use asset graphs in order to study the formation and evolution of financial markets indices, but now covering the years from 2007 to 2010. These years include the recent Suprime Mortgage Crisis, that reached its peak in 2008, and the succeeding Credit Crisis that lasts up to these days.

Asset graphs have been used with a variety of financial data \cite{asset01}-\cite{asset08}, and they often provide a graphic visualization which is complementary to the ones of Minimum Spanning Trees and Planar Maximally Filtered Graphs (see \cite{pmfg06} for examples of both types of graphs). An asset graph is a subset of the whole network based on a distance measure obtained, as an example, from a correlation matrix built from the time series of some variables, where every node is connected to all others. In order to restrict the number of nodes and of edges (connections) between them, a threshold is established, which may be the total number of allowed edges in the graph or a value bellow or above which no connection is to be considered.

In Section 2, correlation matrices based on the time series of benchmark indices of a variety of stock exchanges are built using Spearman rank correlation. The reason why I am using Spearman correlation and not the usual Pearson correlation is because the former is more useful when measuring linear relations between variables, and the latter is better at measuring nonlinear relations between them (there is a comparison between both coefficients in Subsection 2.9). Each correlation matrix so obtained is then used in order to build a measure of distance between nodes (indices), and then threshold values are chosen so that all distances bellow the threshold are used in order to create an asset graph, and the ones above it are discarded. By using a variety of thresholds, up to a limit where random noise becomes very intense, one is then able to witness the formation of clusters, first isolated and then unified as the threshold limit goes up. The upper limit for the threshold is obtained by using randomized data based on the original data so as to keep the original frequency distributions but destroy any temporal relation between them. In order for one to contemplate the true distances between indices, three dimensional maps are built using Principal Component Analysis \cite{multi}.

Section 3 considers the topic of survivability of clusters through time, which is essential, for instance, if one wishes to build risk-minimizing portfolios that are useful for future investments when based in past data. Single-step survival ratio \cite{asset02} and a survivability measure developed in \cite{pruning} are used in order to make this analysis, and the possibility of building asset graphs based on survivability is contemplated.

Section 4 shows a study of some centrality measures of the networks obtained from the asset graphs and on how they change with the value of the threshold used in order to build each graph and also how the node degree evolve in time. Section 5 consists on a general coclusion.

\section{Asset Graphs}

Here I build asset graphs by assigning different threshold values for a distance measure obtained from the correlation between financial market indices. The data consists on the indices of 92 stock market worldwide (for a list of those indices, and their symbols in this article, see Appendix A), collected from the year 2007 to the year 2010. The index values are computed daily, and here represented by $P_t$ for day $t$. Then, I calculate their log-returns, defined by
\begin{equation}
\label{logreturns}
R_t=\ln (P_t)-\ln (P_{t-1})\ ,
\end{equation}
where $P_{t-1}$ is the return on day $t-1$.

The log-returns are then used to calculate the correlations between pairs of indices. From the many correlation measures, I chose Spearman's rank correlation, for it is better at identifying nonlinear correlations between variables. The most common choice would be Pearson's correlation, and I make a comparison between results obtained based on both types of correlations later in this same section. Given a correlation coeficient $c_{ij}$ in a correlation matrix of all indices, a distance measure is defined as
\begin{equation}
\label{distance}
d_{ij}=1-c_{ij}\ .
\end{equation}
As correlations between indices vary from $-1$ (anticorrelated) to $1$ (completely correlated), the distance between them vary from $0$ (totally correlated) to $2$ (completely anticorrelated). Totally uncorrelated indices would have distance $1$ between them.

Based on the distance measures, three dimensional coordinates are assigned to each index using a procedure called Principal Component Analysis \cite{multi}, that minimizes the differences between the true distances and the distances represented in the graph.

As some correlations may be the result of random noise, I ran some simulations based on randomized data, consisting on randomly reordering the time series of each index so as to destroy any true correlations between them but maintain each frequency distribution intact. The result of such simulations (1000 for each period being studied) is a distance value above which correlations are probably due to random noise (this value is usually close to $d=0.6$). Then, asset graphs are built using distance measures as thresholds. As an example, for threshold $T=0.5$, one builds an asset graph where all distances bellow $0.5$ are represented as edges (connections) between nodes (indices). All distances above this threshold are removed, and all indices that don't connect to any other index bellow this threshold is also removed.

In what follows, I show the resulting asset graphs concerning each semester of the years ranging from 2007 to 2010, for thresholds varying from $T=0.2$ to $T=0.5$. These values were chosen so as to help visualization, which becomes harder for higher thresholds and many more connections. The separation into semesters is based on the constatation that correlations between markets may vary somewhat in brief periods of time, and that financial crises tend to concentrate on half of a year (usually the second semester).

\subsection{First Semester, 2007}

Figure 1 shows the three dimensional view of the asset graphs for the first semester of 2007, from $T=0.2$ to $T=0.5$.

At $T=0.1$, there are already strong ties between S\&P and Nasdaq, and between Fance and Germany, forming the seeds of the North American and European clusters. For $T=0.2$, the European cluster grows strong, with the addition of the UK, Switzerland, Italy, Belgium, the Netherlands, and Spain. We also have, for the first time, an African cluster, comprised of Namibia and South Africa. At $T=0.3$, we see the growth of the now American cluster, with the addition of Canada, Mexico, and Brazil. The European cluster receives the addition of Ireland, Austria, Sweden, Denmark, and Finland. A Pacific Asian cluster is formed between Hong Kong, Singapore, and Indonesia.

For $T=0.4$, Argentina joins the American cluster, which is now denser. Denmark, Norway, Portugal, Greece, the Czech Republic, Poland, and Turkey join the European cluster. South Africa connects with the European cluster, and Australia makes connections to both the European and the Pacific Asian clusters. The latter is now comprised of Japan, Hong Kong, South Korea, Malaysia, Singapore, and Indonesia. At $T=0.5$, Chile joins the American cluster, and Luxembourg, Hungary, Estonia, Lithuania, Russia, and Israel join the European cluster. Singapore and Namibia also establish connections with European indices. India and Taiwan join the now Asian cluster, which also becomes denser, and Cyprus connects with Greece. There is also a greater integration of the Asian and African clusters with the European one. At $T=0.6$, the American cluster connects with the European one, and Peru establishes connections with Europe. Philipines connects with both Europe and Asia, and Egypt connects with Europe. For $T=0.7$, noise starts to dominate. Colombia, Romania, Ukraine, and Thailand establish connections with existing clusters. More connections, many of them at random, are made for higher thresholds, and the last connections happen at $T=1.4$.



\vskip 1.8 cm

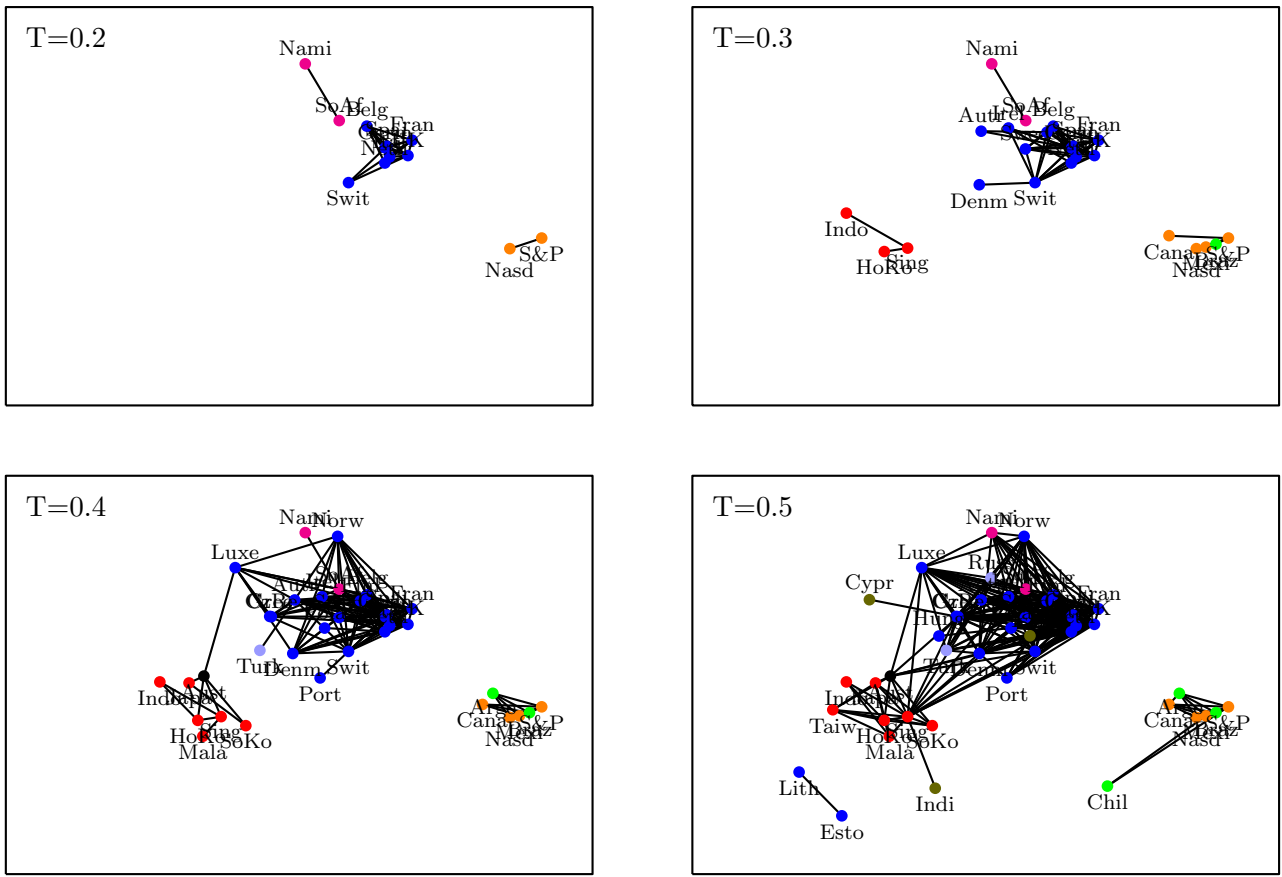
\noindent Fig. 1. Three dimensional view of the asset graphs for the first semester of 2007, with threshold ranging from $T=0.2$ to $T=0.5$.

\vskip 0.3 cm

\subsection{Second Semester, 2007}

Figure 2 shows the three dimensional view of the asset graphs for the second semester of 2007, from $T=0.2$ to $T=0.5$.

At $T=0.1$, we already have a North America cluster, comprised of S\&P and Nasdaq, a well formed European cluster, whose members are the indices from the UK, France, Germany, Italy, Belgium, and the Netherlands, and an African cluster, formed by South Africa and Namibia. For $T=0.2$, only the European cluster grows strong, with the addition of Switzerland, Austria, Sweden, Finland, Norway, and Spain, and the formation of more ties between the previous members of the cluster. For $T=0.3$, the now American cluster is joined by Canada, Mexico, Brazil, and Argentina. The European cluster has now the addition of Ireland, Luxembourg, Denmark, Greece, the Czech Republic, Poland, and Turkey. It also establishes strong connections with South Africa. Peru, Russia, and Cyprus also join the European cluster, although weakly. Peru connects with Denmark, Russia connects with Greece and Turkey, and Cyprus connects with Greece and Turkey. A new cluster, solely comprised of Pacific Asian indices plus the Australian index, is formed, comprised of Japan, Hong Kong, Taiwan, South Korea, Malaysia, Singapore, and Australia.

For $T=0.4$, the first connections appear between the American and the European clusters. The American cluster adds Chile, and connects with Peru. The European cluster is now even denser, and receives Portugal and Hungary. Connections are formed between the European cluster and Israel, and more connections are formed with the African cluster. The Pacific Asian cluster grows with the joining of Thailand and Indonesia, and it also grows denser. A single connection with the European cluster is made by Singapore and Norway, and a connection appears between Singapore and South Africa. At $T=0.5$, a great unification of the world markets occur, and all the networks join at several nodes. Nevertheless, the only indices that join the network are Philipines, connected with the Pacific Asian cluster, and New Zealand, connected with Australia. At $T=0.6$, most of the European connections are already formed, and the cluster receives the addition of Romania. India connects with all clusters at once, and new connections are formed among the existing clusters. At $T=0.7$, a threshold at which noise starts to become important, Colombia, Iceland, Vietnam, and Tunisia join the global network. We also witness the formation of an Arab cluster, comprised of Jordan and the United Arab Emirates, connected with the Pacific Asian cluster by Malaysia, where the Islam is the official religion. The last links are formed at $T=1.4$.

\vskip 0.4 cm



\vskip 2 cm

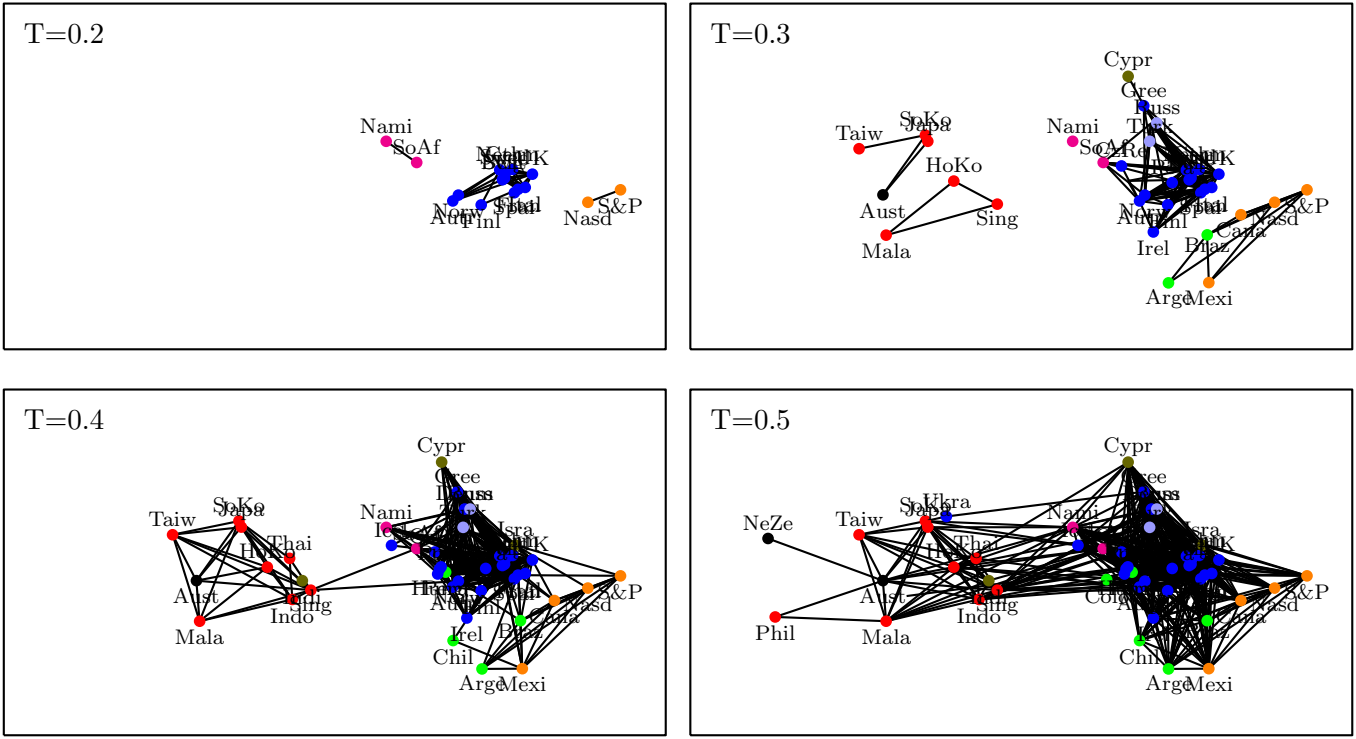
\noindent Fig. 2. Three dimensional view of the asset graphs for the second semester of 2007, with threshold ranging from $T=0.2$ to $T=0.5$.

\vskip 0.3 cm

\subsection{First Semester, 2008}

Figure 3 shows the three dimensional view of the asset graphs for the first semester of 2008, from $T=0.2$ to $T=0.5$.

%

\vskip 2 cm

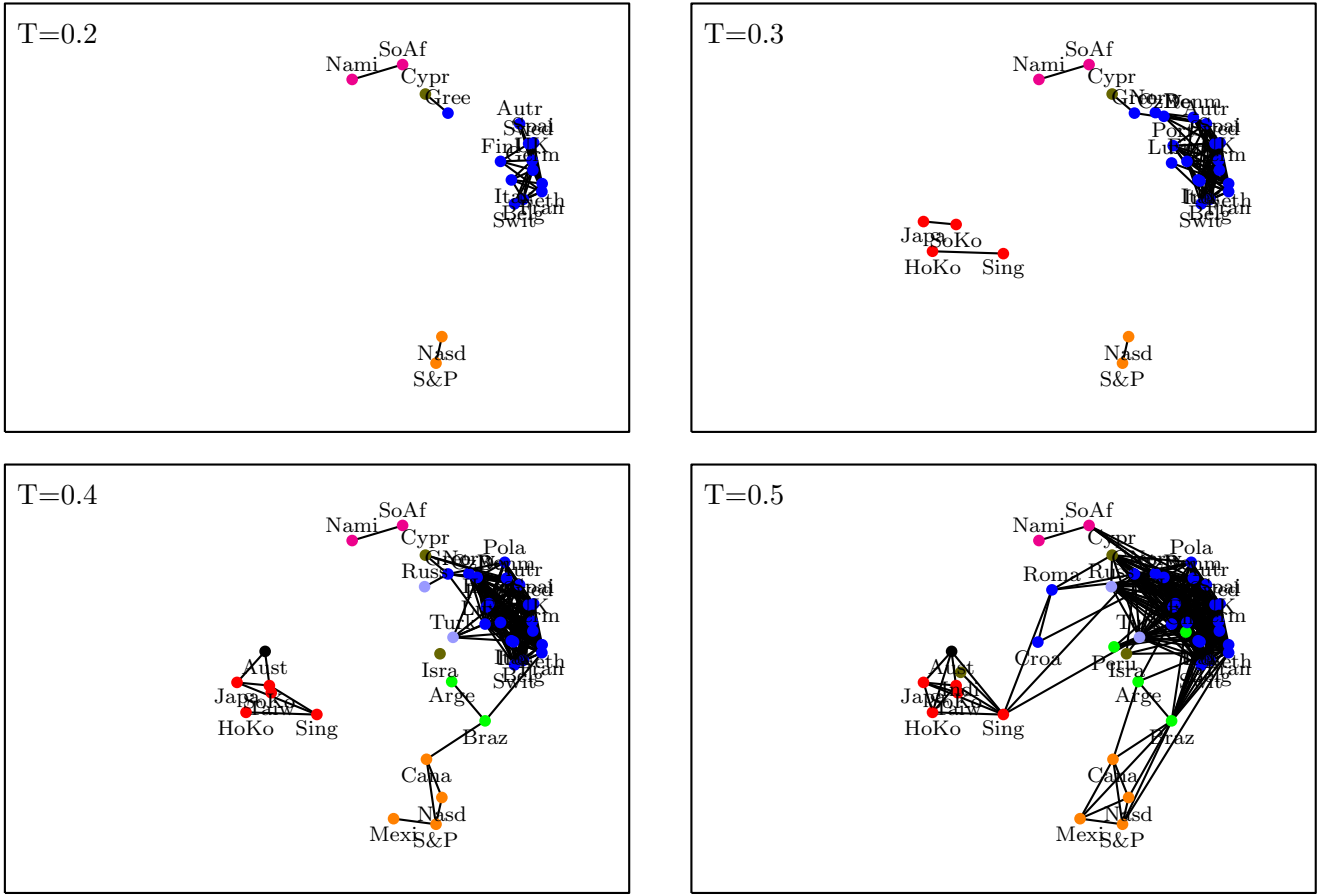
\noindent Fig. 3. Three dimensional view of the asset graphs for the first semester of 2008, with threshold ranging from $T=0.2$ to $T=0.5$.

\vskip 0.3 cm

At $T=0.1$, the North American cluster, for now formed by S\&P and Nasdaq, and the European cluster, comprised of France, Germany, the Netherlands, and Spain, are already present. For $T=0.2$, the European cluster grows with the addition of the UK, Switzerland, Austria, Italy, Belgium, Sweden, and Finland. We also have the connection of Greece and Cyprus, and of Namibia with South Africa. At $T=0.3$, the European cluster grows denser, and adds to itself Ireland, Luxembourg, Denmark, Norway, Portugal, and the Czech Republic. We also witness the formation of the Pacific Asian cluster, comprised of Japan, Hong Kong, South Korea, and Singapore.

For $T=0.4$, Canada, Mexico, Brazil, and Argentina join with S\&P and Nasdaq to form the American cluster. The European cluster grows denser and joins with Greece, Hungary, Poland, Russia, Turkey, and Cyprus. We also have Taiwan and Australia joining the Pacific Asian cluster. At $T=0.5$, the American and European clusters are linked via S\&P and France, and via Brazil, with many European indices. Chile and Peru join the European and not the American cluster, and the European cluster establishes its first connections with Israel, South Africa, and Singapore. Croatia and Romania also join the European cluster, and the Pacific Asian cluster strenghtens its connections, also receiving the addition of India, now.

At $T=0.6$, true globalization occurs, as the American and the European clusters establish many connections, the European cluster strengthens connections with the Pacific Asian one, and also with the African cluster (Namibia and South Africa). New countries that join the clusters are Colombia, Slovenia, and Ukraine (with Europe), and China, Thailand, Malaysia, Philipines, and New Zealand (with Pacific Asia). We also witness the formation of an Arab cluster, first comprised of the United Arab Emirates and Ohman.

At $T=0.7$, a threshold for which noise starts to dominate, we see many more connections between the already participating indices, and the inclusion of Bulgaria, Estonia, Latvia, and Lithuania among the European cluster, of Indonesia in the Pacific Asian cluster, and of Bahrein, Qatar, and Pakistan in the Arab cluster. Saudi Arabia, Egypt, and Kenya, also connect with other indices, but apparently at random. Slovakia, Serbia, Kazakhstan, Palestine, Lebanom, Jordan, Kuwait, Bangladesh, Mongolia, Vietnam, Tanzania, and Mauritius join at $T=0.8$. The last connections are made at $T=1.4$.

\subsection{Second Semester, 2008}

Figure 4 shows the three dimensional view of the asset graphs for the second semester of 2008, from $T=0.2$ to $T=0.5$.



\vskip 2 cm

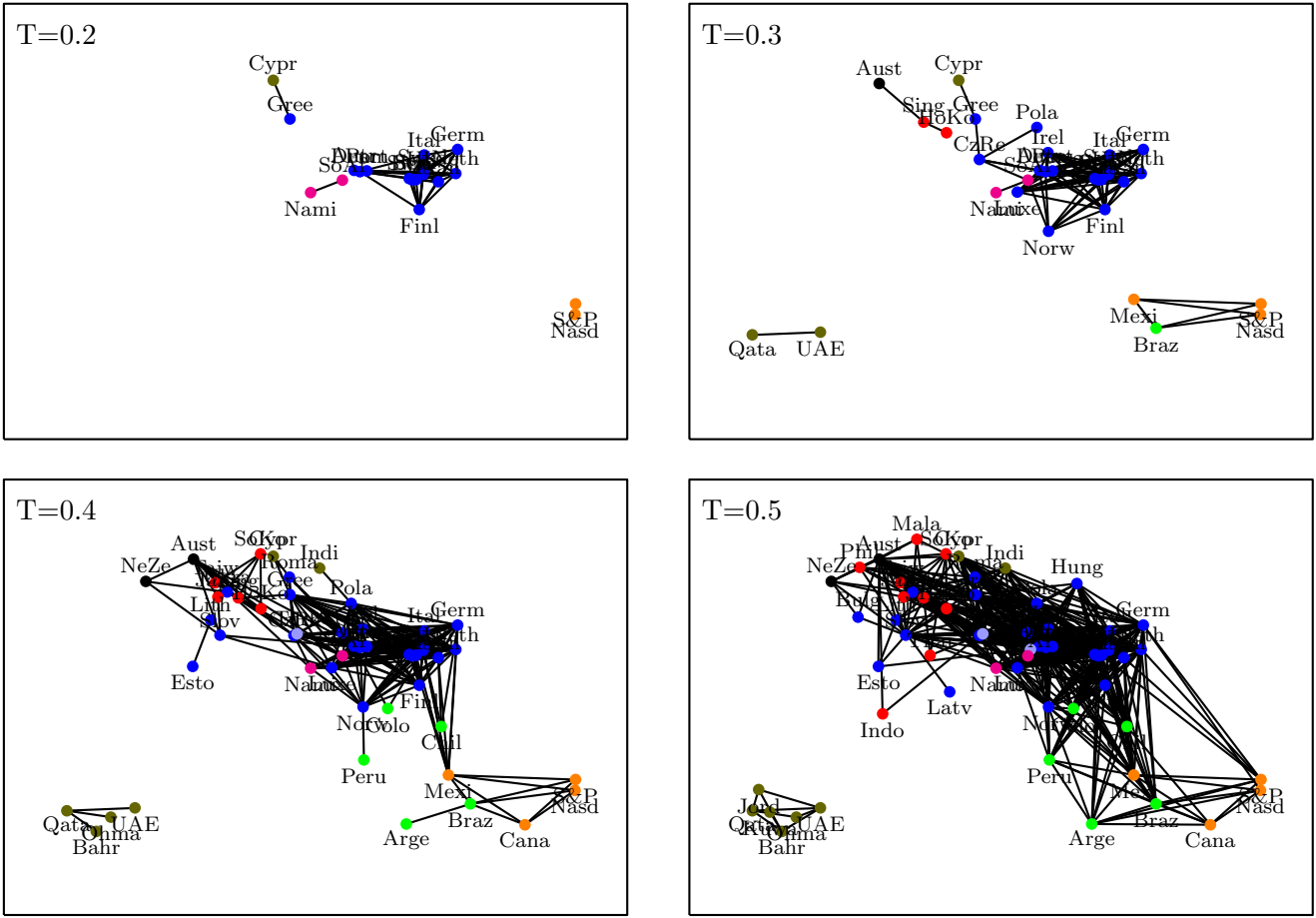
\noindent Fig. 4. Three dimensional view of the asset graphs for the second semester of 2008, with threshold ranging from $T=0.2$ to $T=0.5$.

\vskip 0.3 cm

We are now dealing with a period of crisis, so connections are stronger now. At $T=0.1$, we already have three clusters: the American cluster, formed by S\&P and Nasdaq, the European cluster, comprised of the UK, France, Germany, Switzerland, Italy, the Netherlands, Finland, and Spain, and the African cluster, formed by Namibia and South Africa. At $T=0.2$, the European cluster grows denser and adds to itself Austria, Belgium, Sweden, Denmark, and Portugal. Greece and Cyprus make a separate small cluster, which will soon join the European cluster at $T=0.3$. At this threshold, Mexico and Brazil join the American cluster (Canada does not, for now). Ireland, Luxembourg, Norway, Greece, the Czech Republic, and Poland join the European cluster, and South Africa establishes connections with this same cluster. A Pacific Asian cluster begins to form between Hong Kong, Singapore, and Australia.

For $T=0.4$, Canada joins the American cluster, and Mexico, of the same cluster, establishes connections with Europe. Argentina links with Brazil, and Chile, Colombia, and Peru connect with Europe. Croatia, Slovenia, Romania, Estonia, Lithuania, Ukraine, and Turkey also join the European cluster. India, Japan, Singapore, and Australia also establish connections with Europe. An Arab cluster is formed between Bahrein, Qatar, and Ohman. The Pacific Asian cluster grows with the addition of Japan, Taiwan, South Korea, and New Zealand. The African cluster links to both Europe and Pacific Asia. At $T=0.5$, many connections are made between the American and the European clusters, and more connections exhist between Europe, Pacific Asia, and Africa. New indices that join Europe are the ones from Hungary, Bulgaria, Latvia, and Russia, and Thailand, Malaysia, Indonesia, and Philipines join the Pacific Asian cluster. The Arab cluster is still isolated, but now has the addition of Jordan, Kuwait, and of the United Arab Emirates.

At $T=0.6$, more connections are made between the already participating indices, Israel joins Europe, which is also joined by Serbia and Bosnia and Herzegovina. Montenegro joins with the Pacific Asian cluster, which may already be an effect of random noise. The Arab network now makes many connections with both Europe and Pacific Asia, and Mauritius joins with Australia. A connection appears between Ghana and Botswana, establishing a second African cluster, disconneted from all others. From $T=0.7$ onwards, all clusters are connected, and random noise starts to dominate. The last connections are made at $T=1.3$.

\subsection{First Semester, 2009}

Figure 5 shows the three dimensional view of the asset graphs for the first semester of 2009, from $T=0.2$ to $T=0.5$.



\vskip 2 cm

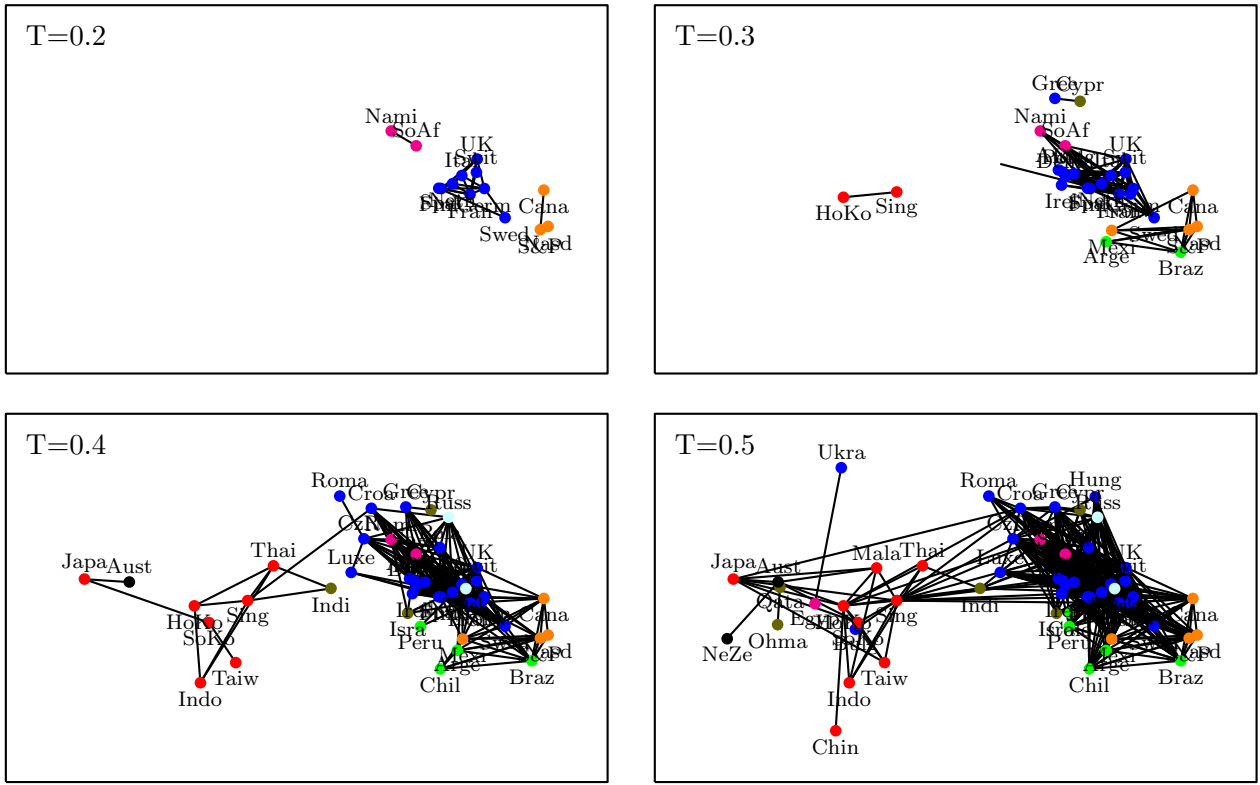
\noindent Fig. 5. Three dimensional view of the asset graphs for the first semester of 2009, with threshold ranging from $T=0.2$ to $T=0.5$.

\vskip 0.3 cm

At $T=0.1$, we have three clusters, the first comprised of the pair S\&P - Nasdaq, the second an European cluster comprised of France, Germany, Italy, the Netherlands, and Spain, and the third formed by the pair Namibia - South Africa. At $T=0.2$, the North American cluster is joined by Canada, and the European cluster is joined by the UK, Switzerland, Sweden, and Finland, and the pair Namibia - South Africa remains isolated. For $T=0.3$, Mexico, Brazil and Argentina join S\&P, Nasdaq and Canada to form the American cluster, the European cluster becomes denser and absorbs Ireland,Denmark, Norway, and Portugal. The South African cluster joins with the European one at many nodes, and Greece forms a cluster with Cyprus, still disconnected from Europe. In Pacific Asia, Hong Kong connects with Singapore.

For $T=0.4$, the American cluster, now joined by Chile, connects with the European one, which becomes even denser and is joined by Peru (a South American country), Austria, Belgium, Luxembourg, Greece, the Czech Republic, Coratia, Poland, Romania, Russia, Turkey, and Israel (a country from the Middle East with strong ties to Europe). The Pacific Asian cluster gorws with the addition of India, Japan, Taiwan, South Korea, Thailand, Indonesia, and Australia. At $T=0.5$, we have more connections between the American and the European clusters. The American cluster is joined by Colombia, and makes connections with the African cluster. The European cluster is joined by Hungary and makes connections with the Pacific Asian cluster, which is joined by China, Malaysia, and Indonesia. New Zealand connects with Australia, and the African cluster connects with the Pacific Asian one. We also have a connections between Egypt and Ukraine, probably the effect of noise.

At $T=0.6$, many more connections are formed. The only indices to join the now Global cluster comprised of the American, European, Pacific Asian and African clusters are those from Iceland (with Europe and Africa), Bosnia and Herzegovina, which joins Europe and Asia via Ukraine, which also joins the Global cluster, Bulgaria, and Kazakhstan. Jordan, Saudi Arabia, Qatar, the United Arab Emirates, and Ohman form a loosely knit Arab cluster which is also connected to the other clusters. Egypt now makes connections with the Global cluster at several nodes. For $T=0,7$, noise starts to take over, and we have many more connections, many of them apparently of random nature, and the joining of Costa Rica, Serbia, Slovenia, Macedonia, Estonia, Kuwait, Bahrein, Vietnam, Philippines, Nigeria, Kenya, and Mauritius. Bermuda, Jamaica, Malta, Slovakia, Palestine, and Lebanon join the Global cluster at $T=0.8$. Panama, Venezuela, Solvakia, Montenegro, Sri Lanka, Bangladesh, Morocco, Tunisia, Ghana, Tanzania, and Botswana only join at $T=0.9$. The network is fully integrated at $T=1.3$.

\subsection{Second Semester, 2009}

Figure 6 shows the three dimensional view of the asset graphs for the second semester of 2009, from $T=0.2$ to $T=0.5$.



\vskip 2 cm

\noindent Fig. 6. Three dimensional view of the asset graphs for the second semester of 2009, with threshold ranging from $T=0.2$ to $T=0.5$.

\vskip 0.3 cm

At $T=0.1$, we have two clusters: the first comprised of S\&P and Nasdaq, and the second, the European one, whose components are the UK, France, Germany, Italy, the Netherlands, and Spain. At $T=0.2$, the American cluster remains unchanged and the European cluster is joined by Switzerland, Belgium, Sweden, Finland, Norway, and Portugal. Greece forms a pair with Cyprus and Namibia forms a pair with South Africa. For $T=0.3$, the American cluster is joined by Canada and connects with Europe through France and Germany, Brazil and Argentina form a pair, and the European cluster is joined by Austria, Denmark, Hungary, Poland, and Russia, which connects to many nodes in the European cluster. Also, a pair is formed between Hong Kong and Singapore.

At $T=0,4$, Brazil, Argentina, and Peru join the American cluster, which makes even more connections with the European one. The latter grows denser and is joined by Luxembourg, the Czech Republic, and Poland and connects with the African cluster (Namibia and South Africa) at several nodes. Estonia and Lithuania form an isolated pair, Qatar and the United Arab Emirates form the seed of an Arab cluster, Japan and Hong Kong connect with Australia, and Hong Kong connects with Thailand, forming an Asian Pacific cluster. There is also a strange connection between Hong Kong and Kazakhstan.

For $T=0.5$, Chile joins the American cluster, which makes more connections to Europe, including Russia, and which also connects with the African cluster. The European cluster grows even denser and is joined by Greece and Cyprus, Romania, Bulgaria, and Turkey. Joined via Bulgaria to the European cluster are the indices from Lithuania (and so Estonia) and Kazakhstan. The Pacific Asian cluster is now joined by India, Taiwan, South Korea, Malaysia, Singapore, and Indonesia, growing to a large cluster. Singapore, Indonesia, and India make some marginal connections with the European cluster, thus weakly connecting the Pacific Asian cluster with Europe, and Hong Kong connects with South Africa. The Arab cluster (Qatar plus the United Arab Emirates) is joined by Egypt and Ohman.

At $T=0.6$, Colombia becomes part of the American cluster, and Venezuela connects with Spain and Portugal. Serbia, Ukraine, Israel, and Saudi Arabia connect with the Global network at many points. Jordan joins the Arab cluster, China connects with Hong Kong, and New Zealand connects with Australia and Bulgaria. At $T=0.7$, a threshold in which noise becomes more important, Iceland, Slovakia, Croatia, Slovenia, Bosnia and Herzegovina, Montenegro, Macedonia, and Latvia, join the European network and, at some nodes, the Global network in general. Palestine, Lebanon, Kuwait, Bahrein, and Pakistan all join the Arab cluster, also establishing some connections with the Global network. Panama, Morocco, Tunisia, Ghana, Nigeria, Tanzania, and Mauritius establish some connections, most probably the result of noise. For larger thresholds, it becomes nearly impossible to distinguish true connections from random ones, and the last connections occur at $T=1.3$.

\subsection{First Semester, 2010}

Figure 7 shows the three dimensional view of the asset graphs for the first semester of 2010, from $T=0.2$ to $T=0.5$.

At $T=0.1$, there already are three clusters: the American cluster, formed by the pair S\&P - Nasdaq, the European cluster, formed by the UK, France, Germany, Italy, Belgium, and the Netherlands, and the African cluster, formed by the pair Namibia - South Africa. At $T=0.2$, the European cluster grows denser and is joined by Switzerland, Austria, Sweden, Denmark, Finland, Norway, Spain, and Portugal. Greece forms a pair with Cyprus and the remaining clusters stay the same.
For $T=0.3$, Canada and Argentina join the American cluster. Ireland, the Czech Republic, Hungary, Poland, Russia, and Turkey join the European cluster, which also connects with the African one. A Pacific Asian cluster is formed, comprised by India, Japan, Hong Kong, Taiwan, South Korea, Singapore, and Australia. At $T=0.4$, the American cluster is joined by Mexico and Brazil, and establishes a first link with Europe through Sweden. Peru links with the American cluster, but also to the European one through Switzerland. Luxembourg and Greece join the European cluster, which connects to the Pacific Asian one, and a pair is formed between Estonia and Lithuania. The Pacific Asian cluster is joined by Indonesia and connects with Kazakhstan.

For $T=0.5$, Chile joins the American network, which makes many more connections with Europe and a single connection with Africa. The European cluster grows even denser and is joined by Croatia, Romania, by the pair Estonia - Lithuania, Ukraine, and Israel. Thailand, Malaysia, Philippines, and New Zealand join the Pacific Asian network, which now connects with Africa and makes many more connections with Europe. At $T=0.6$, many more connections are made within and between clusters. Colombia joins the American cluster. Saudi Arabia, Qatar, the United Arab Emirates, and Ohman connect with Europe, and the last three ones compose an Arab cluster which is also connected with the Pacific Asian one. China joins the Pacific Asian cluster, and Egypt connects with all clusters.

At the beginning of the noisy region, at $T=0.7$, Iceland and Bulgaria join the Global network at several nodes, Serbia and Slovenia join Europe, and Macedonia connects with Egypt. Jordan, Kuwait, and Bahrein join the Arab network, and Vietnam joins the Pacific Asian cluster. For higher thresholds, many connections are established apparently at random. Nigeria, Kenya, Tanzania, and Mauritius join the Global cluster at $T=0.8$, and Venezuela, Palestine, Lebanon,Pakistan, Sri Lanka, Bangladesh, Mongolia, Morocco, Tunisia, Ghana, and Botswana join the Global cluster at $T=0.9$. No new connections are established beyond $T=1.3$.

\vskip 0.1 cm



\vskip 1.9 cm

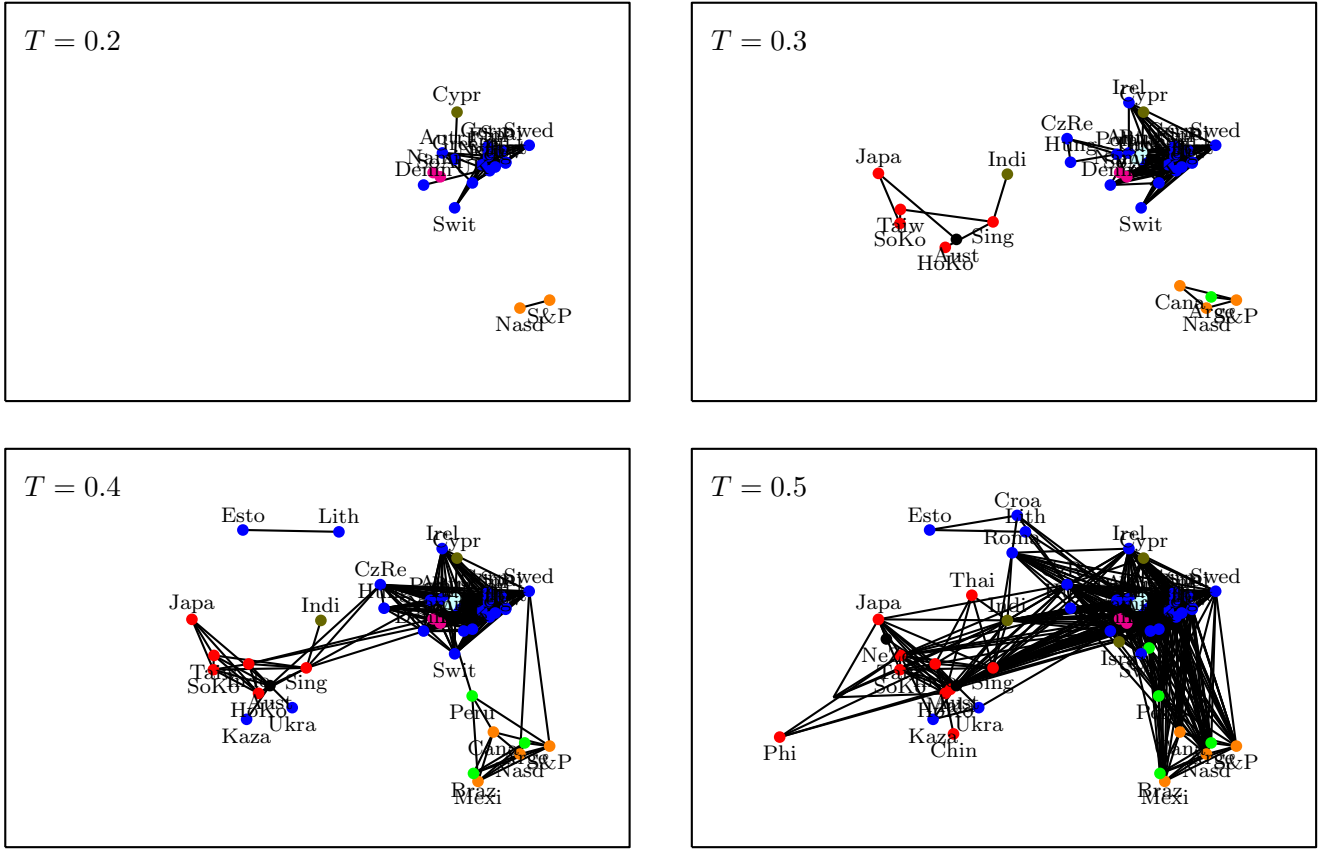
\noindent Fig. 7. Three dimensional view of the asset graphs for the first semester of 2010, with threshold ranging from $T=0.2$ to $T=0.5$.

\vskip 0.3 cm

\subsection{Second Semester, 2010}

Figure 8 shows the three dimensional view of the asset graphs for the second semester of 2010, from $T=0.2$ to $T=0.5$.

The second semester of 2010 was a time of lower volatility and also of lower correlation, so connections between indices are also weaker. At $T=0.1$, the only connections are those between S\&P and Nasdaq, and between France, Germany, Belgium, and the Netherlands, forming the seeds of an American and an European clusters. For $T=0.2$, the American cluster remains the same and the European cluster grows denser and is joined by the UK, Switzerland, Austria, Italy, Finland and Spain. The pair Namibia - South Africa, the African cluster, is formed at this threshold. At $T=0.3$, the European cluster is joined by Ireland, Sweden, Norway, Portugal, and Greece, and makes one single connection with the African cluster.

For $T=0.4$, the American cluster finally grows to include Canada, Mexico, Brazil, and Argentina, and also connects with Europe. The European cluster keeps growing and now includes the Czech Republic, Poland, Russia, and Cyprus (through Greece). The Pacific Asian cluster begins to form with Hong Kong, Taiwan, and Australia. $T=0.5$ brings Colombia and Peru into the American cluster, Luxembourg, Hungary, Turkey, and Israel into the European cluster. The Pacific Asian cluster grows with the addition of Japan, China (connected with Hong Kong), South Korea, Singapore, and Indonesia.

For $T=0.6$, Chile joins the American cluster, and Lithuania, Ukraine, Saudi Arabia, and India connect with the European cluster. Some connections form between the Pacifi Asian and European clusters, and Thailand and Singapore join the Pacific Asian cluster. Noise is already influent at $T=0.7$, where the American network finally establishes connections with the African cluster. Romania and Estonia join the European cluster, Qatar and Ohman connect with the Pacific Asian cluster and not between themselves. Philippines also join the Pacific Asian cluster, and Egypt establishes connections with both the European and the Pacific Asian clusters. Many more connections are made from $T=0.8$ and beyond, but it is practically impossible to separate meaningful connections from random ones, anymore. The last connections occur at $T=1.3$.

\vskip 0.4 cm



\vskip 2.3 cm

\noindent Fig. 8. Three dimensional view of the asset graphs for the second semester of 2010, with threshold ranging from $T=0.2$ to $T=0.5$.

\vskip 0.3 cm

\subsection{Whole data and Pearson correlation}

I now shall analyze two types of criticism the previous results might face. The first one is the use of possibly too few data in order to generate the networks, which might introduce much statistical noise to the results obtained. The second is the use of the Spearman rank correlation instead of the Pearson correlation.

In order to investigate the effects of the amount of data used and the choice of correlation, I now make two asset graphs, both based on the whole data (1015 days) from 2007 to 2010, the first one based on the Spearman rank correlation and the second based on the Pearson correlation. Since the two measures don't have necessarily the same values, I chose $T=0.5$ for the threshold of the Spearman correlation and $T=0.452$ for the threshold of the Pearson correlation. The first threshold is a compromise between visibility and proximity to the noise level, and the second value minimizes the differences between both asset graphs in terms of the number of connections in each graph. In order to facilitate the comparison, the coordinates based on the Spearman correlation are being used for both graphs.

Note that there are nearly no diffferences between both graphs. All clusters are preserved at all levels for other thresholds as well. So, one may conclude that most connections between the indices are basically linear and that the choice between using Pearson or Spearman correlation doesn't modify the major results.



\vskip 2.3 cm

\noindent Fig. 9. Three dimensional view of the asset graphs for the whole period of data (2007 to 2010). The figure on the left is based on the Spearman rank correlation with threshold $T=0.5$, and the figure on the right is based on the Pearson correlation with threshold $T=0.452$.

\vskip 0.3 cm

Also, if one compares Figure 9 (left) with the asset graphs obtained for the separate semesters, one can see the connections are quite stable in time, which is a discussion that shall be pursued further in Section 3.

\subsection{Lagging indices}

A third issue that could be raised is the use of same-day data for calculating correlations, when in truth markets don't operate at the same time. As an example, there is no intersection between the opening hours of the Tokyo Stock Exchange and the New York Stock Exchange. The issue of to lag or to not lag data is an important one, which shall be resolved in some future work. What can be done for now is to lag some indices so that we will compare them to the non-lagged indices of the previous day. In order to decide which indices to lag, we shall take a look at the eigenvector of the second highest eigenvalue of the correlation matrix between the indices. As stated in \cite{clusters1} and in many other works, the eigenvector of the first highest eigenvalue is related with a market mode which explains the common movement of the indices. Such eigenvalue is set far apart from all others, as shown in Figure 10, being much higher than it would be expected from those of a correlation matrix based on noise, and its corresponding eigenvector shows equal contributions from nearly every index being studied. In Figure 10, the noise region is drawn as a shady area. Here, this area was calculated by randomizing each time series of each index so as to eliminate any correlation between them, but keeping their frequency distributions intact. The result was obtained from 1000 simulations using randomized data, but is very similar to the theoretical result obtained using Random Matrix Theory \cite{rmt1}.

\begin{pspicture}(-1.5,-0.5)(7,2.4)
\psset{xunit=1.5,yunit=1.5}
\pspolygon*[linecolor=lightgray](0.050,0)(0.050,1.2)(1.66,1.2)(1.66,0)
\psline{->}(0,0)(8,0)  \rput(8.15,0){$\lambda $} \psline[linecolor=white,linewidth=2pt](6.4,0)(6.6,0) \psline(6.3,-0.05)(6.5,0.05) \psline(6.5,-0.05)(6.7,0.05) \scriptsize \psline(1,-0.05)(1,0.05) \rput(1,-0.15){1} \psline(2,-0.05)(2,0.05) \rput(2,-0.15){2} \psline(3,-0.05)(3,0.05) \rput(3,-0.15){3} \psline(4,-0.05)(4,0.05) \rput(4,-0.15){4} \psline(5,-0.05)(5,0.05) \rput(5,-0.15){5} \psline(6,-0.05)(6,0.05) \rput(6,-0.15){6} \psline(7,-0.05)(7,0.05) \rput(7,-0.15){24}
\psline[linewidth=1pt](0.0403,0)(0.0403,1) \psline[linewidth=1pt](0.0605,0)(0.0605,1) \psline[linewidth=1pt](0.0813,0)(0.0813,1) \psline[linewidth=1pt](0.0840,0)(0.0840,1) \psline[linewidth=1pt](0.0997,0)(0.0997,1) \psline[linewidth=1pt](0.1221,0)(0.1221,1) \psline[linewidth=1pt](0.1295,0)(0.1295,1) \psline[linewidth=1pt](0.1625,0)(0.1625,1) \psline[linewidth=1pt](0.1720,0)(0.1720,1) \psline[linewidth=1pt](0.1763,0)(0.1763,1) \psline[linewidth=1pt](0.1886,0)(0.1886,1) \psline[linewidth=1pt](0.1956,0)(0.1956,1) \psline[linewidth=1pt](0.2088,0)(0.2088,1) \psline[linewidth=1pt](0.2328,0)(0.2328,1) \psline[linewidth=1pt](0.2428,0)(0.2428,1) \psline[linewidth=1pt](0.2470,0)(0.2470,1) \psline[linewidth=1pt](0.2580,0)(0.2580,1) \psline[linewidth=1pt](0.2645,0)(0.2645,1) \psline[linewidth=1pt](0.2806,0)(0.2806,1) \psline[linewidth=1pt](0.3065,0)(0.3065,1) \psline[linewidth=1pt](0.3178,0)(0.3178,1) \psline[linewidth=1pt](0.3300,0)(0.3300,1) \psline[linewidth=1pt](0.3416,0)(0.3416,1) \psline[linewidth=1pt](0.3542,0)(0.3542,1) \psline[linewidth=1pt](0.3641,0)(0.3641,1) \psline[linewidth=1pt](0.3671,0)(0.3671,1) \psline[linewidth=1pt](0.3854,0)(0.3854,1) \psline[linewidth=1pt](0.3970,0)(0.3970,1) \psline[linewidth=1pt](0.4161,0)(0.4161,1) \psline[linewidth=1pt](0.4331,0)(0.4331,1) \psline[linewidth=1pt](0.4436,0)(0.4436,1) \psline[linewidth=1pt](0.4530,0)(0.4530,1) \psline[linewidth=1pt](0.4734,0)(0.4734,1) \psline[linewidth=1pt](0.4878,0)(0.4878,1) \psline[linewidth=1pt](0.4915,0)(0.4915,1) \psline[linewidth=1pt](0.5042,0)(0.5042,1) \psline[linewidth=1pt](0.5071,0)(0.5071,1) \psline[linewidth=1pt](0.5106,0)(0.5106,1) \psline[linewidth=1pt](0.5330,0)(0.5330,1) \psline[linewidth=1pt](0.5335,0)(0.5335,1) \psline[linewidth=1pt](0.5509,0)(0.5509,1) \psline[linewidth=1pt](0.5654,0)(0.5654,1) \psline[linewidth=1pt](0.5769,0)(0.5769,1) \psline[linewidth=1pt](0.5946,0)(0.5946,1) \psline[linewidth=1pt](0.6160,0)(0.6160,1) \psline[linewidth=1pt](0.6292,0)(0.6292,1) \psline[linewidth=1pt](0.6485,0)(0.6485,1) \psline[linewidth=1pt](0.6554,0)(0.6554,1) \psline[linewidth=1pt](0.6579,0)(0.6579,1) \psline[linewidth=1pt](0.6713,0)(0.6713,1) \psline[linewidth=1pt](0.7045,0)(0.7045,1) \psline[linewidth=1pt](0.7118,0)(0.7118,1) \psline[linewidth=1pt](0.7353,0)(0.7353,1) \psline[linewidth=1pt](0.7474,0)(0.7474,1) \psline[linewidth=1pt](0.7615,0)(0.7615,1) \psline[linewidth=1pt](0.7748,0)(0.7748,1) \psline[linewidth=1pt](0.7802,0)(0.7802,1) \psline[linewidth=1pt](0.7971,0)(0.7971,1) \psline[linewidth=1pt](0.8225,0)(0.8225,1) \psline[linewidth=1pt](0.8312,0)(0.8312,1) \psline[linewidth=1pt](0.8504,0)(0.8504,1) \psline[linewidth=1pt](0.8704,0)(0.8704,1) \psline[linewidth=1pt](0.8730,0)(0.8730,1) \psline[linewidth=1pt](0.8881,0)(0.8881,1) \psline[linewidth=1pt](0.9018,0)(0.9018,1) \psline[linewidth=1pt](0.9290,0)(0.9290,1) \psline[linewidth=1pt](0.9470,0)(0.9470,1) \psline[linewidth=1pt](0.9506,0)(0.9506,1) \psline[linewidth=1pt](0.9670,0)(0.9670,1) \psline[linewidth=1pt](0.9946,0)(0.9946,1) \psline[linewidth=1pt](1.0124,0)(1.0124,1) \psline[linewidth=1pt](1.0261,0)(1.0261,1) \psline[linewidth=1pt](1.0326,0)(1.0326,1) \psline[linewidth=1pt](1.0469,0)(1.0469,1) \psline[linewidth=1pt](1.0648,0)(1.0648,1) \psline[linewidth=1pt](1.0786,0)(1.0786,1) \psline[linewidth=1pt](1.1201,0)(1.1201,1) \psline[linewidth=1pt](1.1476,0)(1.1476,1) \psline[linewidth=1pt](1.1713,0)(1.1713,1) \psline[linewidth=1pt](1.1992,0)(1.1992,1) \psline[linewidth=1pt](1.2302,0)(1.2302,1) \psline[linewidth=1pt](1.2366,0)(1.2366,1) \psline[linewidth=1pt](1.2635,0)(1.2635,1) \psline[linewidth=1pt](1.2962,0)(1.2962,1) \psline[linewidth=1pt](1.3228,0)(1.3228,1) \psline[linewidth=1pt](1.3426,0)(1.3426,1) \psline[linewidth=1pt](1.4493,0)(1.4493,1) \psline[linewidth=1pt](1.8616,0)(1.8616,1) \psline[linewidth=1pt](2.2265,0)(2.2265,1) \psline[linewidth=1pt](2.5492,0)(2.5492,1) \psline[linewidth=1pt](5.8262,0)(5.8262,1) \psline[linewidth=1pt](7.2941,0)(7.2941,1)
\end{pspicture}

\noindent Fig. 10. Eigenvalues of the correlation for data from 2007 to 2010,  in order of magnitude. The shaded area corresponds to the eigenvalues predicted for randomized data.

\vskip 0.3 cm

The second highest eigenvalue is set not as far apart from the others in the region considered as noise as the highest one. Its corresponding eigenvector, $e_2$, (Figure 11) reveals a part of the internal structure of the correlations that is peculiar to indices of stock exchanges worldwide, which operate at different intervals of time \cite{clusters1}. In the figure, white bars correspond to positive signs in the eigenvector, and gray bars correspond to negative signs in the eigenvector.

\begin{pspicture}(-0.1,0)(3.5,2.8)
\psset{xunit=0.18,yunit=4.5}
\pspolygon*[linecolor=white](0.5,0)(0.5,0.190)(1.5,0.190)(1.5,0)
\pspolygon*[linecolor=white](1.5,0)(1.5,0.177)(2.5,0.177)(2.5,0)
\pspolygon*[linecolor=white](2.5,0)(2.5,0.146)(3.5,0.146)(3.5,0)
\pspolygon*[linecolor=white](3.5,0)(3.5,0.122)(4.5,0.122)(4.5,0)
\pspolygon*[linecolor=gray](4.5,0)(4.5,0.026)(5.5,0.026)(5.5,0)
\pspolygon*[linecolor=gray](5.5,0)(5.5,0.014)(6.5,0.014)(6.5,0)
\pspolygon*[linecolor=gray](6.5,0)(6.5,0.001)(7.5,0.001)(7.5,0)
\pspolygon*[linecolor=gray](7.5,0)(7.5,0.041)(8.5,0.041)(8.5,0)
\pspolygon*[linecolor=white](8.5,0)(8.5,0.138)(9.5,0.138)(9.5,0)
\pspolygon*[linecolor=white](9.5,0)(9.5,0.116)(10.5,0.116)(10.5,0)
\pspolygon*[linecolor=white](10.5,0)(10.5,0.086)(11.5,0.086)(11.5,0)
\pspolygon*[linecolor=white](11.5,0)(11.5,0.048)(12.5,0.048)(12.5,0)
\pspolygon*[linecolor=gray](12.5,0)(12.5,0.021)(13.5,0.021)(13.5,0)
\pspolygon*[linecolor=white](13.5,0)(13.5,0.043)(14.5,0.043)(14.5,0)
\pspolygon*[linecolor=white](14.5,0)(14.5,0.124)(15.5,0.124)(15.5,0)
\pspolygon*[linecolor=white](15.5,0)(15.5,0.067)(16.5,0.067)(16.5,0)
\pspolygon*[linecolor=white](16.5,0)(16.5,0.132)(17.5,0.132)(17.5,0)
\pspolygon*[linecolor=white](17.5,0)(17.5,0.132)(18.5,0.132)(18.5,0)
\pspolygon*[linecolor=white](18.5,0)(18.5,0.103)(19.5,0.103)(19.5,0)
\pspolygon*[linecolor=white](19.5,0)(19.5,0.044)(20.5,0.044)(20.5,0)
\pspolygon*[linecolor=white](20.5,0)(20.5,0.116)(21.5,0.116)(21.5,0)
\pspolygon*[linecolor=gray](21.5,0)(21.5,0.017)(22.5,0.017)(22.5,0)
\pspolygon*[linecolor=white](22.5,0)(22.5,0.100)(23.5,0.100)(23.5,0)
\pspolygon*[linecolor=white](23.5,0)(23.5,0.117)(24.5,0.117)(24.5,0)
\pspolygon*[linecolor=white](24.5,0)(24.5,0.001)(25.5,0.001)(25.5,0)
\pspolygon*[linecolor=white](25.5,0)(25.5,0.124)(26.5,0.124)(26.5,0)
\pspolygon*[linecolor=white](26.5,0)(26.5,0.037)(27.5,0.037)(27.5,0)
\pspolygon*[linecolor=white](27.5,0)(27.5,0.091)(28.5,0.091)(28.5,0)
\pspolygon*[linecolor=white](28.5,0)(28.5,0.075)(29.5,0.075)(29.5,0)
\pspolygon*[linecolor=gray](29.5,0)(29.5,0.032)(30.5,0.032)(30.5,0)
\pspolygon*[linecolor=white](30.5,0)(30.5,0.119)(31.5,0.119)(31.5,0)
\pspolygon*[linecolor=white](31.5,0)(31.5,0.059)(32.5,0.059)(32.5,0)
\pspolygon*[linecolor=gray](32.5,0)(32.5,0.009)(33.5,0.009)(33.5,0)
\pspolygon*[linecolor=gray](33.5,0)(33.5,0.032)(34.5,0.032)(34.5,0)
\pspolygon*[linecolor=gray](34.5,0)(34.5,0.047)(35.5,0.047)(35.5,0)
\pspolygon*[linecolor=white](35.5,0)(35.5,0.042)(36.5,0.042)(36.5,0)
\pspolygon*[linecolor=gray](36.5,0)(36.5,0.137)(37.5,0.137)(37.5,0)
\pspolygon*[linecolor=gray](37.5,0)(37.5,0.070)(38.5,0.070)(38.5,0)
\pspolygon*[linecolor=gray](38.5,0)(38.5,0.141)(39.5,0.141)(39.5,0)
\pspolygon*[linecolor=gray](39.5,0)(39.5,0.072)(40.5,0.072)(40.5,0)
\pspolygon*[linecolor=gray](40.5,0)(40.5,0.061)(41.5,0.061)(41.5,0)
\pspolygon*[linecolor=gray](41.5,0)(41.5,0.080)(42.5,0.080)(42.5,0)
\pspolygon*[linecolor=white](42.5,0)(42.5,0.039)(43.5,0.039)(43.5,0)
\pspolygon*[linecolor=gray](43.5,0)(43.5,0.079)(44.5,0.079)(44.5,0)
\pspolygon*[linecolor=gray](44.5,0)(44.5,0.140)(45.5,0.140)(45.5,0)
\pspolygon*[linecolor=gray](45.5,0)(45.5,0.141)(46.5,0.141)(46.5,0)
\pspolygon*[linecolor=gray](46.5,0)(46.5,0.110)(47.5,0.110)(47.5,0)
\pspolygon*[linecolor=gray](47.5,0)(47.5,0.149)(48.5,0.149)(48.5,0)
\pspolygon*[linecolor=gray](48.5,0)(48.5,0.129)(49.5,0.129)(49.5,0)
\pspolygon*[linecolor=white](49.5,0)(49.5,0.039)(50.5,0.039)(50.5,0)
\pspolygon*[linecolor=gray](50.5,0)(50.5,0.149)(51.5,0.149)(51.5,0)
\pspolygon*[linecolor=white](51.5,0)(51.5,0.018)(52.5,0.018)(52.5,0)
\pspolygon*[linecolor=gray](52.5,0)(52.5,0.004)(53.5,0.004)(53.5,0)
\pspolygon*[linecolor=white](53.5,0)(53.5,0.039)(54.5,0.039)(54.5,0)
\pspolygon*[linecolor=gray](54.5,0)(54.5,0.033)(55.5,0.033)(55.5,0)
\pspolygon*[linecolor=gray](55.5,0)(55.5,0.065)(56.5,0.065)(56.5,0)
\pspolygon*[linecolor=gray](56.5,0)(56.5,0.121)(57.5,0.121)(57.5,0)
\pspolygon*[linecolor=gray](57.5,0)(57.5,0.063)(58.5,0.063)(58.5,0)
\pspolygon*[linecolor=gray](58.5,0)(58.5,0.084)(59.5,0.084)(59.5,0)
\pspolygon*[linecolor=gray](59.5,0)(59.5,0.070)(60.5,0.070)(60.5,0)
\pspolygon*[linecolor=gray](60.5,0)(60.5,0.154)(61.5,0.154)(61.5,0)
\pspolygon*[linecolor=gray](61.5,0)(61.5,0.139)(62.5,0.139)(62.5,0)
\pspolygon*[linecolor=gray](62.5,0)(62.5,0.121)(63.5,0.121)(63.5,0)
\pspolygon*[linecolor=gray](63.5,0)(63.5,0.059)(64.5,0.059)(64.5,0)
\pspolygon*[linecolor=gray](64.5,0)(64.5,0.087)(65.5,0.087)(65.5,0)
\pspolygon*[linecolor=gray](65.5,0)(65.5,0.048)(66.5,0.048)(66.5,0)
\pspolygon*[linecolor=gray](66.5,0)(66.5,0.029)(67.5,0.029)(67.5,0)
\pspolygon*[linecolor=gray](67.5,0)(67.5,0.213)(68.5,0.213)(68.5,0)
\pspolygon*[linecolor=gray](68.5,0)(68.5,0.182)(69.5,0.182)(69.5,0)
\pspolygon*[linecolor=gray](69.5,0)(69.5,0.111)(70.5,0.111)(70.5,0)
\pspolygon*[linecolor=gray](70.5,0)(70.5,0.033)(71.5,0.033)(71.5,0)
\pspolygon*[linecolor=gray](71.5,0)(71.5,0.176)(72.5,0.176)(72.5,0)
\pspolygon*[linecolor=gray](72.5,0)(72.5,0.176)(73.5,0.176)(73.5,0)
\pspolygon*[linecolor=gray](73.5,0)(73.5,0.115)(74.5,0.115)(74.5,0)
\pspolygon*[linecolor=gray](74.5,0)(74.5,0.091)(75.5,0.091)(75.5,0)
\pspolygon*[linecolor=gray](75.5,0)(75.5,0.164)(76.5,0.164)(76.5,0)
\pspolygon*[linecolor=gray](76.5,0)(76.5,0.155)(77.5,0.155)(77.5,0)
\pspolygon*[linecolor=gray](77.5,0)(77.5,0.153)(78.5,0.153)(78.5,0)
\pspolygon*[linecolor=gray](78.5,0)(78.5,0.186)(79.5,0.186)(79.5,0)
\pspolygon*[linecolor=gray](79.5,0)(79.5,0.208)(80.5,0.208)(80.5,0)
\pspolygon*[linecolor=gray](80.5,0)(80.5,0.179)(81.5,0.179)(81.5,0)
\pspolygon*[linecolor=gray](81.5,0)(81.5,0.057)(82.5,0.057)(82.5,0)
\pspolygon*[linecolor=gray](82.5,0)(82.5,0.040)(83.5,0.040)(83.5,0)
\pspolygon*[linecolor=gray](83.5,0)(83.5,0.138)(84.5,0.138)(84.5,0)
\pspolygon*[linecolor=gray](84.5,0)(84.5,0.017)(85.5,0.017)(85.5,0)
\pspolygon*[linecolor=gray](85.5,0)(85.5,0.041)(86.5,0.041)(86.5,0)
\pspolygon*[linecolor=gray](86.5,0)(86.5,0.057)(87.5,0.057)(87.5,0)
\pspolygon*[linecolor=gray](87.5,0)(87.5,0.006)(88.5,0.006)(88.5,0)
\pspolygon*[linecolor=gray](88.5,0)(88.5,0.007)(89.5,0.007)(89.5,0)
\pspolygon*[linecolor=gray](89.5,0)(89.5,0.013)(90.5,0.013)(90.5,0)
\pspolygon*[linecolor=white](90.5,0)(90.5,0.007)(91.5,0.007)(91.5,0)
\pspolygon*[linecolor=gray](91.5,0)(91.5,0.104)(92.5,0.104)(92.5,0)
\pspolygon(0.5,0)(0.5,0.190)(1.5,0.190)(1.5,0)
\pspolygon(1.5,0)(1.5,0.177)(2.5,0.177)(2.5,0)
\pspolygon(2.5,0)(2.5,0.146)(3.5,0.146)(3.5,0)
\pspolygon(3.5,0)(3.5,0.122)(4.5,0.122)(4.5,0)
\pspolygon(4.5,0)(4.5,0.026)(5.5,0.026)(5.5,0)
\pspolygon(5.5,0)(5.5,0.014)(6.5,0.014)(6.5,0)
\pspolygon(6.5,0)(6.5,0.001)(7.5,0.001)(7.5,0)
\pspolygon(7.5,0)(7.5,0.041)(8.5,0.041)(8.5,0)
\pspolygon(8.5,0)(8.5,0.138)(9.5,0.138)(9.5,0)
\pspolygon(9.5,0)(9.5,0.116)(10.5,0.116)(10.5,0)
\pspolygon(10.5,0)(10.5,0.086)(11.5,0.086)(11.5,0)
\pspolygon(11.5,0)(11.5,0.048)(12.5,0.048)(12.5,0)
\pspolygon(12.5,0)(12.5,0.021)(13.5,0.021)(13.5,0)
\pspolygon(13.5,0)(13.5,0.043)(14.5,0.043)(14.5,0)
\pspolygon(14.5,0)(14.5,0.124)(15.5,0.124)(15.5,0)
\pspolygon(15.5,0)(15.5,0.067)(16.5,0.067)(16.5,0)
\pspolygon(16.5,0)(16.5,0.132)(17.5,0.132)(17.5,0)
\pspolygon(17.5,0)(17.5,0.132)(18.5,0.132)(18.5,0)
\pspolygon(18.5,0)(18.5,0.103)(19.5,0.103)(19.5,0)
\pspolygon(19.5,0)(19.5,0.044)(20.5,0.044)(20.5,0)
\pspolygon(20.5,0)(20.5,0.116)(21.5,0.116)(21.5,0)
\pspolygon(21.5,0)(21.5,0.017)(22.5,0.017)(22.5,0)
\pspolygon(22.5,0)(22.5,0.100)(23.5,0.100)(23.5,0)
\pspolygon(23.5,0)(23.5,0.117)(24.5,0.117)(24.5,0)
\pspolygon(24.5,0)(24.5,0.001)(25.5,0.001)(25.5,0)
\pspolygon(25.5,0)(25.5,0.124)(26.5,0.124)(26.5,0)
\pspolygon(26.5,0)(26.5,0.037)(27.5,0.037)(27.5,0)
\pspolygon(27.5,0)(27.5,0.091)(28.5,0.091)(28.5,0)
\pspolygon(28.5,0)(28.5,0.075)(29.5,0.075)(29.5,0)
\pspolygon(29.5,0)(29.5,0.032)(30.5,0.032)(30.5,0)
\pspolygon(30.5,0)(30.5,0.119)(31.5,0.119)(31.5,0)
\pspolygon(31.5,0)(31.5,0.059)(32.5,0.059)(32.5,0)
\pspolygon(32.5,0)(32.5,0.009)(33.5,0.009)(33.5,0)
\pspolygon(33.5,0)(33.5,0.032)(34.5,0.032)(34.5,0)
\pspolygon(34.5,0)(34.5,0.047)(35.5,0.047)(35.5,0)
\pspolygon(35.5,0)(35.5,0.042)(36.5,0.042)(36.5,0)
\pspolygon(36.5,0)(36.5,0.137)(37.5,0.137)(37.5,0)
\pspolygon(37.5,0)(37.5,0.070)(38.5,0.070)(38.5,0)
\pspolygon(38.5,0)(38.5,0.141)(39.5,0.141)(39.5,0)
\pspolygon(39.5,0)(39.5,0.072)(40.5,0.072)(40.5,0)
\pspolygon(40.5,0)(40.5,0.061)(41.5,0.061)(41.5,0)
\pspolygon(41.5,0)(41.5,0.080)(42.5,0.080)(42.5,0)
\pspolygon(42.5,0)(42.5,0.039)(43.5,0.039)(43.5,0)
\pspolygon(43.5,0)(43.5,0.079)(44.5,0.079)(44.5,0)
\pspolygon(44.5,0)(44.5,0.140)(45.5,0.140)(45.5,0)
\pspolygon(45.5,0)(45.5,0.141)(46.5,0.141)(46.5,0)
\pspolygon(46.5,0)(46.5,0.110)(47.5,0.110)(47.5,0)
\pspolygon(47.5,0)(47.5,0.149)(48.5,0.149)(48.5,0)
\pspolygon(48.5,0)(48.5,0.129)(49.5,0.129)(49.5,0)
\pspolygon(49.5,0)(49.5,0.039)(50.5,0.039)(50.5,0)
\pspolygon(50.5,0)(50.5,0.149)(51.5,0.149)(51.5,0)
\pspolygon(51.5,0)(51.5,0.018)(52.5,0.018)(52.5,0)
\pspolygon(52.5,0)(52.5,0.004)(53.5,0.004)(53.5,0)
\pspolygon(53.5,0)(53.5,0.039)(54.5,0.039)(54.5,0)
\pspolygon(54.5,0)(54.5,0.033)(55.5,0.033)(55.5,0)
\pspolygon(55.5,0)(55.5,0.065)(56.5,0.065)(56.5,0)
\pspolygon(56.5,0)(56.5,0.121)(57.5,0.121)(57.5,0)
\pspolygon(57.5,0)(57.5,0.063)(58.5,0.063)(58.5,0)
\pspolygon(58.5,0)(58.5,0.084)(59.5,0.084)(59.5,0)
\pspolygon(59.5,0)(59.5,0.070)(60.5,0.070)(60.5,0)
\pspolygon(60.5,0)(60.5,0.154)(61.5,0.154)(61.5,0)
\pspolygon(61.5,0)(61.5,0.139)(62.5,0.139)(62.5,0)
\pspolygon(62.5,0)(62.5,0.121)(63.5,0.121)(63.5,0)
\pspolygon(63.5,0)(63.5,0.059)(64.5,0.059)(64.5,0)
\pspolygon(64.5,0)(64.5,0.087)(65.5,0.087)(65.5,0)
\pspolygon(65.5,0)(65.5,0.048)(66.5,0.048)(66.5,0)
\pspolygon(66.5,0)(66.5,0.029)(67.5,0.029)(67.5,0)
\pspolygon(67.5,0)(67.5,0.213)(68.5,0.213)(68.5,0)
\pspolygon(68.5,0)(68.5,0.182)(69.5,0.182)(69.5,0)
\pspolygon(69.5,0)(69.5,0.111)(70.5,0.111)(70.5,0)
\pspolygon(70.5,0)(70.5,0.033)(71.5,0.033)(71.5,0)
\pspolygon(71.5,0)(71.5,0.176)(72.5,0.176)(72.5,0)
\pspolygon(72.5,0)(72.5,0.176)(73.5,0.176)(73.5,0)
\pspolygon(73.5,0)(73.5,0.115)(74.5,0.115)(74.5,0)
\pspolygon(74.5,0)(74.5,0.091)(75.5,0.091)(75.5,0)
\pspolygon(75.5,0)(75.5,0.164)(76.5,0.164)(76.5,0)
\pspolygon(76.5,0)(76.5,0.155)(77.5,0.155)(77.5,0)
\pspolygon(77.5,0)(77.5,0.153)(78.5,0.153)(78.5,0)
\pspolygon(78.5,0)(78.5,0.186)(79.5,0.186)(79.5,0)
\pspolygon(79.5,0)(79.5,0.208)(80.5,0.208)(80.5,0)
\pspolygon(80.5,0)(80.5,0.179)(81.5,0.179)(81.5,0)
\pspolygon(81.5,0)(81.5,0.057)(82.5,0.057)(82.5,0)
\pspolygon(82.5,0)(82.5,0.040)(83.5,0.040)(83.5,0)
\pspolygon(83.5,0)(83.5,0.138)(84.5,0.138)(84.5,0)
\pspolygon(84.5,0)(84.5,0.017)(85.5,0.017)(85.5,0)
\pspolygon(85.5,0)(85.5,0.041)(86.5,0.041)(86.5,0)
\pspolygon(86.5,0)(86.5,0.057)(87.5,0.057)(87.5,0)
\pspolygon(87.5,0)(87.5,0.006)(88.5,0.006)(88.5,0)
\pspolygon(88.5,0)(88.5,0.007)(89.5,0.007)(89.5,0)
\pspolygon(89.5,0)(89.5,0.013)(90.5,0.013)(90.5,0)
\pspolygon(90.5,0)(90.5,0.007)(91.5,0.007)(91.5,0)
\pspolygon(91.5,0)(91.5,0.104)(92.5,0.104)(92.5,0)
\psline{->}(0,0)(95,0) \psline{->}(0,0)(0,0.5) \rput(2.22,0.5){$e_{2}$} \scriptsize \psline(1,-0.02)(1,0.02) \rput(1,-0.06){S\&P} \psline(10,-0.02)(10,0.02) \rput(10,-0.06){Arge} \psline(20,-0.02)(20,0.02) \rput(20,-0.06){Autr} \psline(30,-0.02)(30,0.02) \rput(30,-0.06){Icel} \psline(40,-0.02)(40,0.02) \rput(40,-0.06){BoHe} \psline(50,-0.02)(50,0.02) \rput(50,-0.06){Russ} \psline(60,-0.02)(60,0.02) \rput(60,-0.06){Bahr} \psline(70,-0.02)(70,0.02) \rput(70,-0.06){Chin} \psline(80,-0.02)(80,0.02) \rput(80,-0.06){Aust} \psline(90,-0.02)(90,0.02) \rput(90,-0.06){Botsw} \psline(92,-0.02)(92,0.02) \rput(95,-0.06){Maur} \scriptsize \psline(-0.55,0.1)(0.55,0.1) \rput(-2.22,0.1){$0.1$} \psline(-0.55,0.2)(0.55,0.2) \rput(-2.22,0.2){$0.2$} \psline(-0.55,0.3)(0.55,0.3) \rput(-2.22,0.3){$0.3$} \psline(-0.55,0.4)(0.55,0.4) \rput(-2.22,0.4){$0.4$}
\end{pspicture}

\vskip 0.6 cm

\noindent Fig. 11. Contributions of the stock market indices to eigenvector $e_{2}$, corresponding to the second largest eigenvalue of the correlation matrix. White bars indicate positive values, and gray bars indicate negative values, corresponding to the first and second semesters of 2001. The indices are aligned in the following way: {\bf S\&P}, Nasd, Cana, Mexi, Pana, CoRi, Berm, Jama, Braz, {\bf Arge}, Chil, Colo, Vene, Peru, UK, Irel, Fran, Germ, Swit, {\bf Autr}, Ital, Malt, Belg, Neth, Luxe, Swed, Denm, Finl, Norw, {\bf Icel}, Spai, Port, Gree, CzRe, Slok, Hung, Serb, Croa, Slov, {\bf BoHe}, Mont, Mace, Pola, Roma, Bulg, Esto, Latv, Lith, Ukra, {\bf Russ}, Kaza, Turk, Cypr, Isra, Pale, Leba, Jord, SaAr, Kuwa, {\bf Bahr}, Qata, UAE, Ohma, Paki, Indi, SrLa, Bang, Japa, HoKo, {\bf Chin}, Mong, Taiw, SoKo, Thai, Viet, Mala, Sing, Indo, Phil, {\bf Aust}, NeZe, Moro, Tuni, Egyp, Ghan, Nige, Keny, Tanz, Nami, {\bf Bots}, SoAf, and {\bf Maur}.

\vskip 0.4 cm

All indices of countries that are from Greece to the East, in terms of latitude, with the exceptions of Russia, Turkey, Israel, and South Africa, appear with negative values in eigenvector $e_2$. All these exceptions are indices from stock exchanges that have a non-negative intersection with the opening hours of the New York Stock Exchange. Also, from Greece to the West, the indices from Panama, Costa Rica, Bermuda, Jamaica, Venezuela, Malta, and Iceland, all of them indices with very low correlation with all the others, also appear with negative values in the same eigenvector.

If one laggs all the indices that have negative values in eigenvector $e_2$, which means that one compares their returns at day $t+1$ with day $t$ of all the other indices, one may then create a new returns matrix and thus a new correlation matrix. The eigenvalues of such a matrix are plotted in Figure 12, together with the region associated with noise, the result of yet more 1000 simulations with randomized data.

\begin{pspicture}(-1.1,-0.5)(7,2.4)
\psset{xunit=1.5,yunit=1.5}
\pspolygon*[linecolor=lightgray](0.050,0)(0.050,1.2)(1.65,1.2)(1.65,0)
\psline{->}(0,0)(10,0)  \rput(10.15,0){$\lambda $} \psline[linecolor=white,linewidth=2pt](8.4,0)(8.6,0) \psline(8.3,-0.05)(8.5,0.05) \psline(8.5,-0.05)(8.7,0.05) \scriptsize \psline(1,-0.05)(1,0.05) \rput(1,-0.15){1} \psline(2,-0.05)(2,0.05) \rput(2,-0.15){2} \psline(3,-0.05)(3,0.05) \rput(3,-0.15){3} \psline(4,-0.05)(4,0.05) \rput(4,-0.15){4} \psline(5,-0.05)(5,0.05) \rput(5,-0.15){5} \psline(6,-0.05)(6,0.05) \rput(6,-0.15){6} \psline(7,-0.05)(7,0.05) \rput(7,-0.15){7} \psline(8,-0.05)(8,0.05) \rput(8,-0.15){8} \psline(9,-0.05)(9,0.05) \rput(9,-0.15){22}
\psline[linewidth=1pt](0.0393,0)(0.0393,1) \psline[linewidth=1pt](0.0597,0)(0.0597,1) \psline[linewidth=1pt](0.0821,0)(0.0821,1) \psline[linewidth=1pt](0.0965,0)(0.0965,1) \psline[linewidth=1pt](0.1248,0)(0.1248,1) \psline[linewidth=1pt](0.1326,0)(0.1326,1) \psline[linewidth=1pt](0.1660,0)(0.1660,1) \psline[linewidth=1pt](0.1708,0)(0.1708,1) \psline[linewidth=1pt](0.1818,0)(0.1818,1) \psline[linewidth=1pt](0.1883,0)(0.1883,1) \psline[linewidth=1pt](0.2054,0)(0.2054,1) \psline[linewidth=1pt](0.2190,0)(0.2190,1) \psline[linewidth=1pt](0.2290,0)(0.2290,1) \psline[linewidth=1pt](0.2412,0)(0.2412,1) \psline[linewidth=1pt](0.2524,0)(0.2524,1) \psline[linewidth=1pt](0.2617,0)(0.2617,1) \psline[linewidth=1pt](0.2715,0)(0.2715,1) \psline[linewidth=1pt](0.2753,0)(0.2753,1) \psline[linewidth=1pt](0.3055,0)(0.3055,1) \psline[linewidth=1pt](0.3256,0)(0.3256,1) \psline[linewidth=1pt](0.3285,0)(0.3285,1) \psline[linewidth=1pt](0.3378,0)(0.3378,1) \psline[linewidth=1pt](0.3454,0)(0.3454,1) \psline[linewidth=1pt](0.3509,0)(0.3509,1) \psline[linewidth=1pt](0.3721,0)(0.3721,1) \psline[linewidth=1pt](0.3841,0)(0.3841,1) \psline[linewidth=1pt](0.3954,0)(0.3954,1) \psline[linewidth=1pt](0.4111,0)(0.4111,1) \psline[linewidth=1pt](0.4165,0)(0.4165,1) \psline[linewidth=1pt](0.4180,0)(0.4180,1) \psline[linewidth=1pt](0.4248,0)(0.4248,1) \psline[linewidth=1pt](0.4456,0)(0.4456,1) \psline[linewidth=1pt](0.4589,0)(0.4589,1) \psline[linewidth=1pt](0.4719,0)(0.4719,1) \psline[linewidth=1pt](0.4866,0)(0.4866,1) \psline[linewidth=1pt](0.4982,0)(0.4982,1) \psline[linewidth=1pt](0.5040,0)(0.5040,1) \psline[linewidth=1pt](0.5170,0)(0.5170,1) \psline[linewidth=1pt](0.5353,0)(0.5353,1) \psline[linewidth=1pt](0.5418,0)(0.5418,1) \psline[linewidth=1pt](0.5701,0)(0.5701,1) \psline[linewidth=1pt](0.5754,0)(0.5754,1) \psline[linewidth=1pt](0.5913,0)(0.5913,1) \psline[linewidth=1pt](0.5992,0)(0.5992,1) \psline[linewidth=1pt](0.6116,0)(0.6116,1) \psline[linewidth=1pt](0.6262,0)(0.6262,1) \psline[linewidth=1pt](0.6275,0)(0.6275,1) \psline[linewidth=1pt](0.6374,0)(0.6374,1) \psline[linewidth=1pt](0.6667,0)(0.6667,1) \psline[linewidth=1pt](0.6912,0)(0.6912,1) \psline[linewidth=1pt](0.7047,0)(0.7047,1) \psline[linewidth=1pt](0.7128,0)(0.7128,1) \psline[linewidth=1pt](0.7292,0)(0.7292,1) \psline[linewidth=1pt](0.7391,0)(0.7391,1) \psline[linewidth=1pt](0.7604,0)(0.7604,1) \psline[linewidth=1pt](0.7611,0)(0.7611,1) \psline[linewidth=1pt](0.7782,0)(0.7782,1) \psline[linewidth=1pt](0.7824,0)(0.7824,1) \psline[linewidth=1pt](0.8035,0)(0.8035,1) \psline[linewidth=1pt](0.8294,0)(0.8294,1) \psline[linewidth=1pt](0.8488,0)(0.8488,1) \psline[linewidth=1pt](0.8669,0)(0.8669,1) \psline[linewidth=1pt](0.8793,0)(0.8793,1) \psline[linewidth=1pt](0.8809,0)(0.8809,1) \psline[linewidth=1pt](0.9021,0)(0.9021,1) \psline[linewidth=1pt](0.9194,0)(0.9194,1) \psline[linewidth=1pt](0.9338,0)(0.9338,1) \psline[linewidth=1pt](0.9545,0)(0.9545,1) \psline[linewidth=1pt](0.9635,0)(0.9635,1) \psline[linewidth=1pt](0.9678,0)(0.9678,1) \psline[linewidth=1pt](1.0123,0)(1.0123,1) \psline[linewidth=1pt](1.0194,0)(1.0194,1) \psline[linewidth=1pt](1.0284,0)(1.0284,1) \psline[linewidth=1pt](1.0615,0)(1.0615,1) \psline[linewidth=1pt](1.0787,0)(1.0787,1) \psline[linewidth=1pt](1.0986,0)(1.0986,1) \psline[linewidth=1pt](1.1172,0)(1.1172,1) \psline[linewidth=1pt](1.1442,0)(1.1442,1) \psline[linewidth=1pt](1.1834,0)(1.1834,1) \psline[linewidth=1pt](1.2019,0)(1.2019,1) \psline[linewidth=1pt](1.2058,0)(1.2058,1) \psline[linewidth=1pt](1.2299,0)(1.2299,1) \psline[linewidth=1pt](1.2657,0)(1.2657,1) \psline[linewidth=1pt](1.3216,0)(1.3216,1) \psline[linewidth=1pt](1.3390,0)(1.3390,1) \psline[linewidth=1pt](1.3733,0)(1.3733,1) \psline[linewidth=1pt](1.5737,0)(1.5737,1) \psline[linewidth=1pt](1.9400,0)(1.9400,1) \psline[linewidth=1pt](2.1472,0)(2.1472,1) \psline[linewidth=1pt](2.3002,0)(2.3002,1) \psline[linewidth=1pt](7.6468,0)(7.6468,1) \psline[linewidth=1pt](9.3242,0)(9.3242,1)
\end{pspicture}

\noindent Fig. 12. Eigenvalues of the correlation for data from 2007 to 2010,  in order of magnitude. The shaded area corresponds to the eigenvalues predicted for randomized data.

\vskip 0.3 cm

The first largest eigenvalue is still associated with a market mode, but the second largest eigenvalue, more detached now from the others than in the non-lagged case, means something else. Figure 13 shows the values of the indices in the new eigenvalue $e_2$, associated with the second largest eigenvalue. Once more, white bars correspond to positive signs in the eigenvector, and gray bars correspond to negative signs in the eigenvector.

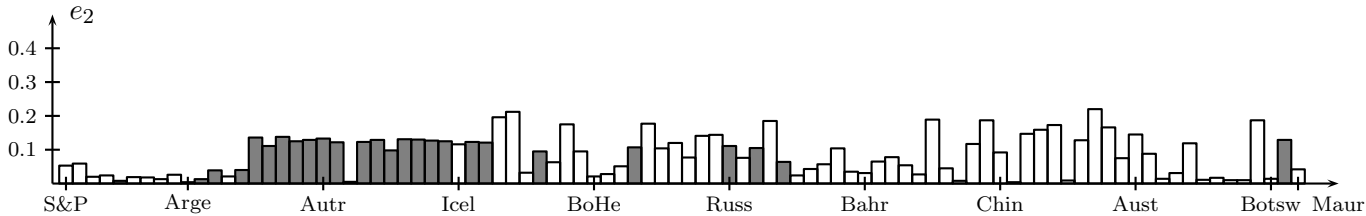
\begin{pspicture}(-0.1,0)(3.5,2.8)
\psset{xunit=0.18,yunit=4.5}
\pspolygon*[linecolor=white](0.5,0)(0.5,0.053)(1.5,0.053)(1.5,0)
\pspolygon*[linecolor=white](1.5,0)(1.5,0.059)(2.5,0.059)(2.5,0)
\pspolygon*[linecolor=white](2.5,0)(2.5,0.020)(3.5,0.020)(3.5,0)
\pspolygon*[linecolor=white](3.5,0)(3.5,0.024)(4.5,0.024)(4.5,0)
\pspolygon*[linecolor=gray](4.5,0)(4.5,0.008)(5.5,0.008)(5.5,0)
\pspolygon*[linecolor=white](5.5,0)(5.5,0.019)(6.5,0.019)(6.5,0)
\pspolygon*[linecolor=white](6.5,0)(6.5,0.018)(7.5,0.018)(7.5,0)
\pspolygon*[linecolor=white](7.5,0)(7.5,0.013)(8.5,0.013)(8.5,0)
\pspolygon*[linecolor=white](8.5,0)(8.5,0.026)(9.5,0.026)(9.5,0)
\pspolygon*[linecolor=white](9.5,0)(9.5,0.004)(10.5,0.004)(10.5,0)
\pspolygon*[linecolor=gray](10.5,0)(10.5,0.013)(11.5,0.013)(11.5,0)
\pspolygon*[linecolor=gray](11.5,0)(11.5,0.039)(12.5,0.039)(12.5,0)
\pspolygon*[linecolor=white](12.5,0)(12.5,0.021)(13.5,0.021)(13.5,0)
\pspolygon*[linecolor=gray](13.5,0)(13.5,0.040)(14.5,0.040)(14.5,0)
\pspolygon*[linecolor=gray](14.5,0)(14.5,0.136)(15.5,0.136)(15.5,0)
\pspolygon*[linecolor=gray](15.5,0)(15.5,0.111)(16.5,0.111)(16.5,0)
\pspolygon*[linecolor=gray](16.5,0)(16.5,0.138)(17.5,0.138)(17.5,0)
\pspolygon*[linecolor=gray](17.5,0)(17.5,0.125)(18.5,0.125)(18.5,0)
\pspolygon*[linecolor=gray](18.5,0)(18.5,0.129)(19.5,0.129)(19.5,0)
\pspolygon*[linecolor=gray](19.5,0)(19.5,0.133)(20.5,0.133)(20.5,0)
\pspolygon*[linecolor=gray](20.5,0)(20.5,0.122)(21.5,0.122)(21.5,0)
\pspolygon*[linecolor=white](21.5,0)(21.5,0.005)(22.5,0.005)(22.5,0)
\pspolygon*[linecolor=gray](22.5,0)(22.5,0.123)(23.5,0.123)(23.5,0)
\pspolygon*[linecolor=gray](23.5,0)(23.5,0.129)(24.5,0.129)(24.5,0)
\pspolygon*[linecolor=gray](24.5,0)(24.5,0.098)(25.5,0.098)(25.5,0)
\pspolygon*[linecolor=gray](25.5,0)(25.5,0.131)(26.5,0.131)(26.5,0)
\pspolygon*[linecolor=gray](26.5,0)(26.5,0.130)(27.5,0.130)(27.5,0)
\pspolygon*[linecolor=gray](27.5,0)(27.5,0.127)(28.5,0.127)(28.5,0)
\pspolygon*[linecolor=gray](28.5,0)(28.5,0.125)(29.5,0.125)(29.5,0)
\pspolygon*[linecolor=white](29.5,0)(29.5,0.116)(30.5,0.116)(30.5,0)
\pspolygon*[linecolor=gray](30.5,0)(30.5,0.123)(31.5,0.123)(31.5,0)
\pspolygon*[linecolor=gray](31.5,0)(31.5,0.121)(32.5,0.121)(32.5,0)
\pspolygon*[linecolor=white](32.5,0)(32.5,0.196)(33.5,0.196)(33.5,0)
\pspolygon*[linecolor=white](33.5,0)(33.5,0.212)(34.5,0.212)(34.5,0)
\pspolygon*[linecolor=white](34.5,0)(34.5,0.032)(35.5,0.032)(35.5,0)
\pspolygon*[linecolor=gray](35.5,0)(35.5,0.095)(36.5,0.095)(36.5,0)
\pspolygon*[linecolor=white](36.5,0)(36.5,0.063)(37.5,0.063)(37.5,0)
\pspolygon*[linecolor=white](37.5,0)(37.5,0.175)(38.5,0.175)(38.5,0)
\pspolygon*[linecolor=white](38.5,0)(38.5,0.095)(39.5,0.095)(39.5,0)
\pspolygon*[linecolor=white](39.5,0)(39.5,0.021)(40.5,0.021)(40.5,0)
\pspolygon*[linecolor=white](40.5,0)(40.5,0.028)(41.5,0.028)(41.5,0)
\pspolygon*[linecolor=white](41.5,0)(41.5,0.051)(42.5,0.051)(42.5,0)
\pspolygon*[linecolor=gray](42.5,0)(42.5,0.107)(43.5,0.107)(43.5,0)
\pspolygon*[linecolor=white](43.5,0)(43.5,0.177)(44.5,0.177)(44.5,0)
\pspolygon*[linecolor=white](44.5,0)(44.5,0.104)(45.5,0.104)(45.5,0)
\pspolygon*[linecolor=white](45.5,0)(45.5,0.120)(46.5,0.120)(46.5,0)
\pspolygon*[linecolor=white](46.5,0)(46.5,0.077)(47.5,0.077)(47.5,0)
\pspolygon*[linecolor=white](47.5,0)(47.5,0.141)(48.5,0.141)(48.5,0)
\pspolygon*[linecolor=white](48.5,0)(48.5,0.144)(49.5,0.144)(49.5,0)
\pspolygon*[linecolor=gray](49.5,0)(49.5,0.111)(50.5,0.111)(50.5,0)
\pspolygon*[linecolor=white](50.5,0)(50.5,0.076)(51.5,0.076)(51.5,0)
\pspolygon*[linecolor=gray](51.5,0)(51.5,0.105)(52.5,0.105)(52.5,0)
\pspolygon*[linecolor=white](52.5,0)(52.5,0.185)(53.5,0.185)(53.5,0)
\pspolygon*[linecolor=gray](53.5,0)(53.5,0.064)(54.5,0.064)(54.5,0)
\pspolygon*[linecolor=white](54.5,0)(54.5,0.024)(55.5,0.024)(55.5,0)
\pspolygon*[linecolor=white](55.5,0)(55.5,0.043)(56.5,0.043)(56.5,0)
\pspolygon*[linecolor=white](56.5,0)(56.5,0.057)(57.5,0.057)(57.5,0)
\pspolygon*[linecolor=white](57.5,0)(57.5,0.104)(58.5,0.104)(58.5,0)
\pspolygon*[linecolor=white](58.5,0)(58.5,0.035)(59.5,0.035)(59.5,0)
\pspolygon*[linecolor=white](59.5,0)(59.5,0.031)(60.5,0.031)(60.5,0)
\pspolygon*[linecolor=white](60.5,0)(60.5,0.065)(61.5,0.065)(61.5,0)
\pspolygon*[linecolor=white](61.5,0)(61.5,0.078)(62.5,0.078)(62.5,0)
\pspolygon*[linecolor=white](62.5,0)(62.5,0.054)(63.5,0.054)(63.5,0)
\pspolygon*[linecolor=white](63.5,0)(63.5,0.027)(64.5,0.027)(64.5,0)
\pspolygon*[linecolor=white](64.5,0)(64.5,0.189)(65.5,0.189)(65.5,0)
\pspolygon*[linecolor=white](65.5,0)(65.5,0.045)(66.5,0.045)(66.5,0)
\pspolygon*[linecolor=white](66.5,0)(66.5,0.008)(67.5,0.008)(67.5,0)
\pspolygon*[linecolor=white](67.5,0)(67.5,0.117)(68.5,0.117)(68.5,0)
\pspolygon*[linecolor=white](68.5,0)(68.5,0.187)(69.5,0.187)(69.5,0)
\pspolygon*[linecolor=white](69.5,0)(69.5,0.092)(70.5,0.092)(70.5,0)
\pspolygon*[linecolor=gray](70.5,0)(70.5,0.004)(71.5,0.004)(71.5,0)
\pspolygon*[linecolor=white](71.5,0)(71.5,0.147)(72.5,0.147)(72.5,0)
\pspolygon*[linecolor=white](72.5,0)(72.5,0.159)(73.5,0.159)(73.5,0)
\pspolygon*[linecolor=white](73.5,0)(73.5,0.173)(74.5,0.173)(74.5,0)
\pspolygon*[linecolor=white](74.5,0)(74.5,0.009)(75.5,0.009)(75.5,0)
\pspolygon*[linecolor=white](75.5,0)(75.5,0.128)(76.5,0.128)(76.5,0)
\pspolygon*[linecolor=white](76.5,0)(76.5,0.220)(77.5,0.220)(77.5,0)
\pspolygon*[linecolor=white](77.5,0)(77.5,0.166)(78.5,0.166)(78.5,0)
\pspolygon*[linecolor=white](78.5,0)(78.5,0.075)(79.5,0.075)(79.5,0)
\pspolygon*[linecolor=white](79.5,0)(79.5,0.145)(80.5,0.145)(80.5,0)
\pspolygon*[linecolor=white](80.5,0)(80.5,0.088)(81.5,0.088)(81.5,0)
\pspolygon*[linecolor=white](81.5,0)(81.5,0.014)(82.5,0.014)(82.5,0)
\pspolygon*[linecolor=white](82.5,0)(82.5,0.031)(83.5,0.031)(83.5,0)
\pspolygon*[linecolor=white](83.5,0)(83.5,0.119)(84.5,0.119)(84.5,0)
\pspolygon*[linecolor=white](84.5,0)(84.5,0.011)(85.5,0.011)(85.5,0)
\pspolygon*[linecolor=white](85.5,0)(85.5,0.017)(86.5,0.017)(86.5,0)
\pspolygon*[linecolor=white](86.5,0)(86.5,0.010)(87.5,0.010)(87.5,0)
\pspolygon*[linecolor=white](87.5,0)(87.5,0.010)(88.5,0.010)(88.5,0)
\pspolygon*[linecolor=white](88.5,0)(88.5,0.187)(89.5,0.187)(89.5,0)
\pspolygon*[linecolor=white](89.5,0)(89.5,0.014)(90.5,0.014)(90.5,0)
\pspolygon*[linecolor=gray](90.5,0)(90.5,0.129)(91.5,0.129)(91.5,0)
\pspolygon*[linecolor=white](91.5,0)(91.5,0.042)(92.5,0.042)(92.5,0)
\pspolygon(0.5,0)(0.5,0.053)(1.5,0.053)(1.5,0)
\pspolygon(1.5,0)(1.5,0.059)(2.5,0.059)(2.5,0)
\pspolygon(2.5,0)(2.5,0.020)(3.5,0.020)(3.5,0)
\pspolygon(3.5,0)(3.5,0.024)(4.5,0.024)(4.5,0)
\pspolygon(4.5,0)(4.5,0.008)(5.5,0.008)(5.5,0)
\pspolygon(5.5,0)(5.5,0.019)(6.5,0.019)(6.5,0)
\pspolygon(6.5,0)(6.5,0.018)(7.5,0.018)(7.5,0)
\pspolygon(7.5,0)(7.5,0.013)(8.5,0.013)(8.5,0)
\pspolygon(8.5,0)(8.5,0.026)(9.5,0.026)(9.5,0)
\pspolygon(9.5,0)(9.5,0.004)(10.5,0.004)(10.5,0)
\pspolygon(10.5,0)(10.5,0.013)(11.5,0.013)(11.5,0)
\pspolygon(11.5,0)(11.5,0.039)(12.5,0.039)(12.5,0)
\pspolygon(12.5,0)(12.5,0.021)(13.5,0.021)(13.5,0)
\pspolygon(13.5,0)(13.5,0.040)(14.5,0.040)(14.5,0)
\pspolygon(14.5,0)(14.5,0.136)(15.5,0.136)(15.5,0)
\pspolygon(15.5,0)(15.5,0.111)(16.5,0.111)(16.5,0)
\pspolygon(16.5,0)(16.5,0.138)(17.5,0.138)(17.5,0)
\pspolygon(17.5,0)(17.5,0.125)(18.5,0.125)(18.5,0)
\pspolygon(18.5,0)(18.5,0.129)(19.5,0.129)(19.5,0)
\pspolygon(19.5,0)(19.5,0.133)(20.5,0.133)(20.5,0)
\pspolygon(20.5,0)(20.5,0.122)(21.5,0.122)(21.5,0)
\pspolygon(21.5,0)(21.5,0.005)(22.5,0.005)(22.5,0)
\pspolygon(22.5,0)(22.5,0.123)(23.5,0.123)(23.5,0)
\pspolygon(23.5,0)(23.5,0.129)(24.5,0.129)(24.5,0)
\pspolygon(24.5,0)(24.5,0.098)(25.5,0.098)(25.5,0)
\pspolygon(25.5,0)(25.5,0.131)(26.5,0.131)(26.5,0)
\pspolygon(26.5,0)(26.5,0.130)(27.5,0.130)(27.5,0)
\pspolygon(27.5,0)(27.5,0.127)(28.5,0.127)(28.5,0)
\pspolygon(28.5,0)(28.5,0.125)(29.5,0.125)(29.5,0)
\pspolygon(29.5,0)(29.5,0.116)(30.5,0.116)(30.5,0)
\pspolygon(30.5,0)(30.5,0.123)(31.5,0.123)(31.5,0)
\pspolygon(31.5,0)(31.5,0.121)(32.5,0.121)(32.5,0)
\pspolygon(32.5,0)(32.5,0.196)(33.5,0.196)(33.5,0)
\pspolygon(33.5,0)(33.5,0.212)(34.5,0.212)(34.5,0)
\pspolygon(34.5,0)(34.5,0.032)(35.5,0.032)(35.5,0)
\pspolygon(35.5,0)(35.5,0.095)(36.5,0.095)(36.5,0)
\pspolygon(36.5,0)(36.5,0.063)(37.5,0.063)(37.5,0)
\pspolygon(37.5,0)(37.5,0.175)(38.5,0.175)(38.5,0)
\pspolygon(38.5,0)(38.5,0.095)(39.5,0.095)(39.5,0)
\pspolygon(39.5,0)(39.5,0.021)(40.5,0.021)(40.5,0)
\pspolygon(40.5,0)(40.5,0.028)(41.5,0.028)(41.5,0)
\pspolygon(41.5,0)(41.5,0.051)(42.5,0.051)(42.5,0)
\pspolygon(42.5,0)(42.5,0.107)(43.5,0.107)(43.5,0)
\pspolygon(43.5,0)(43.5,0.177)(44.5,0.177)(44.5,0)
\pspolygon(44.5,0)(44.5,0.104)(45.5,0.104)(45.5,0)
\pspolygon(45.5,0)(45.5,0.120)(46.5,0.120)(46.5,0)
\pspolygon(46.5,0)(46.5,0.077)(47.5,0.077)(47.5,0)
\pspolygon(47.5,0)(47.5,0.141)(48.5,0.141)(48.5,0)
\pspolygon(48.5,0)(48.5,0.144)(49.5,0.144)(49.5,0)
\pspolygon(49.5,0)(49.5,0.111)(50.5,0.111)(50.5,0)
\pspolygon(50.5,0)(50.5,0.076)(51.5,0.076)(51.5,0)
\pspolygon(51.5,0)(51.5,0.105)(52.5,0.105)(52.5,0)
\pspolygon(52.5,0)(52.5,0.185)(53.5,0.185)(53.5,0)
\pspolygon(53.5,0)(53.5,0.064)(54.5,0.064)(54.5,0)
\pspolygon(54.5,0)(54.5,0.024)(55.5,0.024)(55.5,0)
\pspolygon(55.5,0)(55.5,0.043)(56.5,0.043)(56.5,0)
\pspolygon(56.5,0)(56.5,0.057)(57.5,0.057)(57.5,0)
\pspolygon(57.5,0)(57.5,0.104)(58.5,0.104)(58.5,0)
\pspolygon(58.5,0)(58.5,0.035)(59.5,0.035)(59.5,0)
\pspolygon(59.5,0)(59.5,0.031)(60.5,0.031)(60.5,0)
\pspolygon(60.5,0)(60.5,0.065)(61.5,0.065)(61.5,0)
\pspolygon(61.5,0)(61.5,0.078)(62.5,0.078)(62.5,0)
\pspolygon(62.5,0)(62.5,0.054)(63.5,0.054)(63.5,0)
\pspolygon(63.5,0)(63.5,0.027)(64.5,0.027)(64.5,0)
\pspolygon(64.5,0)(64.5,0.189)(65.5,0.189)(65.5,0)
\pspolygon(65.5,0)(65.5,0.045)(66.5,0.045)(66.5,0)
\pspolygon(66.5,0)(66.5,0.008)(67.5,0.008)(67.5,0)
\pspolygon(67.5,0)(67.5,0.117)(68.5,0.117)(68.5,0)
\pspolygon(68.5,0)(68.5,0.187)(69.5,0.187)(69.5,0)
\pspolygon(69.5,0)(69.5,0.092)(70.5,0.092)(70.5,0)
\pspolygon(70.5,0)(70.5,0.004)(71.5,0.004)(71.5,0)
\pspolygon(71.5,0)(71.5,0.147)(72.5,0.147)(72.5,0)
\pspolygon(72.5,0)(72.5,0.159)(73.5,0.159)(73.5,0)
\pspolygon(73.5,0)(73.5,0.173)(74.5,0.173)(74.5,0)
\pspolygon(74.5,0)(74.5,0.009)(75.5,0.009)(75.5,0)
\pspolygon(75.5,0)(75.5,0.128)(76.5,0.128)(76.5,0)
\pspolygon(76.5,0)(76.5,0.220)(77.5,0.220)(77.5,0)
\pspolygon(77.5,0)(77.5,0.166)(78.5,0.166)(78.5,0)
\pspolygon(78.5,0)(78.5,0.075)(79.5,0.075)(79.5,0)
\pspolygon(79.5,0)(79.5,0.145)(80.5,0.145)(80.5,0)
\pspolygon(80.5,0)(80.5,0.088)(81.5,0.088)(81.5,0)
\pspolygon(81.5,0)(81.5,0.014)(82.5,0.014)(82.5,0)
\pspolygon(82.5,0)(82.5,0.031)(83.5,0.031)(83.5,0)
\pspolygon(83.5,0)(83.5,0.119)(84.5,0.119)(84.5,0)
\pspolygon(84.5,0)(84.5,0.011)(85.5,0.011)(85.5,0)
\pspolygon(85.5,0)(85.5,0.017)(86.5,0.017)(86.5,0)
\pspolygon(86.5,0)(86.5,0.010)(87.5,0.010)(87.5,0)
\pspolygon(87.5,0)(87.5,0.010)(88.5,0.010)(88.5,0)
\pspolygon(88.5,0)(88.5,0.187)(89.5,0.187)(89.5,0)
\pspolygon(89.5,0)(89.5,0.014)(90.5,0.014)(90.5,0)
\pspolygon(90.5,0)(90.5,0.129)(91.5,0.129)(91.5,0)
\pspolygon(91.5,0)(91.5,0.042)(92.5,0.042)(92.5,0)
\psline{->}(0,0)(95,0) \psline{->}(0,0)(0,0.5) \rput(2.22,0.5){$e_{2}$} \scriptsize \psline(1,-0.02)(1,0.02) \rput(1,-0.06){S\&P} \psline(10,-0.02)(10,0.02) \rput(10,-0.06){Arge} \psline(20,-0.02)(20,0.02) \rput(20,-0.06){Autr} \psline(30,-0.02)(30,0.02) \rput(30,-0.06){Icel} \psline(40,-0.02)(40,0.02) \rput(40,-0.06){BoHe} \psline(50,-0.02)(50,0.02) \rput(50,-0.06){Russ} \psline(60,-0.02)(60,0.02) \rput(60,-0.06){Bahr} \psline(70,-0.02)(70,0.02) \rput(70,-0.06){Chin} \psline(80,-0.02)(80,0.02) \rput(80,-0.06){Aust} \psline(90,-0.02)(90,0.02) \rput(90,-0.06){Botsw} \psline(92,-0.02)(92,0.02) \rput(95,-0.06){Maur} \scriptsize \psline(-0.55,0.1)(0.55,0.1) \rput(-2.22,0.1){$0.1$} \psline(-0.55,0.2)(0.55,0.2) \rput(-2.22,0.2){$0.2$} \psline(-0.55,0.3)(0.55,0.3) \rput(-2.22,0.3){$0.3$} \psline(-0.55,0.4)(0.55,0.4) \rput(-2.22,0.4){$0.4$}
\end{pspicture}

\vskip 0.6 cm

\noindent Fig. 13. Contributions of the stock market indices to eigenvector $e_{2}$, corresponding to the second largest eigenvalue of the lagged correlation matrix. White bars indicate positive values, and gray bars indicate negative values, corresponding to the first and second semesters of 2001. The indices are aligned in the following way: {\bf S\&P}, Nasd, Cana, Mexi, Pana, CoRi, Berm, Jama, Braz, {\bf Arge}, Chil, Colo, Vene, Peru, UK, Irel, Fran, Germ, Swit, {\bf Autr}, Ital, Malt, Belg, Neth, Luxe, Swed, Denm, Finl, Norw, {\bf Icel}, Spai, Port, Gree, CzRe, Slok, Hung, Serb, Croa, Slov, {\bf BoHe}, Mont, Mace, Pola, Roma, Bulg, Esto, Latv, Lith, Ukra, {\bf Russ}, Kaza, Turk, Cypr, Isra, Pale, Leba, Jord, SaAr, Kuwa, {\bf Bahr}, Qata, UAE, Ohma, Paki, Indi, SrLa, Bang, Japa, HoKo, {\bf Chin}, Mong, Taiw, SoKo, Thai, Viet, Mala, Sing, Indo, Phil, {\bf Aust}, NeZe, Moro, Tuni, Egyp, Ghan, Nige, Keny, Tanz, Nami, {\bf Bots}, SoAf, and {\bf Maur}.

\vskip 0.4 cm

Eigenvector $e_2$ detects the structure of the European block, which consists on indices from countries from Central Europe, and also of indices from Hungary, Poland, Russia, Turkey, Israel, and South Africa, which basically are part of the European cluster. Also appearing with negative but diminute values in eigenvector $e_2$ are the indices from Panama, Chile, Colombia, Peru, and Mongolia, most of them with marginal connections in the correlation network.

Now, lets analyze the network formed using lagged data. Figure 14 shows the resulting asset graph for lagged data at threshold $T=0.5$ using Spearman correlation in comparison to the netwrok based on the original (non-lagged) data. What one can readily see is that the new coordinates for the lagged indices are more stretched, with the American cluster placed in between the European and the Pacific Asian clusters, and with some of the outlier indices from Europe placed farther than the Pacific Asian ones. The connections, nevertheless, are practically the same, so there are nearly no changes in the way indices connect, even when lagging some of them. This confirms the robustness of the asset graphs that have been built here so far.



\vskip 2.3 cm

\noindent Fig. 14. Three dimensional view of the asset graphs for the whole period of data (2007 to 2010). The figure on the left is based on the non-lagged data, and the figure on the right, on lagged data, both with threshold $T=0.5$.

\vskip 0.3 cm

\subsection{Overview}

Let us now draw some conclusions from the many graphs seen in this section. First of all, we saw that there are two main clusters that maintain their cohesiveness through time and through most leves of threshold, which are the American and the European clusters. The American cluster is basically comprised of the indices S\&P 500 and Nasdaq, which are soon joined by Canada, and then by some Latin American indices. The European cluster is denser, and comprised of a core of Central European indices, then joined by Eastern European indices. We can also identify two other persistent connections: one between South Africa and Namibia, and other between Greece and Cyprus. These are not surprising, for Namibia, whose economy is in early stages of development, is highly dependent both economically and politically on South Africa. Now Cyprus is divided between Greek and Turkish populations, and the Stock Exchange to which the index from Cyprus is assigned is in Nicosia, in the Greek part of the island. We also have a similar connection, at higher thresholds, between Estonia and Lithuania, both former members of the extinct Soviet Union.

The pair South Africa-Namibia is firmly connected with the European cluster, and Israel also has a mildly strong connection with the same cluster. At higher thresholds, a Pacific Asian cluster unfolds, with Australia attached to it, and at still higher thresholds, close to the noise limit, we can see an Arab cluster forming. We don't see the formation of an African cluster until deep inside the thresholds already in the region dominated by noise. The American and European clusters do not form connections until a certain threshold, and Pacific Asia makes connections with Europe at slightly higher values of the threshold. So, what we see are three major clusters, based on geographical and sometimes cultural ties, which maintain some indepedence, with the addition of more indices as the threshold values grow. From the data concerning the second semester of 2008, we can also see that networks shrink in times of high volatility, which means there is higher correlation between the indices, as already stated in \cite{leocorr}. All connections are stable both in what concerns the choice of correlation measure and the lagging of some of the indices so as to acommodate differences in operating hours of the stock exchanges.

\section{Survivability of connections and clusters}

I now shall turn my attention to the survivability of connections between indices, and specifically of clusters of indices, through time. By survivability I mean the endurance of connections in time given a certain threshold. The survival of connections is heavily dependant on the threshold value chosen for the network. For $T=1.5$, as an example, all connections survive 100\% of the time, for all indices are fully connected at this threshold. Now, for lower threshold values, survivability becomes scarce.

The timeline that is being studied is varied in terms of volatility and correlation of indices. Figure 15a shows a graph of the average volatility of the MSCI (Morgan and Stanley Capital International) World Index, based on a portfolio of indices from a variety of developed economies worldwide. The average volatility of this index is calculated by taking the standard deviations of the log-returns of the index for running windows of 100 days each with the windows shifted by one day at a time. Plotted in figure 15b is the average correlation of the 92 indices I am using in this research from correlation matrices calculated in overlapping windows of 100 days of the complete time series from 2007 to 2010. Since I am considering windows of 100 days, I assign the result of each window to the last day of that window. So, the curves in both graphs start 100 working days after the beginning of 2007 (05/22/2007).

What may be clearly noticed is that high volatility is associated with high correlation of financial markets (as stated in \cite{leocorr} and references therein), which means that markets tend to behave similarly in times of crisis.

\begin{pspicture}(-0.3,-0.8)(3.5,5.8)
\psset{xunit=0.07,yunit=1.25}
\psline{->}(0,0)(103,0) \psline{->}(0,0)(0,4) \rput(109,0){day} \rput(15,4){Volatility}
\scriptsize \psline(17.1,-0.08)(17.1,0.08) \rput(17.1,-0.24){01/2008} \psline(43.2,-0.08)(43.2,0.08) \rput(43.2,-0.24){01/2009} \psline(69.3,-0.08)(69.3,0.08) \rput(69.3,-0.24){01/2010} \psline(95.4,-0.08)(95.4,0.08) \rput(92.4,-0.24){01/2011} \psline(-1.4,1)(1.4,1) \rput(-5,1){$0.01$} \psline(-1.4,2)(1.4,2) \rput(-5,2){$0.02$} \psline(-1.4,3)(1.4,3) \rput(-5,3){$0.03$} \rput(-8,4){a)}
\psline[linecolor=red](1,0.61958) (1.1,0.61618) (1.2,0.61504) (1.3,0.62297) (1.4,0.61377) (1.5,0.61377) (1.6,0.61409) (1.7,0.61126) (1.8,0.60976) (1.9,0.60690) (2,0.60652) (2.1,0.60729) (2.2,0.61764) (2.3,0.63546) (2.4,0.63462) (2.5,0.63454) (2.6,0.63983) (2.7,0.64072) (2.8,0.64106) (2.9,0.64471) (3,0.64475) (3.1,0.64325) (3.2,0.64597) (3.3,0.64037) (3.4,0.64562) (3.5,0.64595) (3.6,0.64658) (3.7,0.64603) (3.8,0.64645) (3.9,0.64675) (4,0.64975) (4.1,0.64758) (4.2,0.63947) (4.3,0.63936) (4.4,0.64018) (4.5,0.64042) (4.6,0.64604) (4.7,0.64455) (4.8,0.65623) (4.9,0.65767) (5,0.65768) (5.1,0.60405) (5.2,0.59906) (5.3,0.59754) (5.4,0.59743) (5.5,0.57375) (5.6,0.58001) (5.7,0.58293) (5.8,0.62214) (5.9,0.64154) (6,0.64284) (6.1,0.62571) (6.2,0.62162) (6.3,0.61588) (6.4,0.63420) (6.5,0.62913) (6.6,0.62819) (6.7,0.64066) (6.8,0.67419) (6.9,0.69148) (7,0.69334) (7.1,0.70712) (7.2,0.72134) (7.3,0.73402) (7.4,0.75343) (7.5,0.75552) (7.6,0.75063) (7.7,0.75959) (7.8,0.76105) (7.9,0.76537) (8,0.76581) (8.1,0.78443) (8.2,0.79093) (8.3,0.79064) (8.4,0.80188) (8.5,0.79353) (8.6,0.79607) (8.7,0.80329) (8.8,0.80369) (8.9,0.80969) (9,0.81138) (9.1,0.82372) (9.2,0.82104) (9.3,0.82328) (9.4,0.82335) (9.5,0.82620) (9.6,0.84221) (9.7,0.86113) (9.8,0.86079) (9.9,0.85989) (10,0.85860) (10.1,0.85669) (10.2,0.85680) (10.3,0.85369) (10.4,0.85190) (10.5,0.85683) (10.6,0.85683) (10.7,0.85678) (10.8,0.85653) (10.9,0.85934) (11,0.86111) (11.1,0.86394) (11.2,0.86351) (11.3,0.85782) (11.4,0.85776) (11.5,0.86038) (11.6,0.86653) (11.7,0.86689) (11.8,0.86561) (11.9,0.88150) (12,0.88447) (12.1,0.89144) (12.2,0.88588) (12.3,0.87476) (12.4,0.88579) (12.5,0.88870) (12.6,0.88608) (12.7,0.88981) (12.8,0.90414) (12.9,0.89968) (13,0.90295) (13.1,0.90865) (13.2,0.91651) (13.3,0.91834) (13.4,0.92632) (13.5,0.93606) (13.6,0.94927) (13.7,0.94999) (13.8,0.95936) (13.9,0.95890) (14,0.96765) (14.1,0.97151) (14.2,0.98732) (14.3,0.98785) (14.4,0.99540) (14.5,0.99822) (14.6,0.99714) (14.7,1.01606) (14.8,1.00930) (14.9,1.00966) (15,1.01046) (15.1,1.01254) (15.2,1.01728) (15.3,1.01886) (15.4,1.01677) (15.5,1.01883) (15.6,1.01742) (15.7,1.01641) (15.8,1.00031) (15.9,0.99620) (16,1.01240) (16.1,1.01270) (16.2,1.01128) (16.3,1.01020) (16.4,1.01058) (16.5,1.00907) (16.6,1.00751) (16.7,0.99494) (16.8,0.97109) (16.9,0.95881) (17,0.95844) (17.1,0.94947) (17.2,0.93650) (17.3,0.95016) (17.4,0.93682) (17.5,0.93746) (17.6,0.93686) (17.7,0.92898) (17.8,0.93181) (17.9,0.93307) (18,0.96263) (18.1,0.96011) (18.2,0.96371) (18.3,0.96582) (18.4,0.99848) (18.5,0.99844) (18.6,0.99682) (18.7,1.04025) (18.8,1.03881) (18.9,1.03186) (19,1.03626) (19.1,1.02726) (19.2,1.03416) (19.3,1.04289) (19.4,1.04298) (19.5,1.08990) (19.6,1.07839) (19.7,1.06034) (19.8,1.06026) (19.9,1.05961) (20,1.07465) (20.1,1.07645) (20.2,1.07522) (20.3,1.07078) (20.4,1.07196) (20.5,1.06852) (20.6,1.06988) (20.7,1.07069) (20.8,1.07081) (20.9,1.07923) (21,1.08485) (21.1,1.08477) (21.2,1.08697) (21.3,1.10283) (21.4,1.10571) (21.5,1.10564) (21.6,1.10644) (21.7,1.10933) (21.8,1.11494) (21.9,1.11157) (22,1.13309) (22.1,1.12885) (22.2,1.12866) (22.3,1.13472) (22.4,1.14026) (22.5,1.18782) (22.6,1.19744) (22.7,1.19144) (22.8,1.18142) (22.9,1.18446) (23,1.20586) (23.1,1.20029) (23.2,1.19455) (23.3,1.19450) (23.4,1.18744) (23.5,1.20052) (23.6,1.19268) (23.7,1.19236) (23.8,1.18584) (23.9,1.18656) (24,1.17841) (24.1,1.17562) (24.2,1.16251) (24.3,1.16810) (24.4,1.16216) (24.5,1.16040) (24.6,1.18360) (24.7,1.16693) (24.8,1.17095) (24.9,1.17069) (25,1.17145) (25.1,1.16985) (25.2,1.16466) (25.3,1.16504) (25.4,1.16457) (25.5,1.16299) (25.6,1.15636) (25.7,1.15728) (25.8,1.15445) (25.9,1.14846) (26,1.13447) (26.1,1.13772) (26.2,1.13780) (26.3,1.14086) (26.4,1.13381) (26.5,1.13221) (26.6,1.13252) (26.7,1.13623) (26.8,1.13725) (26.9,1.13735) (27,1.13963) (27.1,1.14191) (27.2,1.14196) (27.3,1.12737) (27.4,1.12729) (27.5,1.12540) (27.6,1.12535) (27.7,1.12590) (27.8,1.12269) (27.9,1.12051) (28,1.09532) (28.1,1.08282) (28.2,1.07663) (28.3,1.08431) (28.4,1.04145) (28.5,1.05043) (28.6,1.05601) (28.7,1.01299) (28.8,1.01661) (28.9,1.01746) (29,1.01238) (29.1,1.01383) (29.2,1.00807) (29.3,1.00641) (29.4,1.00696) (29.5,0.95203) (29.6,0.95027) (29.7,0.96478) (29.8,0.96554) (29.9,0.96613) (30,0.95359) (30.1,0.95958) (30.2,0.95964) (30.3,0.95955) (30.4,0.95830) (30.5,0.95661) (30.6,0.95475) (30.7,0.95415) (30.8,0.95926) (30.9,0.94501) (31,0.94581) (31.1,0.94893) (31.2,0.96235) (31.3,0.94249) (31.4,0.93979) (31.5,0.94039) (31.6,0.93824) (31.7,0.94471) (31.8,0.93641) (31.9,0.93156) (32,0.90743) (32.1,0.91909) (32.2,0.92073) (32.3,0.91577) (32.4,0.89671) (32.5,0.84901) (32.6,0.83464) (32.7,0.84297) (32.8,0.84390) (32.9,0.84542) (33,0.81946) (33.1,0.82646) (33.2,0.82778) (33.3,0.82669) (33.4,0.82800) (33.5,0.81223) (33.6,0.80898) (33.7,0.80953) (33.8,0.81192) (33.9,0.81516) (34,0.81241) (34.1,0.81409) (34.2,0.82496) (34.3,0.81689) (34.4,0.81681) (34.5,0.81660) (34.6,0.78162) (34.7,0.82555) (34.8,0.82028) (34.9,0.85149) (35,0.88340) (35.1,0.88312) (35.2,0.88312) (35.3,0.89376) (35.4,0.95901) (35.5,0.95982) (35.6,1.00050) (35.7,1.02540) (35.8,1.17223) (35.9,1.18189) (36,1.18423) (36.1,1.18234) (36.2,1.19183) (36.3,1.19154) (36.4,1.37776) (36.5,1.39040) (36.6,1.38987) (36.7,1.42600) (36.8,1.42278) (36.9,1.52850) (37,1.55291) (37.1,1.58309) (37.2,1.63769) (37.3,1.70110) (37.4,1.95367) (37.5,1.97522) (37.6,2.09250) (37.7,2.09298) (37.8,2.10027) (37.9,2.14730) (38,2.15157) (38.1,2.23853) (38.2,2.23466) (38.3,2.26603) (38.4,2.29883) (38.5,2.41111) (38.6,2.44378) (38.7,2.46763) (38.8,2.46610) (38.9,2.46580) (39,2.51783) (39.1,2.52624) (39.2,2.58924) (39.3,2.59500) (39.4,2.59565) (39.5,2.61670) (39.6,2.63995) (39.7,2.65618) (39.8,2.65883) (39.9,2.66441) (40,2.66554) (40.1,2.69838) (40.2,2.75927) (40.3,2.78204) (40.4,2.87274) (40.5,2.87994) (40.6,2.88703) (40.7,2.88995) (40.8,2.89237) (40.9,2.97251) (41,2.98119) (41.1,2.98448) (41.2,2.98004) (41.3,2.98075) (41.4,3.03546) (41.5,3.03388) (41.6,3.03710) (41.7,3.03524) (41.8,3.03540) (41.9,3.03658) (42,3.05268) (42.1,3.05036) (42.2,3.05078) (42.3,3.05156) (42.4,3.05329) (42.5,3.04635) (42.6,3.04477) (42.7,3.04416) (42.8,3.04436) (42.9,3.04429) (43,3.05077) (43.1,3.05102) (43.2,3.05022) (43.3,3.06634) (43.4,3.06638) (43.5,3.06490) (43.6,3.06431) (43.7,3.06352) (43.8,3.06454) (43.9,3.06934) (44,3.07119) (44.1,3.08118) (44.2,3.07699) (44.3,3.08280) (44.4,3.08282) (44.5,3.10815) (44.6,3.11711) (44.7,3.10867) (44.8,3.10818) (44.9,3.10628) (45,3.10467) (45.1,3.12101) (45.2,3.12755) (45.3,3.12671) (45.4,3.11051) (45.5,3.11795) (45.6,3.10750) (45.7,3.09909) (45.8,3.05089) (45.9,3.05000) (46,3.06539) (46.1,3.06547) (46.2,3.06069) (46.3,3.06052) (46.4,2.98278) (46.5,3.00130) (46.6,3.00090) (46.7,2.98259) (46.8,2.98829) (46.9,2.94009) (47,2.93136) (47.1,2.91332) (47.2,2.87959) (47.3,2.84116) (47.4,2.72381) (47.5,2.70967) (47.6,2.63080) (47.7,2.64338) (47.8,2.63756) (47.9,2.60300) (48,2.65722) (48.1,2.58117) (48.2,2.59031) (48.3,2.56069) (48.4,2.52723) (48.5,2.43866) (48.6,2.41485) (48.7,2.40230) (48.8,2.40578) (48.9,2.45979) (49,2.41140) (49.1,2.40183) (49.2,2.33192) (49.3,2.33403) (49.4,2.36506) (49.5,2.34376) (49.6,2.31473) (49.7,2.33763) (49.8,2.33227) (49.9,2.32143) (50,2.32863) (50.1,2.27982) (50.2,2.20893) (50.3,2.18986) (50.4,2.08828) (50.5,2.08262) (50.6,2.07669) (50.7,2.07950) (50.8,2.07843) (50.9,1.98127) (51,1.97343) (51.1,1.96714) (51.2,1.95985) (51.3,1.97031) (51.4,1.89741) (51.5,1.89921) (51.6,1.90839) (51.7,1.90856) (51.8,1.90726) (51.9,1.92629) (52,1.90508) (52.1,1.90587) (52.2,1.90350) (52.3,1.90940) (52.4,1.90564) (52.5,1.90501) (52.6,1.92021) (52.7,1.92021) (52.8,1.91969) (52.9,1.92488) (53,1.92176) (53.1,1.92214) (53.2,1.92977) (53.3,1.91024) (53.4,1.91029) (53.5,1.91681) (53.6,1.91486) (53.7,1.91503) (53.8,1.91315) (53.9,1.91501) (54,1.90899) (54.1,1.89039) (54.2,1.88678) (54.3,1.88285) (54.4,1.88287) (54.5,1.82745) (54.6,1.81942) (54.7,1.81748) (54.8,1.81626) (54.9,1.82881) (55,1.82658) (55.1,1.80832) (55.2,1.79156) (55.3,1.77637) (55.4,1.79398) (55.5,1.78629) (55.6,1.79023) (55.7,1.79050) (55.8,1.78049) (55.9,1.78054) (56,1.74407) (56.1,1.74554) (56.2,1.76439) (56.3,1.76441) (56.4,1.76401) (56.5,1.71371) (56.6,1.71530) (56.7,1.71558) (56.8,1.70031) (56.9,1.69020) (57,1.68884) (57.1,1.70640) (57.2,1.70731) (57.3,1.69590) (57.4,1.61155) (57.5,1.60510) (57.6,1.59218) (57.7,1.56212) (57.8,1.56006) (57.9,1.55097) (58,1.47592) (58.1,1.47751) (58.2,1.47287) (58.3,1.46876) (58.4,1.47573) (58.5,1.46904) (58.6,1.46774) (58.7,1.46081) (58.8,1.44774) (58.9,1.36979) (59,1.37217) (59.1,1.37304) (59.2,1.37088) (59.3,1.35488) (59.4,1.32367) (59.5,1.31943) (59.6,1.31357) (59.7,1.25479) (59.8,1.26266) (59.9,1.25946) (60,1.24109) (60.1,1.24273) (60.2,1.21089) (60.3,1.21194) (60.4,1.21585) (60.5,1.22979) (60.6,1.23334) (60.7,1.22787) (60.8,1.22997) (60.9,1.16658) (61,1.16909) (61.1,1.16991) (61.2,1.16994) (61.3,1.15525) (61.4,1.15213) (61.5,1.14342) (61.6,1.13416) (61.7,1.13291) (61.8,1.13394) (61.9,1.10866) (62,1.11223) (62.1,1.10877) (62.2,1.11882) (62.3,1.10545) (62.4,1.10243) (62.5,1.10237) (62.6,1.07230) (62.7,1.09973) (62.8,1.10819) (62.9,1.10086) (63,1.11122) (63.1,1.11034) (63.2,1.10024) (63.3,1.10006) (63.4,1.10059) (63.5,1.09205) (63.6,1.10044) (63.7,1.10050) (63.8,1.09726) (63.9,1.07748) (64,1.07816) (64.1,1.06207) (64.2,1.06211) (64.3,1.06577) (64.4,1.07057) (64.5,1.07254) (64.6,1.09316) (64.7,1.10133) (64.8,1.11989) (64.9,1.08512) (65,1.08200) (65.1,1.08134) (65.2,1.08496) (65.3,1.08397) (65.4,1.06470) (65.5,1.06466) (65.6,1.05870) (65.7,1.06260) (65.8,1.06206) (65.9,1.06848) (66,1.06701) (66.1,1.06456) (66.2,1.04360) (66.3,1.04583) (66.4,1.05387) (66.5,1.04699) (66.6,1.04250) (66.7,1.05409) (66.8,1.05710) (66.9,1.05000) (67,1.06057) (67.1,1.02879) (67.2,1.02763) (67.3,1.02812) (67.4,1.02252) (67.5,1.02864) (67.6,1.02919) (67.7,1.01391) (67.8,1.01379) (67.9,1.01554) (68,1.01684) (68.1,1.01662) (68.2,1.02461) (68.3,1.02369) (68.4,1.01025) (68.5,1.01050) (68.6,1.01018) (68.7,1.01046) (68.8,1.01020) (68.9,1.00991) (69,1.00346) (69.1,1.00279) (69.2,0.99749) (69.3,0.99465) (69.4,0.96402) (69.5,0.95983) (69.6,0.95903) (69.7,0.95327) (69.8,0.94158) (69.9,0.94210) (70,0.94500) (70.1,0.94387) (70.2,0.94466) (70.3,0.94903) (70.4,0.94504) (70.5,0.92666) (70.6,0.93840) (70.7,0.95116) (70.8,0.95897) (70.9,0.95528) (71,0.94952) (71.1,0.94766) (71.2,0.95088) (71.3,0.95336) (71.4,0.95724) (71.5,0.96690) (71.6,0.95463) (71.7,0.99694) (71.8,1.00131) (71.9,0.99896) (72,0.99578) (72.1,0.99479) (72.2,0.98825) (72.3,0.98604) (72.4,0.97982) (72.5,0.99232) (72.6,0.99540) (72.7,0.97019) (72.8,0.96317) (72.9,0.95967) (73,0.94442) (73.1,0.94524) (73.2,0.93979) (73.3,0.94330) (73.4,0.94395) (73.5,0.94652) (73.6,0.93222) (73.7,0.93165) (73.8,0.93607) (73.9,0.92850) (74,0.92765) (74.1,0.92882) (74.2,0.92901) (74.3,0.92542) (74.4,0.91864) (74.5,0.92013) (74.6,0.89871) (74.7,0.88569) (74.8,0.86435) (74.9,0.86424) (75,0.86422) (75.1,0.86440) (75.2,0.85733) (75.3,0.85688) (75.4,0.83018) (75.5,0.83010) (75.6,0.82905) (75.7,0.82954) (75.8,0.82830) (75.9,0.81647) (76,0.81470) (76.1,0.81575) (76.2,0.79737) (76.3,0.80056) (76.4,0.78375) (76.5,0.78253) (76.6,0.78461) (76.7,0.77149) (76.8,0.78310) (76.9,0.78373) (77,0.76385) (77.1,0.76422) (77.2,0.76579) (77.3,0.76721) (77.4,0.76667) (77.5,0.78782) (77.6,0.79082) (77.7,0.79784) (77.8,0.80233) (77.9,0.79908) (78,0.83920) (78.1,0.84461) (78.2,0.87390) (78.3,0.90364) (78.4,1.01935) (78.5,1.02157) (78.6,1.02694) (78.7,1.02695) (78.8,1.05740) (78.9,1.05840) (79,1.05869) (79.1,1.06688) (79.2,1.10563) (79.3,1.11038) (79.4,1.09727) (79.5,1.10478) (79.6,1.10615) (79.7,1.14781) (79.8,1.14691) (79.9,1.14372) (80,1.14555) (80.1,1.14966) (80.2,1.15411) (80.3,1.18309) (80.4,1.19437) (80.5,1.19421) (80.6,1.18683) (80.7,1.20402) (80.8,1.19714) (80.9,1.20159) (81,1.21146) (81.1,1.21152) (81.2,1.20717) (81.3,1.20492) (81.4,1.20265) (81.5,1.20216) (81.6,1.20390) (81.7,1.17517) (81.8,1.17138) (81.9,1.17149) (82,1.21236) (82.1,1.21381) (82.2,1.21270) (82.3,1.21276) (82.4,1.21292) (82.5,1.21506) (82.6,1.22793) (82.7,1.23349) (82.8,1.23475) (82.9,1.23367) (83,1.24064) (83.1,1.24048) (83.2,1.23788) (83.3,1.25256) (83.4,1.25071) (83.5,1.24876) (83.6,1.24720) (83.7,1.26368) (83.8,1.25867) (83.9,1.26320) (84,1.26328) (84.1,1.26203) (84.2,1.26180) (84.3,1.26126) (84.4,1.28423) (84.5,1.28018) (84.6,1.27783) (84.7,1.27708) (84.8,1.27624) (84.9,1.27716) (85,1.28001) (85.1,1.30848) (85.2,1.30946) (85.3,1.30937) (85.4,1.30797) (85.5,1.31362) (85.6,1.31369) (85.7,1.31387) (85.8,1.31840) (85.9,1.31752) (86,1.32284) (86.1,1.32306) (86.2,1.32381) (86.3,1.32477) (86.4,1.32394) (86.5,1.32394) (86.6,1.35094) (86.7,1.35294) (86.8,1.34993) (86.9,1.35045) (87,1.35112) (87.1,1.35206) (87.2,1.35360) (87.3,1.35247) (87.4,1.35953) (87.5,1.34358) (87.6,1.34203) (87.7,1.33664) (87.8,1.33443) (87.9,1.34096) (88,1.31552) (88.1,1.31031) (88.2,1.27981) (88.3,1.26801) (88.4,1.17551) (88.5,1.17323) (88.6,1.16731) (88.7,1.16679) (88.8,1.13843) (88.9,1.13882) (89,1.15440) (89.1,1.14463) (89.2,1.10192) (89.3,1.09900) (89.4,1.09421) (89.5,1.08290) (89.6,1.08737) (89.7,1.05040) (89.8,1.04962) (89.9,1.04974) (90,1.05777) (90.1,1.05916) (90.2,1.05479) (90.3,1.01160) (90.4,0.99322) (90.5,0.99505) (90.6,0.99926) (90.7,0.97534) (90.8,0.97464) (90.9,0.97088) (91,0.96295) (91.1,0.96291) (91.2,0.98897) (91.3,0.98891) (91.4,0.98908) (91.5,0.97805) (91.6,0.97636) (91.7,0.96708) (91.8,0.97239) (91.9,0.97251) (92,0.93330) (92.1,0.92985) (92.2,0.93852) (92.3,0.93847) (92.4,0.93867) (92.5,0.94424) (92.6,0.93087) (92.7,0.92440) (92.8,0.93101) (92.9,0.93521) (93,0.92546) (93.1,0.94326) (93.2,0.95669) (93.3,0.93247) (93.4,0.93286) (93.5,0.93282) (93.6,0.93154) (93.7,0.91316) (93.8,0.91205) (93.9,0.90861) (94,0.90864) (94.1,0.91137) (94.2,0.91118) (94.3,0.91069) (94.4,0.88085) (94.5,0.88462) (94.6,0.88456) (94.7,0.88462) (94.8,0.88447) (94.9,0.88374) (95,0.87534) (95.1,0.82175) (95.2,0.81894) (95.3,0.81817)
\end{pspicture}
\begin{pspicture}(-5.5,-0.8)(3.5,5.8)
\psset{xunit=0.07,yunit=17}
\psline{->}(0,0)(103,0) \psline{->}(0,0)(0,0.3) \rput(109,0){day} \rput(18,0.3){Correlation}
\scriptsize \scriptsize \psline(17.1,-0.006)(17.1,0.006) \rput(17.1,-0.018){01/2008} \scriptsize \psline(43.2,-0.006)(43.2,0.006) \rput(43.2,-0.018){01/2009} \scriptsize \psline(69.3,-0.006)(69.3,0.006) \rput(69.3,-0.018){01/2010} \psline(92.4,-0.006)(92.4,0.006) \rput(92.4,-0.018){01/2011} \psline(-1.4,0.1)(1.4,0.1) \rput(-5,0.1){$0.1$} \psline(-1.4,0.2)(1.4,0.2) \rput(-5,0.2){$0.2$} \rput(-8,0.3){b)}
\psline[linecolor=blue](1,0.1054) (1.1,0.1067) (1.2,0.1060) (1.3,0.1068) (1.4,0.1072) (1.5,0.1078) (1.6,0.1099) (1.7,0.1103) (1.8,0.1135) (1.9,0.1125) (2,0.1124) (2.1,0.1144) (2.2,0.1165) (2.3,0.1202) (2.4,0.1220) (2.5,0.1223) (2.6,0.1225) (2.7,0.1208) (2.8,0.1206) (2.9,0.1229) (3,0.1233) (3.1,0.1244) (3.2,0.1201) (3.3,0.1206) (3.4,0.1202) (3.5,0.1206) (3.6,0.1222) (3.7,0.1226) (3.8,0.1229) (3.9,0.1235) (4,0.1210) (4.1,0.1232) (4.2,0.1227) (4.3,0.1257) (4.4,0.1260) (4.5,0.1255) (4.6,0.1262) (4.7,0.1258) (4.8,0.1260) (4.9,0.1260) (5,0.1158) (5.1,0.1167) (5.2,0.1168) (5.3,0.1236) (5.4,0.1259) (5.5,0.1248) (5.6,0.1235) (5.7,0.1225) (5.8,0.1228) (5.9,0.1299) (6,0.1217) (6.1,0.1256) (6.2,0.1266) (6.3,0.1366) (6.4,0.1364) (6.5,0.1350) (6.6,0.1367) (6.7,0.1445) (6.8,0.1489) (6.9,0.1526) (7,0.1512) (7.1,0.1510) (7.2,0.1541) (7.3,0.1537) (7.4,0.1550) (7.5,0.1512) (7.6,0.1536) (7.7,0.1537) (7.8,0.1540) (7.9,0.1540) (8,0.1545) (8.1,0.1538) (8.2,0.1545) (8.3,0.1541) (8.4,0.1558) (8.5,0.1572) (8.6,0.1550) (8.7,0.1546) (8.8,0.1546) (8.9,0.1551) (9,0.1574) (9.1,0.1598) (9.2,0.1604) (9.3,0.1595) (9.4,0.1592) (9.5,0.1595) (9.6,0.1591) (9.7,0.1598) (9.8,0.1603) (9.9,0.1600) (10,0.1631) (10.1,0.1625) (10.2,0.1625) (10.3,0.1629) (10.4,0.1639) (10.5,0.1628) (10.6,0.1636) (10.7,0.1640) (10.8,0.1641) (10.9,0.1632) (11,0.1640) (11.1,0.1652) (11.2,0.1641) (11.3,0.1645) (11.4,0.1684) (11.5,0.1672) (11.6,0.1667) (11.7,0.1657) (11.8,0.1637) (11.9,0.1605) (12,0.1601) (12.1,0.1612) (12.2,0.1597) (12.3,0.1588) (12.4,0.1569) (12.5,0.1558) (12.6,0.1553) (12.7,0.1594) (12.8,0.1597) (12.9,0.1624) (13,0.1622) (13.1,0.1616) (13.2,0.1606) (13.3,0.1642) (13.4,0.1650) (13.5,0.1689) (13.6,0.1673) (13.7,0.1688) (13.8,0.1686) (13.9,0.1699) (14,0.1712) (14.1,0.1715) (14.2,0.1721) (14.3,0.1732) (14.4,0.1693) (14.5,0.1691) (14.6,0.1684) (14.7,0.1725) (14.8,0.1730) (14.9,0.1716) (15,0.1719) (15.1,0.1727) (15.2,0.1732) (15.3,0.1747) (15.4,0.1696) (15.5,0.1696) (15.6,0.1679) (15.7,0.1697) (15.8,0.1720) (15.9,0.1765) (16,0.1719) (16.1,0.1727) (16.2,0.1705) (16.3,0.1687) (16.4,0.1636) (16.5,0.1638) (16.6,0.1634) (16.7,0.1605) (16.8,0.1540) (16.9,0.1506) (17,0.1460) (17.1,0.1540) (17.2,0.1567) (17.3,0.1540) (17.4,0.1591) (17.5,0.1686) (17.6,0.1671) (17.7,0.1636) (17.8,0.1710) (17.9,0.1693) (18,0.1664) (18.1,0.1663) (18.2,0.1654) (18.3,0.1647) (18.4,0.1695) (18.5,0.1664) (18.6,0.1709) (18.7,0.1707) (18.8,0.1708) (18.9,0.1694) (19,0.1688) (19.1,0.1686) (19.2,0.1667) (19.3,0.1649) (19.4,0.1646) (19.5,0.1673) (19.6,0.1670) (19.7,0.1668) (19.8,0.1650) (19.9,0.1634) (20,0.1639) (20.1,0.1595) (20.2,0.1593) (20.3,0.1602) (20.4,0.1640) (20.5,0.1611) (20.6,0.1607) (20.7,0.1606) (20.8,0.1632) (20.9,0.1652) (21,0.1646) (21.1,0.1670) (21.2,0.1659) (21.3,0.1650) (21.4,0.1701) (21.5,0.1686) (21.6,0.1708) (21.7,0.1705) (21.8,0.1698) (21.9,0.1684) (22,0.1677) (22.1,0.1696) (22.2,0.1699) (22.3,0.1694) (22.4,0.1746) (22.5,0.1751) (22.6,0.1751) (22.7,0.1755) (22.8,0.1734) (22.9,0.1742) (23,0.1726) (23.1,0.1721) (23.2,0.1753) (23.3,0.1756) (23.4,0.1755) (23.5,0.1764) (23.6,0.1730) (23.7,0.1742) (23.8,0.1723) (23.9,0.1717) (24,0.1698) (24.1,0.1677) (24.2,0.1677) (24.3,0.1648) (24.4,0.1635) (24.5,0.1648) (24.6,0.1654) (24.7,0.1662) (24.8,0.1619) (24.9,0.1609) (25,0.1610) (25.1,0.1601) (25.2,0.1598) (25.3,0.1599) (25.4,0.1587) (25.5,0.1578) (25.6,0.1578) (25.7,0.1575) (25.8,0.1588) (25.9,0.1576) (26,0.1526) (26.1,0.1559) (26.2,0.1559) (26.3,0.1557) (26.4,0.1573) (26.5,0.1556) (26.6,0.1557) (26.7,0.1566) (26.8,0.1568) (26.9,0.1570) (27,0.1576) (27.1,0.1611) (27.2,0.1562) (27.3,0.1557) (27.4,0.1571) (27.5,0.1523) (27.6,0.1419) (27.7,0.1429) (27.8,0.1446) (27.9,0.1391) (28,0.1388) (28.1,0.1419) (28.2,0.1420) (28.3,0.1427) (28.4,0.1424) (28.5,0.1390) (28.6,0.1393) (28.7,0.1337) (28.8,0.1352) (28.9,0.1360) (29,0.1354) (29.1,0.1354) (29.2,0.1321) (29.3,0.1278) (29.4,0.1272) (29.5,0.1270) (29.6,0.1241) (29.7,0.1237) (29.8,0.1265) (29.9,0.1280) (30,0.1292) (30.1,0.1282) (30.2,0.1269) (30.3,0.1262) (30.4,0.1270) (30.5,0.1254) (30.6,0.1261) (30.7,0.1276) (30.8,0.1266) (30.9,0.1247) (31,0.1231) (31.1,0.1260) (31.2,0.1243) (31.3,0.1264) (31.4,0.1263) (31.5,0.1215) (31.6,0.1198) (31.7,0.1171) (31.8,0.1169) (31.9,0.1166) (32,0.1163) (32.1,0.1165) (32.2,0.1145) (32.3,0.1146) (32.4,0.1143) (32.5,0.1071) (32.6,0.1061) (32.7,0.1088) (32.8,0.1070) (32.9,0.1075) (33,0.1097) (33.1,0.1096) (33.2,0.1103) (33.3,0.1063) (33.4,0.1080) (33.5,0.1127) (33.6,0.1085) (33.7,0.1082) (33.8,0.1105) (33.9,0.1139) (34,0.1141) (34.1,0.1137) (34.2,0.1247) (34.3,0.1236) (34.4,0.1303) (34.5,0.1336) (34.6,0.1376) (34.7,0.1394) (34.8,0.1390) (34.9,0.1387) (35,0.1379) (35.1,0.1402) (35.2,0.1451) (35.3,0.1457) (35.4,0.1447) (35.5,0.1458) (35.6,0.1476) (35.7,0.1563) (35.8,0.1653) (35.9,0.1642) (36,0.1692) (36.1,0.1668) (36.2,0.1741) (36.3,0.1719) (36.4,0.1822) (36.5,0.1802) (36.6,0.1790) (36.7,0.1808) (36.8,0.1844) (36.9,0.1945) (37,0.2000) (37.1,0.2045) (37.2,0.2023) (37.3,0.2020) (37.4,0.2058) (37.5,0.2073) (37.6,0.2085) (37.7,0.2101) (37.8,0.2138) (37.9,0.2173) (38,0.2161) (38.1,0.2176) (38.2,0.2176) (38.3,0.2206) (38.4,0.2249) (38.5,0.2232) (38.6,0.2221) (38.7,0.2263) (38.8,0.2264) (38.9,0.2331) (39,0.2331) (39.1,0.2325) (39.2,0.2378) (39.3,0.2381) (39.4,0.2402) (39.5,0.2433) (39.6,0.2428) (39.7,0.2480) (39.8,0.2483) (39.9,0.2474) (40,0.2455) (40.1,0.2420) (40.2,0.2422) (40.3,0.2427) (40.4,0.2435) (40.5,0.2403) (40.6,0.2392) (40.7,0.2404) (40.8,0.2408) (40.9,0.2388) (41,0.2372) (41.1,0.2366) (41.2,0.2364) (41.3,0.2367) (41.4,0.2357) (41.5,0.2372) (41.6,0.2381) (41.7,0.2386) (41.8,0.2400) (41.9,0.2398) (42,0.2397) (42.1,0.2391) (42.2,0.2382) (42.3,0.2425) (42.4,0.2425) (42.5,0.2420) (42.6,0.2431) (42.7,0.2450) (42.8,0.2440) (42.9,0.2433) (43,0.2430) (43.1,0.2428) (43.2,0.2419) (43.3,0.2428) (43.4,0.2424) (43.5,0.2425) (43.6,0.2406) (43.7,0.2423) (43.8,0.2418) (43.9,0.2423) (44,0.2435) (44.1,0.2441) (44.2,0.2446) (44.3,0.2390) (44.4,0.2396) (44.5,0.2365) (44.6,0.2362) (44.7,0.2319) (44.8,0.2303) (44.9,0.2305) (45,0.2303) (45.1,0.2316) (45.2,0.2302) (45.3,0.2277) (45.4,0.2268) (45.5,0.2281) (45.6,0.2297) (45.7,0.2283) (45.8,0.2226) (45.9,0.2167) (46,0.2186) (46.1,0.2157) (46.2,0.2213) (46.3,0.2134) (46.4,0.2181) (46.5,0.2108) (46.6,0.2114) (46.7,0.2114) (46.8,0.2110) (46.9,0.2095) (47,0.2014) (47.1,0.1995) (47.2,0.1927) (47.3,0.1966) (47.4,0.1972) (47.5,0.1956) (47.6,0.1931) (47.7,0.1937) (47.8,0.1982) (47.9,0.2010) (48,0.1974) (48.1,0.1971) (48.2,0.1966) (48.3,0.1991) (48.4,0.1995) (48.5,0.1949) (48.6,0.1964) (48.7,0.1998) (48.8,0.1936) (48.9,0.1975) (49,0.1885) (49.1,0.1887) (49.2,0.1890) (49.3,0.1837) (49.4,0.1860) (49.5,0.1846) (49.6,0.1875) (49.7,0.1891) (49.8,0.1847) (49.9,0.1864) (50,0.1915) (50.1,0.1918) (50.2,0.1946) (50.3,0.1930) (50.4,0.1928) (50.5,0.1947) (50.6,0.1972) (50.7,0.1988) (50.8,0.2023) (50.9,0.2047) (51,0.2034) (51.1,0.2019) (51.2,0.2010) (51.3,0.2012) (51.4,0.2028) (51.5,0.2025) (51.6,0.2018) (51.7,0.2050) (51.8,0.2071) (51.9,0.2030) (52,0.2045) (52.1,0.2046) (52.2,0.2025) (52.3,0.2034) (52.4,0.1952) (52.5,0.1958) (52.6,0.1968) (52.7,0.1929) (52.8,0.1896) (52.9,0.1918) (53,0.1913) (53.1,0.1917) (53.2,0.1905) (53.3,0.1968) (53.4,0.1970) (53.5,0.1958) (53.6,0.1925) (53.7,0.1912) (53.8,0.1921) (53.9,0.1933) (54,0.1930) (54.1,0.1960) (54.2,0.1971) (54.3,0.1963) (54.4,0.1976) (54.5,0.1981) (54.6,0.1982) (54.7,0.1951) (54.8,0.1913) (54.9,0.1929) (55,0.1955) (55.1,0.1935) (55.2,0.1889) (55.3,0.1891) (55.4,0.1885) (55.5,0.1898) (55.6,0.1870) (55.7,0.1792) (55.8,0.1800) (55.9,0.1836) (56,0.1804) (56.1,0.1793) (56.2,0.1803) (56.3,0.1803) (56.4,0.1804) (56.5,0.1796) (56.6,0.1798) (56.7,0.1810) (56.8,0.1819) (56.9,0.1854) (57,0.1870) (57.1,0.1886) (57.2,0.1940) (57.3,0.1937) (57.4,0.1930) (57.5,0.1940) (57.6,0.1903) (57.7,0.1956) (57.8,0.1953) (57.9,0.1904) (58,0.1872) (58.1,0.1879) (58.2,0.1866) (58.3,0.1846) (58.4,0.1869) (58.5,0.1836) (58.6,0.1854) (58.7,0.1876) (58.8,0.1868) (58.9,0.1881) (59,0.1803) (59.1,0.1809) (59.2,0.1813) (59.3,0.1815) (59.4,0.1848) (59.5,0.1836) (59.6,0.1846) (59.7,0.1798) (59.8,0.1792) (59.9,0.1755) (60,0.1784) (60.1,0.1767) (60.2,0.1790) (60.3,0.1794) (60.4,0.1798) (60.5,0.1827) (60.6,0.1770) (60.7,0.1798) (60.8,0.1810) (60.9,0.1761) (61,0.1748) (61.1,0.1757) (61.2,0.1752) (61.3,0.1747) (61.4,0.1746) (61.5,0.1744) (61.6,0.1755) (61.7,0.1770) (61.8,0.1780) (61.9,0.1792) (62,0.1797) (62.1,0.1789) (62.2,0.1867) (62.3,0.1918) (62.4,0.1957) (62.5,0.1943) (62.6,0.1975) (62.7,0.1982) (62.8,0.1953) (62.9,0.1889) (63,0.1880) (63.1,0.1863) (63.2,0.1863) (63.3,0.1869) (63.4,0.1788) (63.5,0.1804) (63.6,0.1804) (63.7,0.1806) (63.8,0.1809) (63.9,0.1812) (64,0.1838) (64.1,0.1850) (64.2,0.1801) (64.3,0.1786) (64.4,0.1790) (64.5,0.1802) (64.6,0.1794) (64.7,0.1780) (64.8,0.1765) (64.9,0.1750) (65,0.1730) (65.1,0.1683) (65.2,0.1707) (65.3,0.1744) (65.4,0.1726) (65.5,0.1733) (65.6,0.1733) (65.7,0.1680) (65.8,0.1680) (65.9,0.1684) (66,0.1683) (66.1,0.1695) (66.2,0.1662) (66.3,0.1655) (66.4,0.1647) (66.5,0.1650) (66.6,0.1665) (66.7,0.1674) (66.8,0.1672) (66.9,0.1681) (67,0.1649) (67.1,0.1628) (67.2,0.1627) (67.3,0.1547) (67.4,0.1582) (67.5,0.1598) (67.6,0.1602) (67.7,0.1590) (67.8,0.1509) (67.9,0.1507) (68,0.1526) (68.1,0.1579) (68.2,0.1593) (68.3,0.1604) (68.4,0.1606) (68.5,0.1573) (68.6,0.1579) (68.7,0.1551) (68.8,0.1525) (68.9,0.1522) (69,0.1520) (69.1,0.1597) (69.2,0.1585) (69.3,0.1588) (69.4,0.1602) (69.5,0.1561) (69.6,0.1504) (69.7,0.1506) (69.8,0.1510) (69.9,0.1532) (70,0.1534) (70.1,0.1512) (70.2,0.1494) (70.3,0.1474) (70.4,0.1495) (70.5,0.1497) (70.6,0.1480) (70.7,0.1494) (70.8,0.1453) (70.9,0.1430) (71,0.1432) (71.1,0.1426) (71.2,0.1455) (71.3,0.1460) (71.4,0.1469) (71.5,0.1457) (71.6,0.1454) (71.7,0.1452) (71.8,0.1443) (71.9,0.1448) (72,0.1429) (72.1,0.1437) (72.2,0.1445) (72.3,0.1354) (72.4,0.1308) (72.5,0.1236) (72.6,0.1237) (72.7,0.1195) (72.8,0.1187) (72.9,0.1210) (73,0.1217) (73.1,0.1230) (73.2,0.1249) (73.3,0.1250) (73.4,0.1246) (73.5,0.1264) (73.6,0.1268) (73.7,0.1264) (73.8,0.1288) (73.9,0.1304) (74,0.1381) (74.1,0.1350) (74.2,0.1353) (74.3,0.1367) (74.4,0.1360) (74.5,0.1371) (74.6,0.1319) (74.7,0.1386) (74.8,0.1401) (74.9,0.1371) (75,0.1384) (75.1,0.1384) (75.2,0.1477) (75.3,0.1524) (75.4,0.1554) (75.5,0.1532) (75.6,0.1507) (75.7,0.1502) (75.8,0.1515) (75.9,0.1544) (76,0.1655) (76.1,0.1627) (76.2,0.1622) (76.3,0.1669) (76.4,0.1742) (76.5,0.1729) (76.6,0.1774) (76.7,0.1740) (76.8,0.1767) (76.9,0.1800) (77,0.1791) (77.1,0.1793) (77.2,0.1784) (77.3,0.1802) (77.4,0.1800) (77.5,0.1849) (77.6,0.1839) (77.7,0.1827) (77.8,0.1831) (77.9,0.1867) (78,0.1865) (78.1,0.1842) (78.2,0.1802) (78.3,0.1801) (78.4,0.1803) (78.5,0.1816) (78.6,0.1814) (78.7,0.1829) (78.8,0.1851) (78.9,0.1887) (79,0.1875) (79.1,0.1878) (79.2,0.1845) (79.3,0.1871) (79.4,0.1885) (79.5,0.1869) (79.6,0.1862) (79.7,0.1851) (79.8,0.1896) (79.9,0.1907) (80,0.1891) (80.1,0.1895) (80.2,0.1925) (80.3,0.1925) (80.4,0.1920) (80.5,0.1922) (80.6,0.1924) (80.7,0.1922) (80.8,0.1902) (80.9,0.1938) (81,0.1936) (81.1,0.1957) (81.2,0.1959) (81.3,0.1931) (81.4,0.1934) (81.5,0.1929) (81.6,0.1933) (81.7,0.1934) (81.8,0.1940) (81.9,0.1942) (82,0.1925) (82.1,0.1922) (82.2,0.1959) (82.3,0.1997) (82.4,0.1998) (82.5,0.1990) (82.6,0.1993) (82.7,0.2010) (82.8,0.2017) (82.9,0.2033) (83,0.2033) (83.1,0.2030) (83.2,0.2033) (83.3,0.2011) (83.4,0.2020) (83.5,0.2031) (83.6,0.2016) (83.7,0.2023) (83.8,0.2066) (83.9,0.2056) (84,0.2060) (84.1,0.2009) (84.2,0.2019) (84.3,0.2018) (84.4,0.2021) (84.5,0.2029) (84.6,0.2023) (84.7,0.2028) (84.8,0.1988) (84.9,0.1988) (85,0.1982) (85.1,0.1980) (85.2,0.1983) (85.3,0.1920) (85.4,0.1864) (85.5,0.1809) (85.6,0.1836) (85.7,0.1862) (85.8,0.1863) (85.9,0.1858) (86,0.1833) (86.1,0.1735) (86.2,0.1776) (86.3,0.1772) (86.4,0.1725) (86.5,0.1639) (86.6,0.1640) (86.7,0.1598) (86.8,0.1613) (86.9,0.1582) (87,0.1556) (87.1,0.1556) (87.2,0.1572) (87.3,0.1576) (87.4,0.1546) (87.5,0.1555) (87.6,0.1465) (87.7,0.1463) (87.8,0.1476) (87.9,0.1471) (88,0.1428) (88.1,0.1426) (88.2,0.1435) (88.3,0.1408) (88.4,0.1430) (88.5,0.1434) (88.6,0.1412) (88.7,0.1412) (88.8,0.1397) (88.9,0.1397) (89,0.1338) (89.1,0.1328) (89.2,0.1355) (89.3,0.1296) (89.4,0.1288) (89.5,0.1274) (89.6,0.1322) (89.7,0.1367) (89.8,0.1387) (89.9,0.1338) (90,0.1324) (90.1,0.1340) (90.2,0.1335) (90.3,0.1348) (90.4,0.1365) (90.5,0.1360) (90.6,0.1323) (90.7,0.1324) (90.8,0.1321) (90.9,0.1325) (91,0.1262) (91.1,0.1265) (91.2,0.1246) (91.3,0.1244) (91.4,0.1247) (91.5,0.1250) (91.6,0.1267) (91.7,0.1250) (91.8,0.1252) (91.9,0.1248) (92,0.1245) (92.1,0.1241) (92.2,0.1261) (92.3,0.1199)
\end{pspicture}

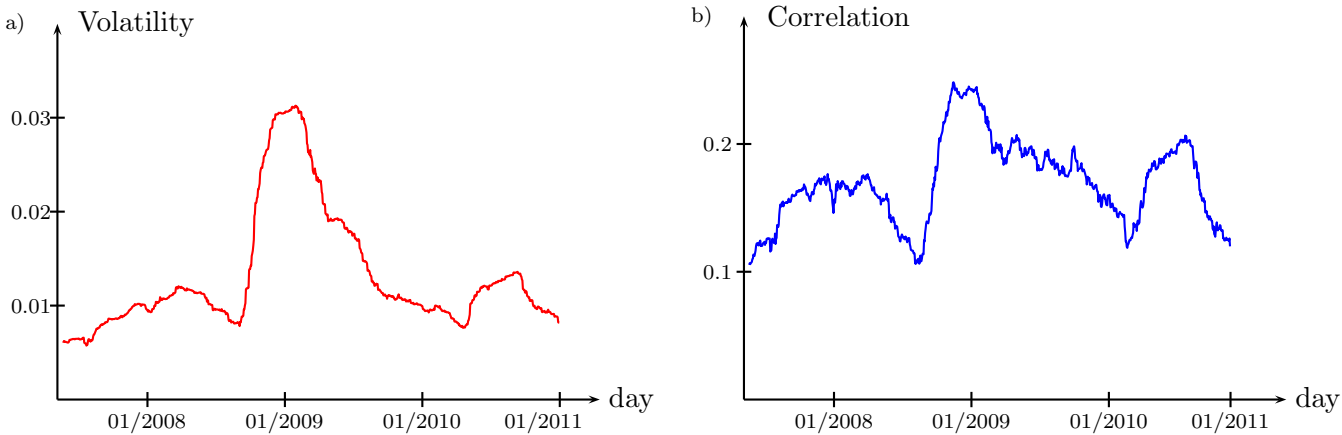
\noindent Fig. 15. a) Average volatility of the MSCI World Index in time. b) Average correlation of the 92 indices in time.

\vskip 0.3 cm

\subsection{Single step survival ratio}

Measuring survivability of connections in a network is not a difficult task, but there are many ways to do that and they are not all equivalent. The simplest measure is to consider the {\sl single step survival ratio} \cite{asset02}, here adapted to asset graphs, defined as the fraction of edges (connections) common to two consecutive asset graphs, divided by the number of edges of the asset graph. This can be easily calculated by defining the adjacency matrix $A$ \cite{Newman}, which is a matrix whose elements $a_{ij}$ are one if there is a connection between indices $i$ and $j$ and zero otherwise. In the present case, an asset graph may be represented as the distance matrix between all indices, but limited by a threshold value $T$. So, the adjacency matrix associated with this distance matrix with a cut-off has ones for distances $d_{ij}\leq T$ and zeros for $d_{ij}>T$. Now, it is fairly easy to define the single step survival ratio: it is the sum of all elements of the product of an adjacency matrix at time $t$ with an adjacency matrix at time $t+1$, divided by the number of edges at time $t$:
\begin{equation}
\label{singlestep}
S(t)=\frac{1}{M}\sum_{i,j=1}^NA_{ij}(t)A_{ij}(t+1)\ ,
\end{equation}
where $A_{ij}(t)$ is the element $ij$ of matrix A(t), $A_{ij}(t+1)$ is the element $ij$ of matrix $A(t+1)$, and $N$ is the total number of nodes in the correlation matrix (this is the same as the scalar product of both matrices divided by $M$). Since I am using a distance threshold and not the number of edges to establish an asset graph, the number of edges is not constant in time. So, I am using $M$ as the number of connections in the asset graph related with $A(t)$.

The results depend on the threshold $T$, and Figure 16 shows the results for thresholds going from $T=0.1$ to $T=0.5$ and $T=0.6$, just below and above the noise threshold. For high threshold values up to $T=1.5$, $S(t)$ assumes values that increasingly lead to $S(t)=1$ for all values of $t$, for $T=1.5$ is the distance at which all nodes are connected and so all connections survive all the time.



\vskip -0.5 cm

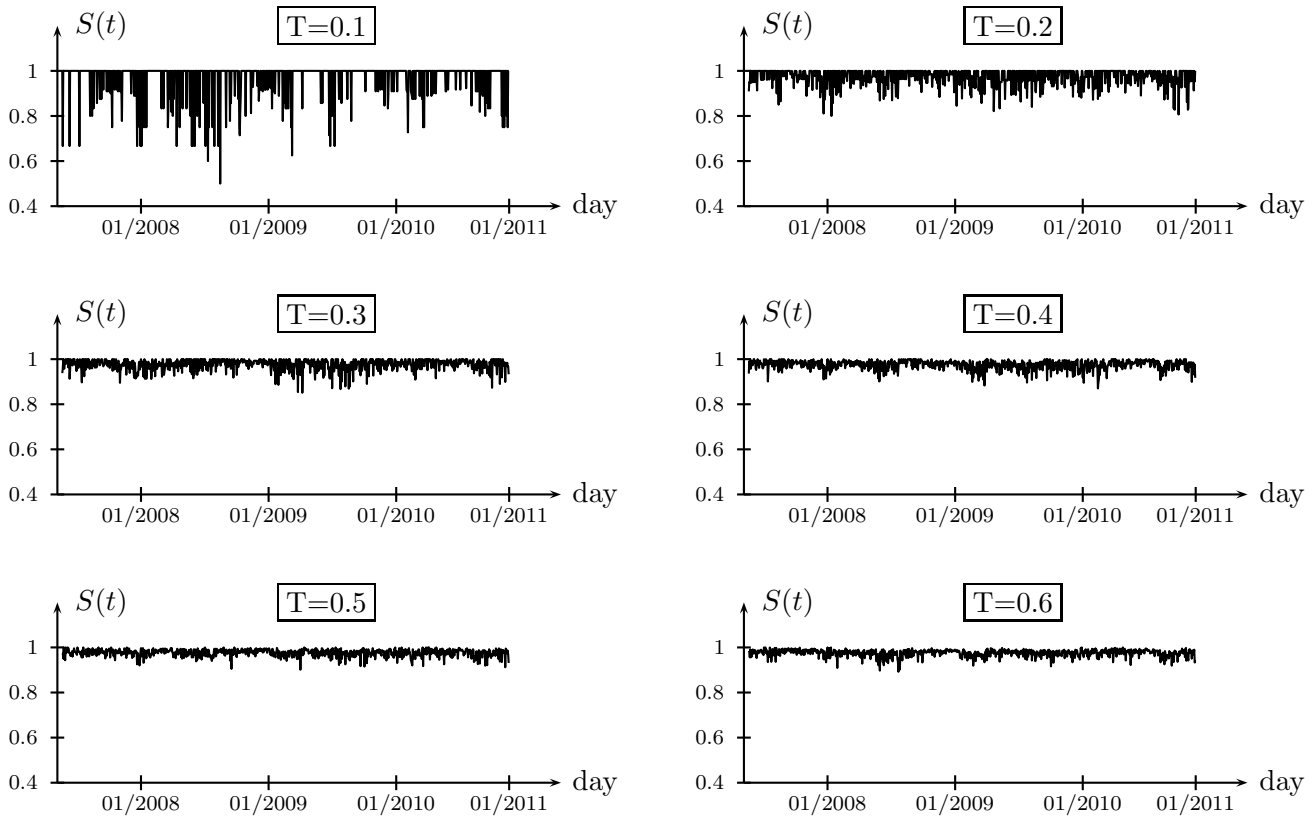
\noindent Fig. 16. Single step survival ratio for various asset graphs based on a diversity of thresholds for running windows of 100 days with steps of one day at a time.

\vskip 0.3 cm

As expected, the single step survival ratio varies much for $T=0.1$ and $T=0.2$, for one has, for these thresholds, very few connections, although strong ones. As the value of the threshold grows, one has many more connections, and then it is expected that the single step survival ratio goes slowly towards the unit value. If one considers $T=0.5$, which borders the noise limit, one then has strong variations of the single step survival ratio at 10/22/2009, 11/03/2009, 04/29/2010, 10/22/2010, and 12/20/2010, which do not coincide with any particular important event. When analyzing this, one should bear in mind that the single step survival ratio is being calculated over windows of 100 days, and so it is not very sensitive when pinpointing events that happen in one or few days.

If one now wished to compare the single step survival ratio of real data with the one of randomized data, the first striking difference between them is that, for randomized data, there are virtually no connections bellow $T=0.7$. So, one can only compare the results for connections above this distance value. Figure 17 shows the comparison of the single step survival ratio for real (graph on the left) and randomized (graph on the right) data for a typical simulation where all time series were randomly shuffled. One thing to notice is that the single step survival ratio for randomized data oscillates around $0.87$. The other is that it doesn't come close to the value $1$, so that the single step survival ratio of randomized data is never large, and it is also much more volatile (high standard deviation). The fact that the survival ratio is so high for randomized data is that it is an effect of my use of overlapping windows of 100 days with shifts of one day at a time. This also applies to the real data, but does not explain by itself the high survival ratio in that case.

\vskip 0.4 cm

\begin{pspicture}(-0.5,0)(3.5,4)
\psset{xunit=0.065,yunit=3}
\psline{->}(0,0.4)(103,0.4) \psline{->}(0,0.4)(0,1.2) \rput(110,0.4){day} \rput(9,1.2){$S(t)$} \rput(55,1.2){\boxed{Real}}
\scriptsize \scriptsize \psline(17.1,0.367)(17.1,0.433) \rput(17.1,0.3){01/2008} \scriptsize \psline(43.2,0.367)(43.2,0.433) \rput(43.2,0.3){01/2009} \scriptsize \psline(69.3,0.367)(69.3,0.433) \rput(69.3,0.3){01/2010} \psline(92.4,0.367)(92.4,0.433) \rput(92.4,0.3){01/2011} \psline(-1.4,0.4)(1.4,0.4) \rput(-7,0.4){$0.4$} \psline(-1.4,0.6)(1.4,0.6) \rput(-7,0.6){$0.6$} \psline(-1.4,0.8)(1.4,0.8) \rput(-7,0.8){$0.8$} \psline(-1.4,1)(1.4,1) \rput(-5,1){$1$}
\psline (1,0.971) (1.1,0.977) (1.2,0.947) (1.3,0.974) (1.4,0.970) (1.5,0.976) (1.6,0.949) (1.7,0.974) (1.8,0.986) (1.9,0.966) (2,0.985) (2.1,0.977) (2.2,0.972) (2.3,0.983) (2.4,0.970) (2.5,0.979) (2.6,0.970) (2.7,0.977) (2.8,0.976) (2.9,0.981) (3,0.981) (3.1,0.964) (3.2,0.965) (3.3,0.984) (3.4,0.977) (3.5,0.983) (3.6,0.979) (3.7,0.981) (3.8,0.973) (3.9,0.967) (4,0.963) (4.1,0.986) (4.2,0.972) (4.3,0.985) (4.4,0.978) (4.5,0.987) (4.6,0.973) (4.7,0.966) (4.8,0.974) (4.9,0.919) (5,0.976) (5.1,0.956) (5.2,0.971) (5.3,0.980) (5.4,0.969) (5.5,0.969) (5.6,0.965) (5.7,0.967) (5.8,0.981) (5.9,0.977) (6,0.983) (6.1,0.982) (6.2,0.978) (6.3,0.995) (6.4,0.979) (6.5,0.961) (6.6,0.989) (6.7,0.990) (6.8,0.985) (6.9,0.976) (7,0.969) (7.1,0.978) (7.2,0.985) (7.3,0.984) (7.4,0.968) (7.5,0.986) (7.6,0.986) (7.7,0.988) (7.8,0.983) (7.9,0.984) (8,0.979) (8.1,0.978) (8.2,0.977) (8.3,0.972) (8.4,0.991) (8.5,0.975) (8.6,0.976) (8.7,0.987) (8.8,0.988) (8.9,0.983) (9,0.988) (9.1,0.990) (9.2,0.980) (9.3,0.983) (9.4,0.977) (9.5,0.987) (9.6,0.977) (9.7,0.984) (9.8,0.986) (9.9,0.986) (10,0.991) (10.1,0.980) (10.2,0.981) (10.3,0.982) (10.4,0.983) (10.5,0.981) (10.6,0.981) (10.7,0.985) (10.8,0.981) (10.9,0.980) (11,0.981) (11.1,0.982) (11.2,0.980) (11.3,0.969) (11.4,0.994) (11.5,0.974) (11.6,0.974) (11.7,0.968) (11.8,0.970) (11.9,0.979) (12,0.989) (12.1,0.977) (12.2,0.969) (12.3,0.972) (12.4,0.973) (12.5,0.983) (12.6,0.972) (12.7,0.984) (12.8,0.976) (12.9,0.967) (13,0.982) (13.1,0.976) (13.2,0.981) (13.3,0.989) (13.4,0.982) (13.5,0.978) (13.6,0.961) (13.7,0.979) (13.8,0.979) (13.9,0.964) (14,0.978) (14.1,0.976) (14.2,0.982) (14.3,0.971) (14.4,0.978) (14.5,0.987) (14.6,0.970) (14.7,0.987) (14.8,0.983) (14.9,0.987) (15,0.983) (15.1,0.982) (15.2,0.974) (15.3,0.953) (15.4,0.990) (15.5,0.967) (15.6,0.978) (15.7,0.990) (15.8,0.987) (15.9,0.979) (16,0.984) (16.1,0.976) (16.2,0.982) (16.3,0.944) (16.4,0.981) (16.5,0.985) (16.6,0.971) (16.7,0.947) (16.8,0.961) (16.9,0.967) (17,0.973) (17.1,0.984) (17.2,0.959) (17.3,0.982) (17.4,0.981) (17.5,0.990) (17.6,0.958) (17.7,0.956) (17.8,0.992) (17.9,0.969) (18,0.967) (18.1,0.977) (18.2,0.988) (18.3,0.971) (18.4,0.987) (18.5,0.985) (18.6,0.990) (18.7,0.986) (18.8,0.981) (18.9,0.983) (19,0.976) (19.1,0.960) (19.2,0.979) (19.3,0.979) (19.4,0.974) (19.5,0.988) (19.6,0.972) (19.7,0.973) (19.8,0.975) (19.9,0.974) (20,0.967) (20.1,0.978) (20.2,0.981) (20.3,0.986) (20.4,0.983) (20.5,0.978) (20.6,0.979) (20.7,0.972) (20.8,0.990) (20.9,0.990) (21,0.989) (21.1,0.981) (21.2,0.972) (21.3,0.982) (21.4,0.983) (21.5,0.983) (21.6,0.985) (21.7,0.985) (21.8,0.971) (21.9,0.969) (22,0.987) (22.1,0.988) (22.2,0.982) (22.3,0.976) (22.4,0.985) (22.5,0.976) (22.6,0.980) (22.7,0.975) (22.8,0.981) (22.9,0.974) (23,0.978) (23.1,0.976) (23.2,0.989) (23.3,0.970) (23.4,0.987) (23.5,0.974) (23.6,0.982) (23.7,0.975) (23.8,0.985) (23.9,0.973) (24,0.977) (24.1,0.978) (24.2,0.967) (24.3,0.982) (24.4,0.979) (24.5,0.987) (24.6,0.971) (24.7,0.968) (24.8,0.985) (24.9,0.978) (25,0.987) (25.1,0.982) (25.2,0.988) (25.3,0.984) (25.4,0.967) (25.5,0.967) (25.6,0.983) (25.7,0.987) (25.8,0.975) (25.9,0.962) (26,0.975) (26.1,0.992) (26.2,0.978) (26.3,0.976) (26.4,0.974) (26.5,0.981) (26.6,0.980) (26.7,0.977) (26.8,0.978) (26.9,0.976) (27,0.982) (27.1,0.959) (27.2,0.973) (27.3,0.978) (27.4,0.968) (27.5,0.917) (27.6,0.955) (27.7,0.964) (27.8,0.907) (27.9,0.972) (28,0.964) (28.1,0.972) (28.2,0.979) (28.3,0.962) (28.4,0.925) (28.5,0.968) (28.6,0.947) (28.7,0.968) (28.8,0.952) (28.9,0.968) (29,0.961) (29.1,0.947) (29.2,0.938) (29.3,0.959) (29.4,0.958) (29.5,0.948) (29.6,0.973) (29.7,0.969) (29.8,0.980) (29.9,0.964) (30,0.940) (30.1,0.957) (30.2,0.969) (30.3,0.953) (30.4,0.947) (30.5,0.975) (30.6,0.975) (30.7,0.967) (30.8,0.969) (30.9,0.956) (31,0.974) (31.1,0.963) (31.2,0.964) (31.3,0.971) (31.4,0.918) (31.5,0.965) (31.6,0.942) (31.7,0.951) (31.8,0.976) (31.9,0.940) (32,0.971) (32.1,0.953) (32.2,0.968) (32.3,0.972) (32.4,0.941) (32.5,0.966) (32.6,0.963) (32.7,0.978) (32.8,0.963) (32.9,0.970) (33,0.970) (33.1,0.980) (33.2,0.946) (33.3,0.974) (33.4,0.976) (33.5,0.980) (33.6,0.948) (33.7,0.931) (33.8,0.976) (33.9,0.973) (34,0.954) (34.1,0.942) (34.2,0.973) (34.3,0.938) (34.4,0.970) (34.5,0.949) (34.6,0.973) (34.7,0.973) (34.8,0.966) (34.9,0.963) (35,0.958) (35.1,0.980) (35.2,0.973) (35.3,0.972) (35.4,0.953) (35.5,0.973) (35.6,0.958) (35.7,0.977) (35.8,0.979) (35.9,0.943) (36,0.970) (36.1,0.935) (36.2,0.975) (36.3,0.968) (36.4,0.978) (36.5,0.953) (36.6,0.957) (36.7,0.968) (36.8,0.978) (36.9,0.989) (37,0.974) (37.1,0.977) (37.2,0.960) (37.3,0.963) (37.4,0.969) (37.5,0.983) (37.6,0.974) (37.7,0.990) (37.8,0.980) (37.9,0.980) (38,0.966) (38.1,0.975) (38.2,0.967) (38.3,0.988) (38.4,0.983) (38.5,0.969) (38.6,0.975) (38.7,0.980) (38.8,0.977) (38.9,0.982) (39,0.957) (39.1,0.973) (39.2,0.985) (39.3,0.974) (39.4,0.978) (39.5,0.980) (39.6,0.979) (39.7,0.989) (39.8,0.977) (39.9,0.979) (40,0.979) (40.1,0.954) (40.2,0.988) (40.3,0.988) (40.4,0.982) (40.5,0.974) (40.6,0.977) (40.7,0.985) (40.8,0.988) (40.9,0.980) (41,0.991) (41.1,0.979) (41.2,0.989) (41.3,0.982) (41.4,0.990) (41.5,0.978) (41.6,0.985) (41.7,0.979) (41.8,0.989) (41.9,0.985) (42,0.985) (42.1,0.985) (42.2,0.984) (42.3,0.988) (42.4,0.982) (42.5,0.983) (42.6,0.986) (42.7,0.992) (42.8,0.990) (42.9,0.992) (43,0.982) (43.1,0.988) (43.2,0.990) (43.3,0.990) (43.4,0.992) (43.5,0.983) (43.6,0.988) (43.7,0.986) (43.8,0.981) (43.9,0.989) (44,0.988) (44.1,0.983) (44.2,0.969) (44.3,0.987) (44.4,0.978) (44.5,0.974) (44.6,0.968) (44.7,0.978) (44.8,0.989) (44.9,0.988) (45,0.984) (45.1,0.984) (45.2,0.981) (45.3,0.981) (45.4,0.983) (45.5,0.982) (45.6,0.972) (45.7,0.961) (45.8,0.951) (45.9,0.983) (46,0.970) (46.1,0.975) (46.2,0.957) (46.3,0.979) (46.4,0.966) (46.5,0.972) (46.6,0.976) (46.7,0.981) (46.8,0.972) (46.9,0.954) (47,0.953) (47.1,0.958) (47.2,0.986) (47.3,0.978) (47.4,0.954) (47.5,0.968) (47.6,0.947) (47.7,0.974) (47.8,0.979) (47.9,0.979) (48,0.972) (48.1,0.972) (48.2,0.972) (48.3,0.964) (48.4,0.972) (48.5,0.973) (48.6,0.975) (48.7,0.963) (48.8,0.968) (48.9,0.951) (49,0.968) (49.1,0.973) (49.2,0.952) (49.3,0.955) (49.4,0.979) (49.5,0.953) (49.6,0.984) (49.7,0.941) (49.8,0.988) (49.9,0.985) (50,0.987) (50.1,0.977) (50.2,0.963) (50.3,0.977) (50.4,0.974) (50.5,0.976) (50.6,0.983) (50.7,0.986) (50.8,0.987) (50.9,0.982) (51,0.964) (51.1,0.976) (51.2,0.986) (51.3,0.985) (51.4,0.990) (51.5,0.979) (51.6,0.987) (51.7,0.986) (51.8,0.965) (51.9,0.972) (52,0.978) (52.1,0.967) (52.2,0.978) (52.3,0.955) (52.4,0.982) (52.5,0.986) (52.6,0.954) (52.7,0.949) (52.8,0.981) (52.9,0.984) (53,0.979) (53.1,0.979) (53.2,0.981) (53.3,0.991) (53.4,0.981) (53.5,0.965) (53.6,0.969) (53.7,0.989) (53.8,0.976) (53.9,0.983) (54,0.979) (54.1,0.989) (54.2,0.978) (54.3,0.978) (54.4,0.989) (54.5,0.987) (54.6,0.965) (54.7,0.970) (54.8,0.985) (54.9,0.984) (55,0.979) (55.1,0.963) (55.2,0.981) (55.3,0.981) (55.4,0.989) (55.5,0.968) (55.6,0.974) (55.7,0.989) (55.8,0.984) (55.9,0.983) (56,0.975) (56.1,0.986) (56.2,0.978) (56.3,0.975) (56.4,0.982) (56.5,0.983) (56.6,0.980) (56.7,0.988) (56.8,0.982) (56.9,0.990) (57,0.985) (57.1,0.977) (57.2,0.997) (57.3,0.974) (57.4,0.980) (57.5,0.957) (57.6,0.987) (57.7,0.986) (57.8,0.972) (57.9,0.969) (58,0.980) (58.1,0.977) (58.2,0.962) (58.3,0.984) (58.4,0.965) (58.5,0.988) (58.6,0.983) (58.7,0.980) (58.8,0.986) (58.9,0.949) (59,0.983) (59.1,0.983) (59.2,0.980) (59.3,0.984) (59.4,0.981) (59.5,0.989) (59.6,0.968) (59.7,0.970) (59.8,0.965) (59.9,0.976) (60,0.963) (60.1,0.985) (60.2,0.980) (60.3,0.985) (60.4,0.984) (60.5,0.967) (60.6,0.979) (60.7,0.986) (60.8,0.963) (60.9,0.979) (61,0.985) (61.1,0.988) (61.2,0.977) (61.3,0.977) (61.4,0.977) (61.5,0.987) (61.6,0.981) (61.7,0.964) (61.8,0.978) (61.9,0.985) (62,0.967) (62.1,0.981) (62.2,0.996) (62.3,0.988) (62.4,0.987) (62.5,0.965) (62.6,0.984) (62.7,0.980) (62.8,0.953) (62.9,0.969) (63,0.978) (63.1,0.986) (63.2,0.973) (63.3,0.951) (63.4,0.982) (63.5,0.985) (63.6,0.973) (63.7,0.975) (63.8,0.983) (63.9,0.985) (64,0.981) (64.1,0.969) (64.2,0.961) (64.3,0.981) (64.4,0.967) (64.5,0.984) (64.6,0.983) (64.7,0.953) (64.8,0.975) (64.9,0.965) (65,0.969) (65.1,0.976) (65.2,0.984) (65.3,0.978) (65.4,0.969) (65.5,0.981) (65.6,0.960) (65.7,0.977) (65.8,0.981) (65.9,0.966) (66,0.981) (66.1,0.956) (66.2,0.963) (66.3,0.962) (66.4,0.982) (66.5,0.976) (66.6,0.979) (66.7,0.973) (66.8,0.980) (66.9,0.970) (67,0.961) (67.1,0.976) (67.2,0.936) (67.3,0.965) (67.4,0.968) (67.5,0.976) (67.6,0.962) (67.7,0.923) (67.8,0.974) (67.9,0.978) (68,0.980) (68.1,0.980) (68.2,0.967) (68.3,0.976) (68.4,0.941) (68.5,0.981) (68.6,0.954) (68.7,0.955) (68.8,0.957) (68.9,0.973) (69,0.961) (69.1,0.986) (69.2,0.961) (69.3,0.970) (69.4,0.960) (69.5,0.959) (69.6,0.981) (69.7,0.985) (69.8,0.979) (69.9,0.980) (70,0.954) (70.1,0.962) (70.2,0.953) (70.3,0.976) (70.4,0.978) (70.5,0.951) (70.6,0.975) (70.7,0.956) (70.8,0.957) (70.9,0.967) (71,0.956) (71.1,0.981) (71.2,0.982) (71.3,0.963) (71.4,0.962) (71.5,0.980) (71.6,0.978) (71.7,0.962) (71.8,0.962) (71.9,0.961) (72,0.977) (72.1,0.980) (72.2,0.943) (72.3,0.954) (72.4,0.932) (72.5,0.958) (72.6,0.947) (72.7,0.957) (72.8,0.961) (72.9,0.980) (73,0.956) (73.1,0.975) (73.2,0.977) (73.3,0.972) (73.4,0.974) (73.5,0.974) (73.6,0.974) (73.7,0.979) (73.8,0.976) (73.9,0.980) (74,0.987) (74.1,0.970) (74.2,0.968) (74.3,0.971) (74.4,0.975) (74.5,0.955) (74.6,0.960) (74.7,0.982) (74.8,0.959) (74.9,0.969) (75,0.979) (75.1,0.968) (75.2,0.976) (75.3,0.975) (75.4,0.979) (75.5,0.936) (75.6,0.943) (75.7,0.980) (75.8,0.969) (75.9,0.989) (76,0.990) (76.1,0.965) (76.2,0.957) (76.3,0.980) (76.4,0.988) (76.5,0.971) (76.6,0.979) (76.7,0.938) (76.8,0.968) (76.9,0.980) (77,0.980) (77.1,0.980) (77.2,0.965) (77.3,0.977) (77.4,0.961) (77.5,0.990) (77.6,0.974) (77.7,0.976) (77.8,0.982) (77.9,0.987) (78,0.962) (78.1,0.967) (78.2,0.984) (78.3,0.979) (78.4,0.982) (78.5,0.992) (78.6,0.973) (78.7,0.982) (78.8,0.976) (78.9,0.977) (79,0.980) (79.1,0.929) (79.2,0.968) (79.3,0.973) (79.4,0.971) (79.5,0.973) (79.6,0.978) (79.7,0.975) (79.8,0.985) (79.9,0.980) (80,0.985) (80.1,0.981) (80.2,0.989) (80.3,0.978) (80.4,0.973) (80.5,0.983) (80.6,0.980) (80.7,0.983) (80.8,0.976) (80.9,0.988) (81,0.981) (81.1,0.982) (81.2,0.973) (81.3,0.981) (81.4,0.985) (81.5,0.974) (81.6,0.982) (81.7,0.987) (81.8,0.983) (81.9,0.970) (82,0.975) (82.1,0.972) (82.2,0.984) (82.3,0.989) (82.4,0.983) (82.5,0.988) (82.6,0.986) (82.7,0.978) (82.8,0.982) (82.9,0.984) (83,0.983) (83.1,0.952) (83.2,0.986) (83.3,0.984) (83.4,0.983) (83.5,0.975) (83.6,0.986) (83.7,0.984) (83.8,0.975) (83.9,0.986) (84,0.978) (84.1,0.985) (84.2,0.980) (84.3,0.989) (84.4,0.989) (84.5,0.976) (84.6,0.985) (84.7,0.965) (84.8,0.982) (84.9,0.978) (85,0.983) (85.1,0.980) (85.2,0.947) (85.3,0.956) (85.4,0.935) (85.5,0.962) (85.6,0.979) (85.7,0.982) (85.8,0.973) (85.9,0.971) (86,0.929) (86.1,0.980) (86.2,0.968) (86.3,0.958) (86.4,0.932) (86.5,0.973) (86.6,0.952) (86.7,0.978) (86.8,0.920) (86.9,0.948) (87,0.973) (87.1,0.958) (87.2,0.976) (87.3,0.958) (87.4,0.977) (87.5,0.925) (87.6,0.963) (87.7,0.973) (87.8,0.966) (87.9,0.953) (88,0.965) (88.1,0.963) (88.2,0.964) (88.3,0.972) (88.4,0.972) (88.5,0.964) (88.6,0.966) (88.7,0.960) (88.8,0.976) (88.9,0.946) (89,0.966) (89.1,0.950) (89.2,0.939) (89.3,0.956) (89.4,0.968) (89.5,0.976) (89.6,0.977) (89.7,0.964) (89.8,0.947) (89.9,0.971) (90,0.964) (90.1,0.968) (90.2,0.956) (90.3,0.983) (90.4,0.972) (90.5,0.962) (90.6,0.973) (90.7,0.967) (90.8,0.966) (90.9,0.951) (91,0.978) (91.1,0.971) (91.2,0.972) (91.3,0.988) (91.4,0.978) (91.5,0.978) (91.6,0.965) (91.7,0.981) (91.8,0.986) (91.9,0.977) (92,0.974) (92.1,0.983) (92.2,0.948) (92.3,0.923)
\end{pspicture}
\begin{pspicture}(-5.5,0)(3.5,4)
\psset{xunit=0.065,yunit=3}
\psline{->}(0,0.4)(103,0.4) \psline{->}(0,0.4)(0,1.2) \rput(110,0.4){day} \rput(9,1.2){$S(t)$} \rput(55,1.2){\boxed{Randomized}}
\scriptsize \scriptsize \psline(17.1,0.367)(17.1,0.433) \rput(17.1,0.3){01/2008} \scriptsize \psline(43.2,0.367)(43.2,0.433) \rput(43.2,0.3){01/2009} \scriptsize \psline(69.3,0.367)(69.3,0.433) \rput(69.3,0.3){01/2010} \psline(92.4,0.367)(92.4,0.433) \rput(92.4,0.3){01/2011} \psline(-1.4,0.4)(1.4,0.4) \rput(-7,0.4){$0.4$} \psline(-1.4,0.6)(1.4,0.6) \rput(-7,0.6){$0.6$} \psline(-1.4,0.8)(1.4,0.8) \rput(-7,0.8){$0.8$} \psline(-1.4,1)(1.4,1) \rput(-5,1){$1$}
\psline (1,0.891) (1.1,0.892) (1.2,0.906) (1.3,0.928) (1.4,0.877) (1.5,0.913) (1.6,0.952) (1.7,0.914) (1.8,0.880) (1.9,0.863) (2,0.876) (2.1,0.865) (2.2,0.933) (2.3,0.874) (2.4,0.913) (2.5,0.862) (2.6,0.825) (2.7,0.882) (2.8,0.866) (2.9,0.926) (3,0.796) (3.1,0.889) (3.2,0.874) (3.3,0.906) (3.4,0.850) (3.5,0.903) (3.6,0.895) (3.7,0.917) (3.8,0.851) (3.9,0.879) (4,0.874) (4.1,0.838) (4.2,0.887) (4.3,0.893) (4.4,0.886) (4.5,0.857) (4.6,0.877) (4.7,0.852) (4.8,0.913) (4.9,0.866) (5,0.928) (5.1,0.776) (5.2,0.912) (5.3,0.858) (5.4,0.836) (5.5,0.889) (5.6,0.906) (5.7,0.875) (5.8,0.883) (5.9,0.848) (6,0.913) (6.1,0.868) (6.2,0.877) (6.3,0.825) (6.4,0.840) (6.5,0.872) (6.6,0.806) (6.7,0.840) (6.8,0.907) (6.9,0.860) (7,0.917) (7.1,0.888) (7.2,0.888) (7.3,0.856) (7.4,0.819) (7.5,0.825) (7.6,0.885) (7.7,0.899) (7.8,0.843) (7.9,0.845) (8,0.862) (8.1,0.855) (8.2,0.910) (8.3,0.893) (8.4,0.836) (8.5,0.881) (8.6,0.911) (8.7,0.884) (8.8,0.899) (8.9,0.861) (9,0.855) (9.1,0.880) (9.2,0.892) (9.3,0.866) (9.4,0.930) (9.5,0.899) (9.6,0.873) (9.7,0.858) (9.8,0.805) (9.9,0.854) (10,0.816) (10.1,0.872) (10.2,0.838) (10.3,0.885) (10.4,0.882) (10.5,0.881) (10.6,0.820) (10.7,0.845) (10.8,0.906) (10.9,0.913) (11,0.861) (11.1,0.869) (11.2,0.869) (11.3,0.854) (11.4,0.867) (11.5,0.856) (11.6,0.904) (11.7,0.861) (11.8,0.821) (11.9,0.865) (12,0.904) (12.1,0.926) (12.2,0.889) (12.3,0.905) (12.4,0.851) (12.5,0.856) (12.6,0.884) (12.7,0.896) (12.8,0.909) (12.9,0.930) (13,0.794) (13.1,0.842) (13.2,0.879) (13.3,0.863) (13.4,0.901) (13.5,0.845) (13.6,0.953) (13.7,0.891) (13.8,0.874) (13.9,0.827) (14,0.853) (14.1,0.874) (14.2,0.899) (14.3,0.865) (14.4,0.853) (14.5,0.879) (14.6,0.911) (14.7,0.896) (14.8,0.904) (14.9,0.923) (15,0.877) (15.1,0.867) (15.2,0.830) (15.3,0.830) (15.4,0.854) (15.5,0.907) (15.6,0.910) (15.7,0.926) (15.8,0.828) (15.9,0.915) (16,0.895) (16.1,0.878) (16.2,0.903) (16.3,0.861) (16.4,0.915) (16.5,0.884) (16.6,0.882) (16.7,0.865) (16.8,0.846) (16.9,0.900) (17,0.887) (17.1,0.849) (17.2,0.885) (17.3,0.879) (17.4,0.881) (17.5,0.829) (17.6,0.849) (17.7,0.873) (17.8,0.827) (17.9,0.897) (18,0.920) (18.1,0.867) (18.2,0.812) (18.3,0.896) (18.4,0.814) (18.5,0.866) (18.6,0.827) (18.7,0.843) (18.8,0.820) (18.9,0.813) (19,0.895) (19.1,0.862) (19.2,0.897) (19.3,0.874) (19.4,0.813) (19.5,0.841) (19.6,0.933) (19.7,0.868) (19.8,0.752) (19.9,0.891) (20,0.812) (20.1,0.841) (20.2,0.814) (20.3,0.866) (20.4,0.791) (20.5,0.873) (20.6,0.835) (20.7,0.852) (20.8,0.889) (20.9,0.838) (21,0.875) (21.1,0.905) (21.2,0.889) (21.3,0.899) (21.4,0.905) (21.5,0.899) (21.6,0.882) (21.7,0.824) (21.8,0.867) (21.9,0.889) (22,0.876) (22.1,0.809) (22.2,0.826) (22.3,0.901) (22.4,0.841) (22.5,0.936) (22.6,0.859) (22.7,0.860) (22.8,0.807) (22.9,0.963) (23,0.846) (23.1,0.846) (23.2,0.895) (23.3,0.879) (23.4,0.909) (23.5,0.910) (23.6,0.859) (23.7,0.914) (23.8,0.860) (23.9,0.896) (24,0.817) (24.1,0.913) (24.2,0.854) (24.3,0.897) (24.4,0.864) (24.5,0.820) (24.6,0.865) (24.7,0.853) (24.8,0.910) (24.9,0.875) (25,0.904) (25.1,0.899) (25.2,0.830) (25.3,0.911) (25.4,0.896) (25.5,0.866) (25.6,0.844) (25.7,0.835) (25.8,0.793) (25.9,0.857) (26,0.895) (26.1,0.884) (26.2,0.842) (26.3,0.882) (26.4,0.853) (26.5,0.892) (26.6,0.876) (26.7,0.927) (26.8,0.884) (26.9,0.826) (27,0.828) (27.1,0.837) (27.2,0.839) (27.3,0.930) (27.4,0.933) (27.5,0.846) (27.6,0.886) (27.7,0.872) (27.8,0.900) (27.9,0.870) (28,0.925) (28.1,0.905) (28.2,0.853) (28.3,0.904) (28.4,0.872) (28.5,0.880) (28.6,0.873) (28.7,0.917) (28.8,0.910) (28.9,0.861) (29,0.925) (29.1,0.876) (29.2,0.887) (29.3,0.857) (29.4,0.917) (29.5,0.868) (29.6,0.872) (29.7,0.878) (29.8,0.853) (29.9,0.867) (30,0.887) (30.1,0.847) (30.2,0.854) (30.3,0.871) (30.4,0.925) (30.5,0.861) (30.6,0.814) (30.7,0.869) (30.8,0.837) (30.9,0.870) (31,0.915) (31.1,0.901) (31.2,0.858) (31.3,0.807) (31.4,0.863) (31.5,0.895) (31.6,0.867) (31.7,0.825) (31.8,0.798) (31.9,0.917) (32,0.879) (32.1,0.900) (32.2,0.832) (32.3,0.882) (32.4,0.840) (32.5,0.910) (32.6,0.869) (32.7,0.903) (32.8,0.836) (32.9,0.892) (33,0.843) (33.1,0.871) (33.2,0.824) (33.3,0.843) (33.4,0.871) (33.5,0.947) (33.6,0.860) (33.7,0.851) (33.8,0.892) (33.9,0.839) (34,0.833) (34.1,0.800) (34.2,0.841) (34.3,0.867) (34.4,0.775) (34.5,0.863) (34.6,0.843) (34.7,0.886) (34.8,0.852) (34.9,0.886) (35,0.831) (35.1,0.895) (35.2,0.868) (35.3,0.830) (35.4,0.864) (35.5,0.844) (35.6,0.843) (35.7,0.811) (35.8,0.821) (35.9,0.864) (36,0.787) (36.1,0.882) (36.2,0.848) (36.3,0.847) (36.4,0.825) (36.5,0.911) (36.6,0.862) (36.7,0.826) (36.8,0.925) (36.9,0.883) (37,0.874) (37.1,0.907) (37.2,0.829) (37.3,0.890) (37.4,0.908) (37.5,0.933) (37.6,0.819) (37.7,0.832) (37.8,0.898) (37.9,0.859) (38,0.874) (38.1,0.907) (38.2,0.870) (38.3,0.874) (38.4,0.847) (38.5,0.872) (38.6,0.854) (38.7,0.819) (38.8,0.835) (38.9,0.872) (39,0.844) (39.1,0.897) (39.2,0.884) (39.3,0.872) (39.4,0.869) (39.5,0.849) (39.6,0.789) (39.7,0.897) (39.8,0.805) (39.9,0.827) (40,0.892) (40.1,0.867) (40.2,0.859) (40.3,0.876) (40.4,0.822) (40.5,0.907) (40.6,0.782) (40.7,0.841) (40.8,0.874) (40.9,0.844) (41,0.886) (41.1,0.862) (41.2,0.880) (41.3,0.878) (41.4,0.867) (41.5,0.895) (41.6,0.860) (41.7,0.887) (41.8,0.911) (41.9,0.893) (42,0.745) (42.1,0.936) (42.2,0.885) (42.3,0.945) (42.4,0.876) (42.5,0.883) (42.6,0.885) (42.7,0.892) (42.8,0.872) (42.9,0.882) (43,0.839) (43.1,0.807) (43.2,0.846) (43.3,0.896) (43.4,0.900) (43.5,0.802) (43.6,0.868) (43.7,0.875) (43.8,0.863) (43.9,0.860) (44,0.913) (44.1,0.849) (44.2,0.880) (44.3,0.867) (44.4,0.845) (44.5,0.843) (44.6,0.888) (44.7,0.829) (44.8,0.903) (44.9,0.817) (45,0.904) (45.1,0.927) (45.2,0.928) (45.3,0.907) (45.4,0.917) (45.5,0.893) (45.6,0.934) (45.7,0.927) (45.8,0.810) (45.9,0.851) (46,0.862) (46.1,0.846) (46.2,0.864) (46.3,0.951) (46.4,0.836) (46.5,0.867) (46.6,0.844) (46.7,0.913) (46.8,0.878) (46.9,0.862) (47,0.821) (47.1,0.892) (47.2,0.860) (47.3,0.858) (47.4,0.821) (47.5,0.875) (47.6,0.821) (47.7,0.848) (47.8,0.857) (47.9,0.868) (48,0.861) (48.1,0.835) (48.2,0.878) (48.3,0.848) (48.4,0.939) (48.5,0.851) (48.6,0.886) (48.7,0.849) (48.8,0.881) (48.9,0.839) (49,0.836) (49.1,0.879) (49.2,0.794) (49.3,0.817) (49.4,0.856) (49.5,0.879) (49.6,0.848) (49.7,0.910) (49.8,0.821) (49.9,0.856) (50,0.839) (50.1,0.884) (50.2,0.854) (50.3,0.843) (50.4,0.800) (50.5,0.839) (50.6,0.815) (50.7,0.861) (50.8,0.914) (50.9,0.864) (51,0.897) (51.1,0.868) (51.2,0.870) (51.3,0.857) (51.4,0.921) (51.5,0.824) (51.6,0.880) (51.7,0.902) (51.8,0.833) (51.9,0.882) (52,0.851) (52.1,0.890) (52.2,0.885) (52.3,0.890) (52.4,0.929) (52.5,0.898) (52.6,0.882) (52.7,0.840) (52.8,0.860) (52.9,0.862) (53,0.925) (53.1,0.854) (53.2,0.899) (53.3,0.896) (53.4,0.913) (53.5,0.881) (53.6,0.827) (53.7,0.846) (53.8,0.824) (53.9,0.946) (54,0.896) (54.1,0.837) (54.2,0.868) (54.3,0.925) (54.4,0.948) (54.5,0.911) (54.6,0.871) (54.7,0.878) (54.8,0.867) (54.9,0.879) (55,0.841) (55.1,0.870) (55.2,0.894) (55.3,0.908) (55.4,0.786) (55.5,0.920) (55.6,0.859) (55.7,0.837) (55.8,0.906) (55.9,0.888) (56,0.840) (56.1,0.816) (56.2,0.859) (56.3,0.859) (56.4,0.856) (56.5,0.832) (56.6,0.810) (56.7,0.899) (56.8,0.902) (56.9,0.867) (57,0.853) (57.1,0.912) (57.2,0.903) (57.3,0.908) (57.4,0.814) (57.5,0.930) (57.6,0.839) (57.7,0.910) (57.8,0.885) (57.9,0.867) (58,0.806) (58.1,0.849) (58.2,0.865) (58.3,0.935) (58.4,0.883) (58.5,0.882) (58.6,0.902) (58.7,0.835) (58.8,0.860) (58.9,0.798) (59,0.928) (59.1,0.846) (59.2,0.906) (59.3,0.876) (59.4,0.849) (59.5,0.870) (59.6,0.857) (59.7,0.848) (59.8,0.812) (59.9,0.864) (60,0.947) (60.1,0.862) (60.2,0.942) (60.3,0.863) (60.4,0.865) (60.5,0.844) (60.6,0.912) (60.7,0.872) (60.8,0.868) (60.9,0.828) (61,0.888) (61.1,0.885) (61.2,0.837) (61.3,0.871) (61.4,0.898) (61.5,0.854) (61.6,0.839) (61.7,0.871) (61.8,0.788) (61.9,0.882) (62,0.772) (62.1,0.885) (62.2,0.838) (62.3,0.823) (62.4,0.806) (62.5,0.889) (62.6,0.800) (62.7,0.940) (62.8,0.888) (62.9,0.898) (63,0.846) (63.1,0.847) (63.2,0.894) (63.3,0.872) (63.4,0.927) (63.5,0.915) (63.6,0.910) (63.7,0.820) (63.8,0.793) (63.9,0.850) (64,0.840) (64.1,0.803) (64.2,0.842) (64.3,0.838) (64.4,0.840) (64.5,0.875) (64.6,0.843) (64.7,0.837) (64.8,0.938) (64.9,0.844) (65,0.862) (65.1,0.904) (65.2,0.848) (65.3,0.816) (65.4,0.846) (65.5,0.826) (65.6,0.888) (65.7,0.910) (65.8,0.850) (65.9,0.817) (66,0.822) (66.1,0.826) (66.2,0.859) (66.3,0.852) (66.4,0.871) (66.5,0.872) (66.6,0.965) (66.7,0.889) (66.8,0.907) (66.9,0.860) (67,0.844) (67.1,0.802) (67.2,0.845) (67.3,0.908) (67.4,0.893) (67.5,0.902) (67.6,0.845) (67.7,0.865) (67.8,0.929) (67.9,0.856) (68,0.830) (68.1,0.897) (68.2,0.884) (68.3,0.849) (68.4,0.837) (68.5,0.878) (68.6,0.817) (68.7,0.849) (68.8,0.961) (68.9,0.857) (69,0.884) (69.1,0.871) (69.2,0.871) (69.3,0.843) (69.4,0.824) (69.5,0.841) (69.6,0.846) (69.7,0.895) (69.8,0.829) (69.9,0.922) (70,0.857) (70.1,0.829) (70.2,0.871) (70.3,0.943) (70.4,0.895) (70.5,0.878) (70.6,0.828) (70.7,0.915) (70.8,0.911) (70.9,0.861) (71,0.866) (71.1,0.892) (71.2,0.867) (71.3,0.890) (71.4,0.874) (71.5,0.917) (71.6,0.906) (71.7,0.849) (71.8,0.931) (71.9,0.840) (72,0.885) (72.1,0.909) (72.2,0.861) (72.3,0.908) (72.4,0.900) (72.5,0.872) (72.6,0.825) (72.7,0.882) (72.8,0.833) (72.9,0.932) (73,0.864) (73.1,0.887) (73.2,0.853) (73.3,0.905) (73.4,0.842) (73.5,0.859) (73.6,0.857) (73.7,0.909) (73.8,0.888) (73.9,0.863) (74,0.878) (74.1,0.823) (74.2,0.842) (74.3,0.912) (74.4,0.823) (74.5,0.916) (74.6,0.873) (74.7,0.850) (74.8,0.880) (74.9,0.849) (75,0.898) (75.1,0.902) (75.2,0.929) (75.3,0.907) (75.4,0.876) (75.5,0.835) (75.6,0.837) (75.7,0.863) (75.8,0.902) (75.9,0.896) (76,0.822) (76.1,0.892) (76.2,0.867) (76.3,0.863) (76.4,0.923) (76.5,0.859) (76.6,0.906) (76.7,0.837) (76.8,0.889) (76.9,0.824) (77,0.860) (77.1,0.796) (77.2,0.809) (77.3,0.869) (77.4,0.885) (77.5,0.857) (77.6,0.866) (77.7,0.867) (77.8,0.874) (77.9,0.898) (78,0.886) (78.1,0.931) (78.2,0.894) (78.3,0.843) (78.4,0.923) (78.5,0.877) (78.6,0.865) (78.7,0.902) (78.8,0.880) (78.9,0.802) (79,0.863) (79.1,0.863) (79.2,0.903) (79.3,0.906) (79.4,0.838) (79.5,0.929) (79.6,0.870) (79.7,0.916) (79.8,0.864) (79.9,0.868) (80,0.890) (80.1,0.910) (80.2,0.950) (80.3,0.879) (80.4,0.849) (80.5,0.879) (80.6,0.918) (80.7,0.884) (80.8,0.873) (80.9,0.845) (81,0.933) (81.1,0.802) (81.2,0.822) (81.3,0.840) (81.4,0.889) (81.5,0.922) (81.6,0.816) (81.7,0.854) (81.8,0.904) (81.9,0.827) (82,0.876) (82.1,0.898) (82.2,0.872) (82.3,0.874) (82.4,0.844) (82.5,0.833) (82.6,0.922) (82.7,0.819) (82.8,0.920) (82.9,0.914) (83,0.823) (83.1,0.800) (83.2,0.886) (83.3,0.915) (83.4,0.847) (83.5,0.806) (83.6,0.869) (83.7,0.837) (83.8,0.824) (83.9,0.914) (84,0.844) (84.1,0.830) (84.2,0.843) (84.3,0.904) (84.4,0.833) (84.5,0.909) (84.6,0.867) (84.7,0.857) (84.8,0.922) (84.9,0.867) (85,0.916) (85.1,0.977) (85.2,0.933) (85.3,0.908) (85.4,0.931) (85.5,0.837) (85.6,0.871) (85.7,0.907) (85.8,0.903) (85.9,0.910) (86,0.922) (86.1,0.940) (86.2,0.875) (86.3,0.929) (86.4,0.821) (86.5,0.886) (86.6,0.878) (86.7,0.906) (86.8,0.873) (86.9,0.825) (87,0.904) (87.1,0.932) (87.2,0.893) (87.3,0.860) (87.4,0.944) (87.5,0.847) (87.6,0.844) (87.7,0.881) (87.8,0.933) (87.9,0.908) (88,0.837) (88.1,0.853) (88.2,0.902) (88.3,0.864) (88.4,0.875) (88.5,0.926) (88.6,0.930) (88.7,0.885) (88.8,0.928) (88.9,0.855) (89,0.893) (89.1,0.804) (89.2,0.852) (89.3,0.918) (89.4,0.852) (89.5,0.933) (89.6,0.902) (89.7,0.925) (89.8,0.901) (89.9,0.888) (90,0.884) (90.1,0.927) (90.2,0.907) (90.3,0.924) (90.4,0.928) (90.5,0.908) (90.6,0.921) (90.7,0.917) (90.8,0.862) (90.9,0.911) (91,0.879) (91.1,0.908) (91.2,0.930) (91.3,0.869) (91.4,0.874) (91.5,0.870) (91.6,0.824) (91.7,0.876) (91.8,0.864) (91.9,0.804) (92,0.918) (92.1,0.904) (92.2,0.876) (92.3,0.899)
\end{pspicture}

\vskip -0.5 cm

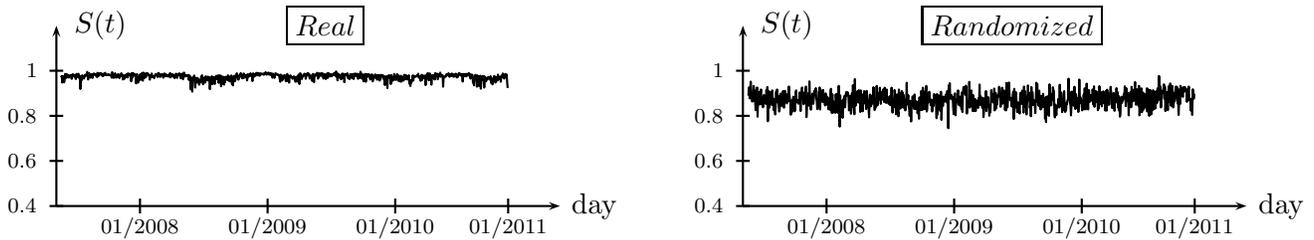
\noindent Fig. 17. Single step survival ratio for $T=0.8$ for running windows of 100 days with steps of one day at a time. The graph on the left is based on real data, and the graph on the right is based on randomized data.

\vskip 0.3 cm

\subsection{Multi step survival ratio}

A natural generalization of the single step survival ratio is the multi step survival ratio \cite{asset02}, which is the percentage of edges that exist in a certain number of windows, divided by the total number of edges. Here I shall adapt this definition to the case where asset graphs are built based on threshold values for the distance, in which case the number of edges may vary in each window. The definition here of a multi step survival ratio for $p$ windows is
\begin{equation}
\label{multistep}
S_p(t)=\frac{1}{M}\sum_{i,j=1}^NA_{ij}(t)A_{ij}(t+1)\cdots A_{ij}(t+p)\ ,
\end{equation}
where $A_{ij}(t)$ is the element $ij$ of matrix A(t), $N$ is the total number of nodes in the correlation matrix, and $M$ is the total number of edges in time $t$.

It is expected that the multi step survival ratio drops as the number of windows $p$ grow, for it is increasingly difficult for a connection (edge) to survive at all intervals being considered. It is also expected that for a large enough value of the thresholds, where all nodes are connected to all others, this survival ratio will always have the value 1. Figure 18 (left graph) shows the multi step survival ratios for thresholds ranging from $T=0.1$ to $T=1.5$ (fully connected network) as functions of the number of windows.

Note the high concentration from thresholds $T=0.5$ to $T=0.8$ (in red), and also with $T=1.1$. It is at these values that one has the most information about connections yet with a bearable amount of noise. For values bellow $T=0.5$, there are two few connections, although strong ones, which makes it easier for the multi step survival ratio to drop faster, and for very high thresholds, there are so many connections that it is easy for a particular edge to survive all the time.

The graph on the left of Figure 18 shows the multi step survival ratio for a typical simulation with randomized data, obtained by shufling randomly the time series of the indices. One difference is that there are no asset graphs for distance thresholds bellow $T=0.7$, for there are no connections formed bellow this threshold for randomized data (only very rarely). Other difference is that the multi step survival ratio for randomized data decays nearly exponentially, as it was to be expected from data that should not be related.

It is also interesting to compare this multi step survival ratio with the one obtained from a minimum spanning tree \cite{asset09} \cite{pruning}. Since edges built using a minimum spanning tree algorithm are much less stable than for asset graphs, it is much easier for those connections to change over time, leading to a near exponential decay of the multi step survival ratio, which does not occur for connections based on asset graphs.



\vskip 1.2 cm

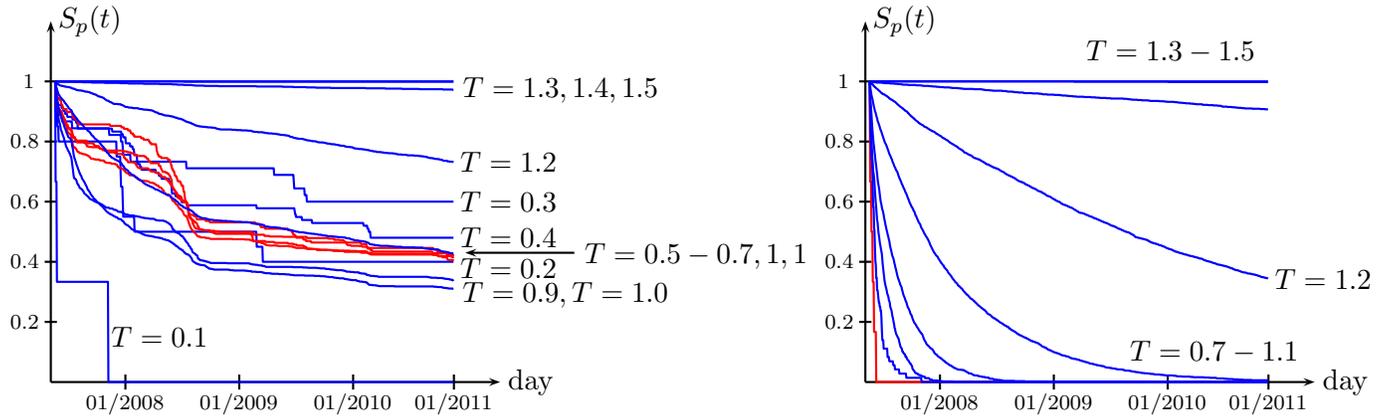
\noindent Fig. 18. Left: multi step survival ratios for thresholds ranging from $T=0.1$ to $T=1.5$ as functions of the number of windows; the curves related with thresholds from $T=0.5$ to $T=0.8$ are in red. Right: the same for randomized data.

\vskip 0.3 cm

\subsection{Survivability}

The multi step survival ratio depends on the starting point of the data being considered and also on the time shift between windows. So, it is a measure that, like the single step survival ratio, depends on time. An alternative measure derived in \cite{pruning} considers each day as a starting point and then tests the multi step survival for a number $D$ of days (5, for example) of an asset graph built with a threshold $T$. Then the percentage of tests in which a particular edge survives is used in order to define a survivability measure. As an example, the connection between nodes $i$ and $j$ for five consecutive days is tested with many different starting points in running windows of length $\Delta t$. If the connection survives in 30\% of the tests, then its survivability is $s=0.3$. So, survivability is a global (in terms of time) measure of how stable a connection between two nodes is through a certain period of time. Dividing the number of surviving edges by the total number of edges of the asset graph, one obtains the survivability ratio, $S_r$, that goes from 0 to 1.

Performing an analysis of survivability for the data concerning the 92 stock market indices from 2007 to 2010, one obtains the results in Figure 19. In this figure, the survivability ratio is plotted against the percentage of tests in which that number of connections survived.

Once again, there is a concentration of curves from $T=0.5$ to $T=0.8$, which is the region around the noise limit. For these threshold values, the percentage of connections that survive 100\% of the time is around 44\%  which, compared with the survivability of connections in a Minimum Spanning Tree \cite{pruning} is, not surprisingly, very high. For $T=0.1$, there is just one connection (from the original 2) that survive 100\% of the time: the one between France and Germany. This number grows and, by $T=0.5$, the connections between S\&P and Nasdaq, between the UK, Ireland, France, Germany, Switzerland, Austria, Italy, Belgium, the Netherlands, Sweden, Denmark, Finland, Norway, Spain, Portugal, and Poland, between Greece and Cyprus, and beween Japan, Hong Kong, Singapore, and Australia survive 100\% of the time. So, again one can detect a small American cluster, a large European cluster, and a Pacific Asian one (including Australia). The number grows for higher threshold values until every index is connected to every other index, which happens by $T=1.5$, and all connections survive 100\% of the time.

Taking $T=0.7$, as another example or 100\% survivability, one has an American cluster, including indices from North and South America, a large European cluster (which includes Israel and South Africa), a small Pacific Asian cluster, the pairs Greece - Cyprus and Namibia - South Africa, and connections between the North American cluster with the other two. By fixing the threshold value and varying the survivability ratio, one may obtain a sequence of networks that grow as the survivability ratio drops. This procedure may be used in order to build ``survivability networks'', which may bring some information on the stability of connections.

\begin{pspicture}(-0.5,0.4)(3.5,5.6)
\psset{xunit=4,yunit=4}
\psline{->}(0,0)(1.2,0) \psline{->}(0,0)(0,1.2) \rput(1.3,0){\% } \rput(0.1,1.2){$S_r$} \scriptsize \psline(0.2,-0.025)(0.2,0.025) \rput(0.2,-0.075){$0.2$} \scriptsize \psline(0.4,-0.025)(0.4,0.025) \rput(0.4,-0.075){$0.4$} \psline(0.6,-0.025)(0.6,0.025) \rput(0.6,-0.075){$0.6$} \psline(0.8,-0.025)(0.8,0.025) \rput(0.8,-0.075){$0.8$} \psline(1,-0.025)(1,0.025) \rput(1,-0.075){$1$} \psline(-0.025,0.2)(0.025,0.2) \rput(-0.12,0.2){$0.2$} \psline(-0.025,0.4)(0.025,0.4) \rput(-0.12,0.4){$0.4$} \psline(-0.025,0.6)(0.025,0.6) \rput(-0.12,0.6){$0.6$} \psline(-0.025,0.8)(0.025,0.8) \rput(-0.12,0.8){$0.8$} \psline(-0.025,1)(0.025,1) \rput(-0.12,1){$1$} \rput(-0.05,-0.05){0}
\normalsize
\psline[linecolor=blue] (0,1) (1,1) \rput(1.45,1){$T=0.1,1.3,1.4,1.5$}
\psline[linecolor=blue] (0,1.000) (0.1,0.941) (0.2,0.588) (0.3,0.588) (0.4,0.588) (0.5,0.588) (0.6,0.471) (0.7,0.471) (0.8,0.471) (0.9,0.471) (1,0.471)
\psline[linecolor=blue] (0,1.000) (0.1,0.927) (0.2,0.805) (0.3,0.805) (0.4,0.780) (0.5,0.780) (0.6,0.707) (0.7,0.659) (0.8,0.659) (0.9,0.659) (1,0.659) \rput(1.2,0.659){$T=0.3$}
\psline[linecolor=blue] (0,1.000) (0.1,0.905) (0.2,0.811) (0.3,0.663) (0.4,0.632) (0.5,0.632) (0.6,0.600) (0.7,0.568) (0.8,0.516) (0.9,0.516) (1,0.516) \rput(1.2,0.53){$T=0.4$}
\psline[linecolor=red] (0,1.000) (0.1,0.898) (0.2,0.856) (0.3,0.717) (0.4,0.556) (0.5,0.551) (0.6,0.508) (0.7,0.487) (0.8,0.455) (0.9,0.455) (1,0.439) \rput(1.5,0.439){$T=0.2,0.5-0.8,1.1$}
\psline[linecolor=red] (0,1.000) (0.1,0.865) (0.2,0.832) (0.3,0.691) (0.4,0.543) (0.5,0.516) (0.6,0.503) (0.7,0.480) (0.8,0.470) (0.9,0.470) (1,0.444)
\psline[linecolor=red] (0,1.000) (0.1,0.842) (0.2,0.781) (0.3,0.680) (0.4,0.540) (0.5,0.529) (0.6,0.521) (0.7,0.511) (0.8,0.481) (0.9,0.481) (1,0.454)
\psline[linecolor=red] (0,1.000) (0.1,0.794) (0.2,0.730) (0.3,0.635) (0.4,0.510) (0.5,0.503) (0.6,0.480) (0.7,0.470) (0.8,0.455) (0.9,0.455) (1,0.442)
\psline[linecolor=blue] (0,1.000) (0.1,0.657) (0.2,0.606) (0.3,0.547) (0.4,0.434) (0.5,0.428) (0.6,0.419) (0.7,0.409) (0.8,0.387) (0.9,0.384) (1,0.369) \rput(1.3,0.340){$T=0.9,1.0$}
\psline[linecolor=blue] (0,1.000) (0.1,0.640) (0.2,0.534) (0.3,0.482) (0.4,0.398) (0.5,0.391) (0.6,0.378) (0.7,0.369) (0.8,0.343) (0.9,0.338) (1,0.330)
\psline[linecolor=blue] (0,1.000) (0.1,0.831) (0.2,0.701) (0.3,0.621) (0.4,0.555) (0.5,0.543) (0.6,0.516) (0.7,0.499) (0.8,0.470) (0.9,0.462) (1,0.440)
\psline[linecolor=blue] (0,1.000) (0.1,0.961) (0.2,0.917) (0.3,0.885) (0.4,0.850) (0.5,0.838) (0.6,0.817) (0.7,0.796) (0.8,0.772) (0.9,0.758) (1,0.736) \rput(1.2,0.758){$T=1.2$}
\psline[linecolor=blue] (0,1.000) (0.1,0.997) (0.2,0.994) (0.3,0.991) (0.4,0.987) (0.5,0.985) (0.6,0.982) (0.7,0.979) (0.8,0.978) (0.9,0.976) (1,0.973)
\psline[linecolor=blue] (0,1.000) (0.1,1.000) (0.2,1.000) (0.3,1.000) (0.4,0.999) (0.5,0.999) (0.6,0.999) (0.7,0.999) (0.8,0.999) (0.9,0.999) (1,0.999)
\psline[linecolor=blue] (0,1.000) (1,1.000)
\end{pspicture}
\begin{pspicture}(-6.7,0.4)(3.5,5.6)
\psset{xunit=4,yunit=4}
\psline{->}(0,0)(1.2,0) \psline{->}(0,0)(0,1.2) \rput(1.3,0){\% } \rput(0.1,1.2){$S_r$} \scriptsize \psline(0.2,-0.025)(0.2,0.025) \rput(0.2,-0.075){$0.2$} \scriptsize \psline(0.4,-0.025)(0.4,0.025) \rput(0.4,-0.075){$0.4$} \psline(0.6,-0.025)(0.6,0.025) \rput(0.6,-0.075){$0.6$} \psline(0.8,-0.025)(0.8,0.025) \rput(0.8,-0.075){$0.8$} \psline(1,-0.025)(1,0.025) \rput(1,-0.075){$1$} \psline(-0.025,0.2)(0.025,0.2) \rput(-0.12,0.2){$0.2$} \psline(-0.025,0.4)(0.025,0.4) \rput(-0.12,0.4){$0.4$} \psline(-0.025,0.6)(0.025,0.6) \rput(-0.12,0.6){$0.6$} \psline(-0.025,0.8)(0.025,0.8) \rput(-0.12,0.8){$0.8$} \psline(-0.025,1)(0.025,1) \rput(-0.12,1){$1$} \rput(-0.05,-0.05){0}
\normalsize
\rput(1.3,0.1){$T=0.7-1.1$}
\psline[linecolor=red] (0,1.000) (0.1,0.683) (0.2,0.383) (0.3,0.250) (0.4,0.117) (0.5,0.067) (0.6,0.067) (0.7,0.033) (0.8,0.017) (0.9,0.017) (1,0.017)
\psline[linecolor=blue] (0,1.000) (0.1,0.499) (0.2,0.282) (0.3,0.150) (0.4,0.082) (0.5,0.039) (0.6,0.020) (0.7,0.007) (0.8,0.004) (0.9,0.002) (1,0.002)
\psline[linecolor=blue] (0,1.000) (0.1,0.471) (0.2,0.225) (0.3,0.106) (0.4,0.045) (0.5,0.021) (0.6,0.012) (0.7,0.003) (0.8,0.001) (0.9,0.001) (1,0.001)
\psline[linecolor=blue] (0,1.000) (0.1,0.615) (0.2,0.381) (0.3,0.233) (0.4,0.140) (0.5,0.081) (0.6,0.052) (0.7,0.033) (0.8,0.023) (0.9,0.012) (1,0.009)
\psline[linecolor=blue] (0,1.000) (0.1,0.902) (0.2,0.817) (0.3,0.734) (0.4,0.656) (0.5,0.588) (0.6,0.527) (0.7,0.477) (0.8,0.428) (0.9,0.388) (1,0.351) \rput(1.2,0.351){$T=1.2$}
\psline[linecolor=blue] (0,1.000) (0.1,0.988) (0.2,0.980) (0.3,0.972) (0.4,0.965) (0.5,0.955) (0.6,0.946) (0.7,0.936) (0.8,0.929) (0.9,0.919) (1,0.912) \rput(1.2,0.912){$T=1.3$}
\psline[linecolor=blue] (0,1.000) (0.1,1.000) (0.2,1.000) (0.3,1.000) (0.4,0.999) (0.5,0.999) (0.6,0.999) (0.7,0.999) (0.8,0.999) (0.9,0.999) (1,0.998) \rput(1.3,1){$T=1.4,1.5$}
\psline[linecolor=blue] (0,1.000) (1,1.000)
\end{pspicture}

\vskip 1 cm

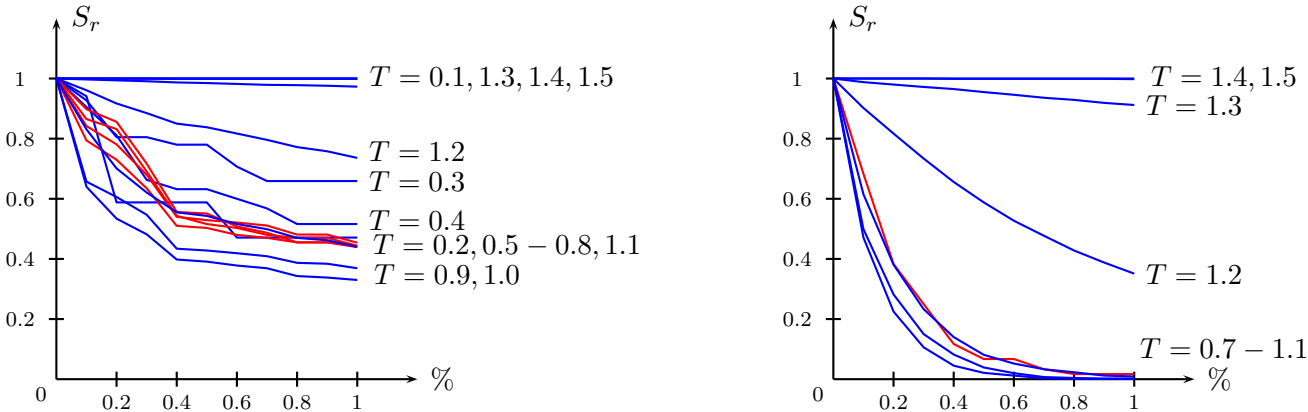
\noindent Fig. 19. Survivability ratios for thresholds ranging from $T=0.1$ to $T=1.5$ as functions of the percentage of surviving windows. On the left, it is plotted the survivability for the network obtained from the stock market indices; on the right, the survivability for a network built with randomized data.

\vskip 0.3 cm

Also in Figure 19, to the right, I plot the survivability ratio for a network obtained by randomly shuffling the time series of the 92 indices being studied. The  curves are results of single simulations, for the computations are quite long, but one does not lose too much information here by doing this. From $T=0.1$ to $T=0.7$, there are no connections in the network and so one cannot even define a survivability ratio. From $T=0.8$ to $T=1.1$, the survivability curves drop exponentially, and for values higher than $T=1.2$, it quickly approaches the line $S=1$, for all connections are active all the time for large enough values of the threshold. The difference between the results obtained with randomized data and with real data are striking: survivability is often much higher for real data, showing there must be information on the connections formed by their correlations.

Figure 20 (graph on the left) analyzes the relationship between distance and survivability for $T=0.5$. The horizontal axis is the distance between nodes as calculated by (\ref{distance}), and the vertical axis is the percentage of time a connection survives. What one can readily see is that, like in the case of Mininum Spanning Trees, there is no linear relation between the two variables. Looking at the data, one could see that around 93\% of the connections have survivability zero, specially those that represent distances higher than $0.4$. Also, around 3\% of the connections have survivability 1 (100\% ), none of them with a distance higher than $0.45$. There is also a lump of connections that survive between 30\% and 60\% of the time with distances varying from $0.2$ to $0.4$ (around 3\% of the total number of connections).



\vskip 1 cm

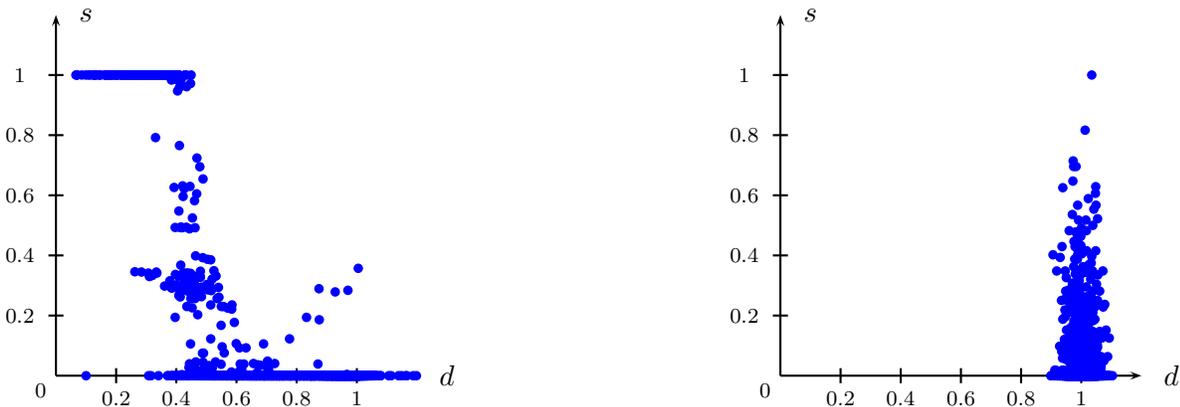
\noindent Fig. 20. Left: scatter plot of the distance $d$ between nodes and the percentage of time a connection survives, $s$, for $T=0.5$. Right: scatter plot of the distance $d$ between nodes and the percentage of time a connection survives, $s$, for $T=0.8$ and randomized data.

\vskip 0.3 cm

The graph on the right in Figure 20 shows the relationship between distance and survavibility ratio based on randomized data and a threshold $T=0.8$ (for thresholds bellow $T=0.6$ have no connections for randomized data). The difference is clear, as distances concentrate around a certain value and connections tend to survive little.

So, concluding this section, we saw that there are remarkable differences between real data and randomized data in terms of survivability, be it measured by the single step survival ratio, the multi step survival ratio or the survivability ratio. For real data, the curves for thresholds ranging from $T=0.5$ to $T=0.8$ tend to agglomerate, being that the region where one can get the most information about the network with an acceptable level of noise.

We could also see that some connections, and by extension, some clusters remain constant in time. Those clusters are heavily associated with geoographic position and on cultural and ethnical ties. Those connections that were more intense also tended to last longer, although there are many exceptions to this rule. Next, the effects of the choice of threshold and of time in centrality measures of network theory are studied.

\section{Centrality Measures}

Centrality measures are of key importance in network theory. Whereas a node has many connections, or it is on the shortest path between many pairs of nodes, or if it lies within a heavily connected region, are of great importance for models of propagation of information, of diseases, and on the vulnerability of a network to attacks, as examples. So, many measures of how central a node is in a network were developed, and each one associates a different concept to the word ``centrality''. Here, we shall see how one of those measures change over time, and what is the influence of the choice of threshold when building an asset graph on those centrality measures.

\subsection{Node degree}

The node degree of a node is the number of connections it has in the network. Most of the stocks have low node degree, and some of them, called hubs, have a large node degree and are generally nodes that are more important in the dynamics of a network. Considering the adjacency matrix $A$ of a network, the node degree of node $(i,j)$ can be defined mathematically as
\begin{equation}
\label{nodedegree}
Nd=\sum_{j=1}^NA_{ij}\ .
\end{equation}

Node degrees vary considerably according to the threshold value of an asset graph. It is expected that, for small threshold values, where very few connections are formed, most node degrees are zero, and, for high threshold values, where all connections are formed, all nodes will have the highest node degree that is possible. As the maximum node degree vary from 1 to 92 as thresholds grow, I divided all node degrees by the maximum node degree for better comparison of the asset graphs.

Considering that the asset graphs are formed only by those nodes which are connected to any other node, then the asset graphs for low threshold values are much smaller. Taking this into account, one must eliminate from the network all nodes which have node degree zero. Figure 21 illustrates frequency distributions of the normalized node degrees for thresholds going from $T=0.1$ to $T=1.1$, and for randomized data for $T=1.0$.

Extreme values for the threshold lead to distributions that are heavily concentrated. Again, the distributions close to the noise threshold seem to be the ones that give the most information. For thresholds down to $T=0.7$, indices alternate among the top highest node degrees in an apparently random way. From $T=0.6$ down to $T=0.1$, Central European indices occupy the top positions, particularly France, the UK, the Netherlands, and Germany.



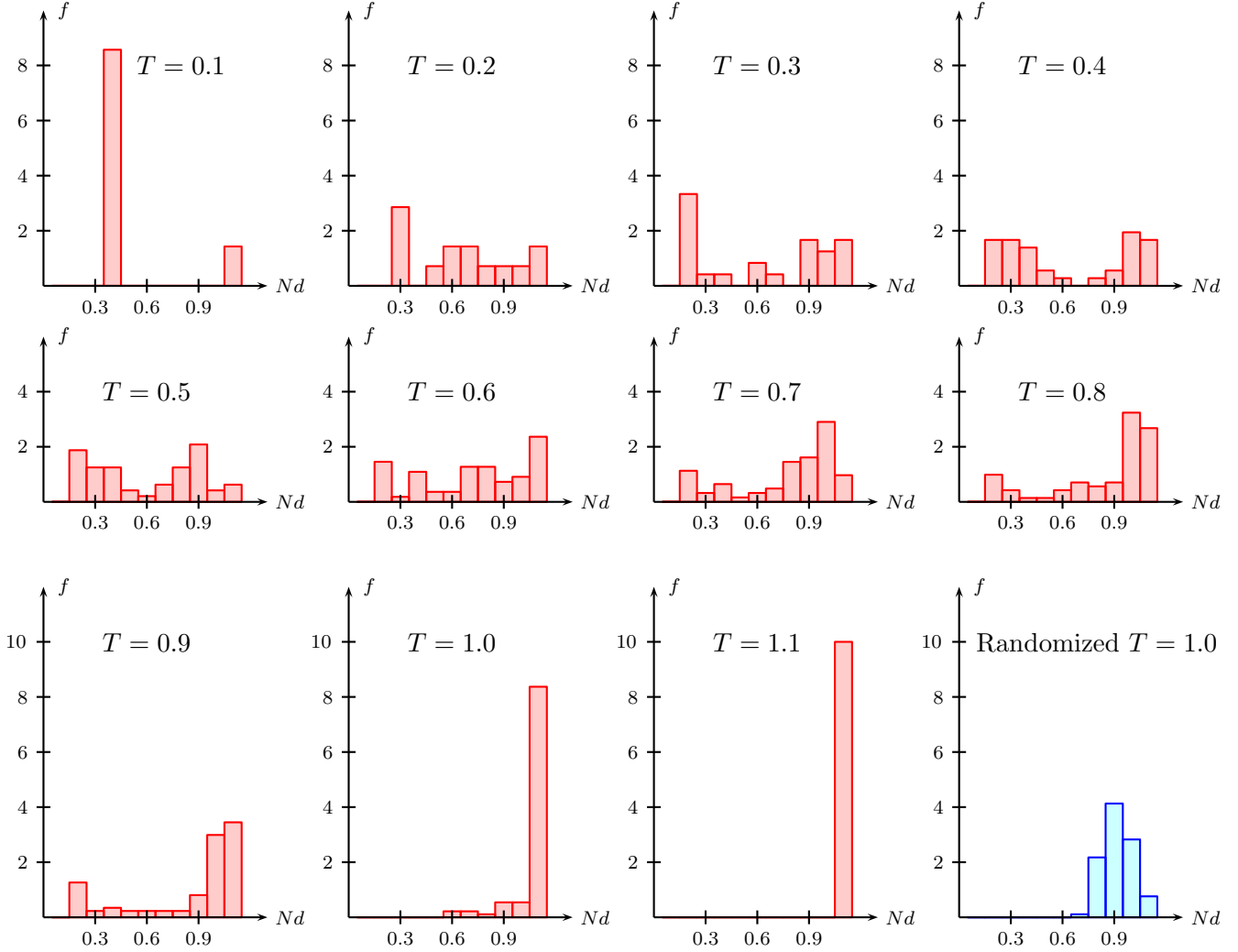
\noindent Fig. 21. Histograms of the normalized node degrees of asset graphs based on thresholds varying from $T=0.1$ to $T=1.1$, plus histogram for randomized data at $T=1.0$.

\vskip 0.3 cm

From Figure 21, one can also see that the resulting histograms are definitely not exponentially decreasing, so they cannot be represented by power laws of the type $p_k=ck^{-\alpha }$, where $p_k$ is the frequency distribution for the value $k$, and $c$ and $\alpha $ are constants. Power laws are the hallmark of a diversity of complex systems, as in the study of earthquakes, the world wide web, networks of scientific citations, of film actors, of social interactions, protein interactions, and many other topics \cite{Newman}. Networks whose centrality measures follow this type of distribution are often called scale-free networks, and this behavior can best be visualized if one plots a graph of the cumulative frequency distribution of a centrality in terms of the centrality values, both in logarithmic representation \cite{Newman}. Figure 22 shows the logarithm of the cumulative frequency distributions as functions of the logarithm from $T=0.1$ to $T=1.1$ (graph on the left), and for randomized data (graph on the right), from $T=0.9$ to $T=1.2$ (for $T=0.9$, the curve is on the vertical axis).

So, although the cumulative frequency distribution of the node degrees for the asset graphs is not the one of a scale-free network, they are definitely not the same as the cumulative frequency distribution of randomized data. A cautionary note must be added here. As the networks that are being considered in this analysis are very small compared with scale-free networks like the internet, or pure scale-free networks, which must be infinite in size. So, conclusions about these asset graphs networks, particularly the ones with smaller thresholds and a very small number of nodes, are not reliable.

\begin{pspicture}(-1.5,-4.3)(5,1.7)
\psset{xunit=2,yunit=2}
\psline{->}(-0.5,0)(2.5,0) \psline{->}(0,-2)(0,0.5) \scriptsize \rput(2.9,0){$log(Nd)$} \rput(0.65,0.5){$\log(Cum.\ distr.)$} \psline(0.5,-0.05)(0.5,0.05) \rput(0.5,-0.15){$0.5$} \psline(1,-0.05)(1,0.05) \rput(1,-0.15){$1$} \psline(1.5,-0.05)(1.5,0.05) \rput(1.5,-0.15){$1.5$} \psline(2,-0.05)(2,0.05) \rput(2,-0.15){$2$} \psline(-0.05,-1.5)(0.05,-1.5) \rput(-0.2,-1.5){$-1.5$} \psline(-0.05,-1)(0.05,-1) \rput(-0.2,-1){$-1$} \psline(-0.05,-0.5)(0.05,-0.5) \rput(-0.2,-0.5){$-0.5$} 
\psline[linecolor=red] (0.602,-0.845) (0.000,-0.544) (0.000,-0.368) (0.000,-0.243) (0.000,-0.146) (0.000,-0.067) (0.000,0.000)
\psline[linecolor=red] (0.954,-1.146) (0.954,-0.845) (0.903,-0.669) (0.845,-0.544) (0.778,-0.447) (0.699,-0.368) (0.699,-0.301) (0.602,-0.243) (0.602,-0.192) (0.477,-0.146) (0.000,-0.105) (0.000,-0.067) (0.000,-0.032) (0.000,0.000)
\psline[linecolor=red] (1.146,-1.380) (1.146,-1.079) (1.114,-0.903) (1.114,-0.778) (1.079,-0.681) (1.079,-0.602) (1.079,-0.535) (1.041,-0.477) (1.041,-0.426) (1.041,-0.380) (1.000,-0.339) (0.903,-0.301) (0.845,-0.266) (0.778,-0.234) (0.602,-0.204) (0.301,-0.176) (0.000,-0.150) (0.000,-0.125) (0.000,-0.101) (0.000,-0.079) (0.000,-0.058) (0.000,-0.038) (0.000,-0.018) (0.000,0.000)
\psline[linecolor=red] (1.255,-1.556) (1.255,-1.255) (1.255,-1.079) (1.230,-0.954) (1.230,-0.857) (1.230,-0.778) (1.204,-0.711) (1.204,-0.653) (1.204,-0.602) (1.176,-0.556) (1.176,-0.515) (1.176,-0.477) (1.176,-0.442) (1.146,-0.410) (1.114,-0.380) (1.079,-0.352) (0.954,-0.326) (0.845,-0.301) (0.778,-0.278) (0.699,-0.255) (0.699,-0.234) (0.602,-0.214) (0.602,-0.195) (0.602,-0.176) (0.477,-0.158) (0.477,-0.141) (0.301,-0.125) (0.301,-0.109) (0.301,-0.094) (0.301,-0.079) (0.000,-0.065) (0.000,-0.051) (0.000,-0.038) (0.000,-0.025) (0.000,-0.012) (0.000,0.000)
\psline[linecolor=red] (1.505,-1.681) (1.491,-1.380) (1.462,-1.204) (1.447,-1.079) (1.415,-0.982) (1.398,-0.903) (1.398,-0.836) (1.398,-0.778) (1.398,-0.727) (1.380,-0.681) (1.380,-0.640) (1.362,-0.602) (1.362,-0.567) (1.362,-0.535) (1.362,-0.505) (1.342,-0.477) (1.322,-0.451) (1.322,-0.426) (1.322,-0.402) (1.322,-0.380) (1.322,-0.359) (1.279,-0.339) (1.255,-0.320) (1.255,-0.301) (1.114,-0.283) (1.079,-0.266) (1.000,-0.250) (0.954,-0.234) (0.954,-0.219) (0.954,-0.204) (0.954,-0.190) (0.845,-0.176) (0.845,-0.163) (0.778,-0.150) (0.699,-0.137) (0.699,-0.125) (0.699,-0.113) (0.602,-0.101) (0.602,-0.090) (0.301,-0.079) (0.301,-0.068) (0.301,-0.058) (0.301,-0.048) (0.301,-0.038) (0.000,-0.028) (0.000,-0.018) (0.000,-0.009) (0.000,0.000)
\psline[linecolor=red] (1.591,-1.740) (1.580,-1.439) (1.580,-1.263) (1.580,-1.138) (1.580,-1.041) (1.580,-0.962) (1.568,-0.895) (1.568,-0.837) (1.568,-0.786) (1.568,-0.740) (1.556,-0.699) (1.556,-0.661) (1.556,-0.626) (1.544,-0.594) (1.544,-0.564) (1.505,-0.536) (1.505,-0.510) (1.505,-0.485) (1.491,-0.462) (1.477,-0.439) (1.462,-0.418) (1.447,-0.398) (1.431,-0.379) (1.431,-0.360) (1.415,-0.342) (1.415,-0.325) (1.398,-0.309) (1.380,-0.293) (1.380,-0.278) (1.362,-0.263) (1.362,-0.249) (1.342,-0.235) (1.342,-0.222) (1.322,-0.209) (1.322,-0.196) (1.322,-0.184) (1.279,-0.172) (1.204,-0.161) (1.146,-0.149) (1.146,-0.138) (1.000,-0.128) (0.954,-0.117) (0.954,-0.107) (0.954,-0.097) (0.903,-0.087) (0.903,-0.078) (0.845,-0.068) (0.477,-0.059) (0.477,-0.050) (0.477,-0.041) (0.000,-0.033) (0.000,-0.024) (0.000,-0.016) (0.000,-0.008) (0.000,0.000)
\psline[linecolor=red] (1.724,-1.792) (1.716,-1.491) (1.708,-1.315) (1.699,-1.190) (1.699,-1.093) (1.681,-1.014) (1.672,-0.947) (1.672,-0.889) (1.672,-0.838) (1.672,-0.792) (1.672,-0.751) (1.672,-0.713) (1.663,-0.678) (1.663,-0.646) (1.663,-0.616) (1.653,-0.588) (1.653,-0.562) (1.653,-0.537) (1.653,-0.514) (1.643,-0.491) (1.643,-0.470) (1.633,-0.450) (1.633,-0.431) (1.633,-0.412) (1.623,-0.394) (1.623,-0.377) (1.613,-0.361) (1.602,-0.345) (1.602,-0.330) (1.591,-0.315) (1.580,-0.301) (1.580,-0.287) (1.580,-0.274) (1.580,-0.261) (1.556,-0.248) (1.556,-0.236) (1.531,-0.224) (1.531,-0.213) (1.519,-0.201) (1.519,-0.190) (1.519,-0.180) (1.519,-0.169) (1.505,-0.159) (1.491,-0.149) (1.462,-0.139) (1.447,-0.130) (1.415,-0.120) (1.398,-0.111) (1.322,-0.102) (1.176,-0.093) (1.146,-0.085) (1.114,-0.076) (1.041,-0.068) (0.954,-0.060) (0.778,-0.052) (0.699,-0.044) (0.602,-0.037) (0.477,-0.029) (0.477,-0.022) (0.477,-0.014) (0.301,-0.007) (0.000,0.000)
\psline[linecolor=red] (1.792,-1.851) (1.778,-1.550) (1.771,-1.374) (1.771,-1.249) (1.763,-1.152) (1.756,-1.073) (1.756,-1.006) (1.756,-0.948) (1.756,-0.897) (1.756,-0.851) (1.756,-0.810) (1.756,-0.772) (1.756,-0.737) (1.748,-0.705) (1.748,-0.675) (1.748,-0.647) (1.748,-0.621) (1.748,-0.596) (1.748,-0.573) (1.740,-0.550) (1.740,-0.529) (1.732,-0.509) (1.732,-0.490) (1.732,-0.471) (1.732,-0.453) (1.732,-0.436) (1.724,-0.420) (1.724,-0.404) (1.724,-0.389) (1.724,-0.374) (1.724,-0.360) (1.724,-0.346) (1.724,-0.333) (1.724,-0.320) (1.716,-0.307) (1.708,-0.295) (1.708,-0.283) (1.708,-0.271) (1.708,-0.260) (1.708,-0.249) (1.708,-0.238) (1.699,-0.228) (1.681,-0.218) (1.672,-0.208) (1.663,-0.198) (1.663,-0.189) (1.653,-0.179) (1.623,-0.170) (1.623,-0.161) (1.613,-0.152) (1.613,-0.144) (1.568,-0.135) (1.568,-0.127) (1.556,-0.119) (1.544,-0.111) (1.505,-0.103) (1.491,-0.095) (1.462,-0.088) (1.431,-0.080) (1.301,-0.073) (1.146,-0.066) (1.079,-0.059) (1.041,-0.052) (0.845,-0.045) (0.602,-0.038) (0.602,-0.032) (0.301,-0.025) (0.301,-0.019) (0.301,-0.012) (0.000,-0.006) (0.000,0.000)
\psline[linecolor=red] (1.863,-1.940) (1.863,-1.638) (1.857,-1.462) (1.851,-1.337) (1.851,-1.241) (1.851,-1.161) (1.845,-1.094) (1.845,-1.036) (1.845,-0.985) (1.845,-0.940) (1.845,-0.898) (1.845,-0.860) (1.839,-0.826) (1.839,-0.793) (1.839,-0.763) (1.839,-0.735) (1.839,-0.709) (1.833,-0.684) (1.833,-0.661) (1.833,-0.638) (1.833,-0.617) (1.826,-0.597) (1.826,-0.578) (1.820,-0.559) (1.820,-0.542) (1.820,-0.525) (1.820,-0.508) (1.820,-0.492) (1.820,-0.477) (1.820,-0.462) (1.813,-0.448) (1.813,-0.434) (1.813,-0.421) (1.813,-0.408) (1.813,-0.395) (1.813,-0.383) (1.813,-0.371) (1.813,-0.360) (1.813,-0.348) (1.813,-0.337) (1.813,-0.327) (1.813,-0.316) (1.806,-0.306) (1.806,-0.296) (1.806,-0.286) (1.806,-0.277) (1.806,-0.267) (1.806,-0.258) (1.799,-0.249) (1.799,-0.241) (1.792,-0.232) (1.792,-0.224) (1.792,-0.215) (1.792,-0.207) (1.785,-0.199) (1.771,-0.191) (1.763,-0.184) (1.756,-0.176) (1.740,-0.169) (1.740,-0.161) (1.732,-0.154) (1.732,-0.147) (1.724,-0.140) (1.699,-0.133) (1.690,-0.127) (1.591,-0.120) (1.580,-0.113) (1.505,-0.107) (1.491,-0.101) (1.415,-0.094) (1.415,-0.088) (1.230,-0.082) (1.204,-0.076) (1.176,-0.070) (1.041,-0.064) (0.903,-0.059) (0.699,-0.053) (0.602,-0.047) (0.477,-0.042) (0.477,-0.036) (0.477,-0.031) (0.301,-0.026) (0.301,-0.020) (0.000,-0.015) (0.000,-0.010) (0.000,-0.005) (0.000,0.000)
\psline[linecolor=red] (1.959,-1.964) (1.959,-1.663) (1.954,-1.487) (1.949,-1.362) (1.949,-0.788) (1.944,-0.760) (1.944,-0.472) (1.940,-0.459) (1.940,-0.330) (1.934,-0.320) (1.934,-0.223) (1.929,-0.216) (1.929,-0.178) (1.924,-0.171) (1.924,-0.131) (1.919,-0.125) (1.919,-0.089) (1.914,-0.083) (1.914,-0.077) (1.903,-0.072) (1.903,-0.066) (1.892,-0.061) (1.886,-0.055) (1.875,-0.050) (1.857,-0.045) (1.857,-0.040) (1.851,-0.034) (1.813,-0.029) (1.806,-0.024) (1.785,-0.019) (1.699,-0.014) (1.699,-0.010) (1.623,-0.005) (1.602,0.000)
\psline[linecolor=red] (1.959,-1.964) (1.959,-0.010) (1.954,-0.005) (1.954,0.000)
\psline[linecolor=red] (1.959,-1.964) (1.959,0.000)
\end{pspicture}
\begin{pspicture}(-4,-4.3)(5,1.7)
\psset{xunit=2,yunit=2}
\psline{->}(-0.5,0)(2.5,0) \psline{->}(0,-2)(0,0.5) \scriptsize \rput(2.9,0){$log(Nd)$} \rput(0.65,0.5){$\log(Cum.\ distr.)$} \psline(0.5,-0.05)(0.5,0.05) \rput(0.5,-0.15){$0.5$} \psline(1,-0.05)(1,0.05) \rput(1,-0.15){$1$} \psline(1.5,-0.05)(1.5,0.05) \rput(1.5,-0.15){$1.5$} \psline(2,-0.05)(2,0.05) \rput(2,-0.15){$2$} \psline(-0.05,-1.5)(0.05,-1.5) \rput(-0.2,-1.5){$-1.5$} \psline(-0.05,-1)(0.05,-1) \rput(-0.2,-1){$-1$} \psline(-0.05,-0.5)(0.05,-0.5) \rput(-0.2,-0.5){$-0.5$}
\psline[linecolor=blue] (0.000,-0.301) (0.000,0.000)
\psline[linecolor=blue] (1.771,-1.964) (1.748,-1.663) (1.740,-1.487) (1.740,-1.362) (1.740,-1.265) (1.740,-1.186) (1.732,-1.119) (1.724,-1.061) (1.724,-1.010) (1.716,-0.964) (1.716,-0.922) (1.716,-0.885) (1.708,-0.850) (1.708,-0.818) (1.708,-0.788) (1.708,-0.760) (1.708,-0.733) (1.708,-0.709) (1.699,-0.685) (1.699,-0.663) (1.699,-0.642) (1.690,-0.621) (1.690,-0.602) (1.690,-0.584) (1.690,-0.566) (1.690,-0.549) (1.681,-0.532) (1.681,-0.517) (1.681,-0.501) (1.681,-0.487) (1.681,-0.472) (1.681,-0.459) (1.681,-0.445) (1.672,-0.432) (1.672,-0.420) (1.672,-0.407) (1.672,-0.396) (1.672,-0.384) (1.672,-0.373) (1.672,-0.362) (1.672,-0.351) (1.663,-0.341) (1.663,-0.330) (1.663,-0.320) (1.663,-0.311) (1.663,-0.301) (1.663,-0.292) (1.653,-0.283) (1.653,-0.274) (1.653,-0.265) (1.653,-0.256) (1.653,-0.248) (1.653,-0.240) (1.643,-0.231) (1.643,-0.223) (1.643,-0.216) (1.643,-0.208) (1.643,-0.200) (1.643,-0.193) (1.633,-0.186) (1.633,-0.178) (1.633,-0.171) (1.633,-0.164) (1.633,-0.158) (1.633,-0.151) (1.633,-0.144) (1.623,-0.138) (1.623,-0.131) (1.623,-0.125) (1.623,-0.119) (1.623,-0.113) (1.613,-0.106) (1.613,-0.100) (1.613,-0.095) (1.613,-0.089) (1.613,-0.083) (1.602,-0.077) (1.591,-0.072) (1.591,-0.066) (1.591,-0.061) (1.591,-0.055) (1.591,-0.050) (1.591,-0.045) (1.591,-0.040) (1.580,-0.034) (1.580,-0.029) (1.580,-0.024) (1.580,-0.019) (1.580,-0.014) (1.568,-0.010) (1.556,-0.005) (1.531,0.000)
\psline[linecolor=blue] (1.959,-1.964) (1.959,-0.010) (1.954,-0.005) (1.954,0.000)
\psline[linecolor=blue] (1.959,-1.964) (1.959,0.000)
\end{pspicture}

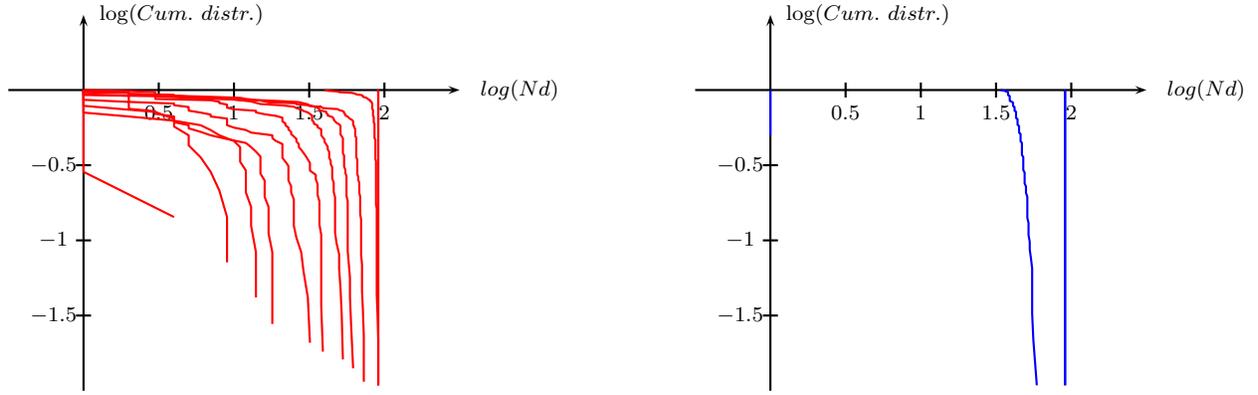
\noindent Fig. 22. Cumulative frequency distribution of the node degree in terms of the centrality values, both in logarithmic representation.

\vskip 0.3 cm

In Figure 23, I show the evolution of the average node degree as a function of time, $\overline{Nd}(t)$, which is the mean of the node degrees of the nodes belonging to each asset graph (the number of nodes may change in time). In the calculation of the graph, I have used windows of 100 days, shifted one day at a time. It does make data overlap, but gives a better number of points for the plot. The curves are, as expected, very similar to the graph of the average correlation in time and, as a consequence, to the graph of the volatility of the world market as measured by the MSCI World Index. The aggreeement between the average node degree and the average correlation time series may be measured by their Pearson correlation, which goes from 0.38 for $T=0.1$ to 0.95 for $T=0.8$. Figure 24 shows the time series of the average node degree for $T=0.7$ in terms of the average correlation, $<C>(t)$, time series.



\vskip 1.8 cm

\subsection{Node strength}

The strength of a node is the sum of the correlations of the node with all other nodes to which it is connected. If $C$ is the matrix that stores the correlations between $n$ nodes that are linked in the asset graph, then the node strength is given by
\begin{equation}
Ns^j=\sum_{i=1}^nC_{ij}\ ,
\end{equation}
where $C_{ij}$ is an element of matrix $C$.

For the asset graphs of the world stock indices, the Central European indices appear as the most central according to node strength for all threshold values. The topmost indices are those from the Netherlands, France, Austria, Belgium, Austria, the UK, Germany, and Italy.

Figure 25 displays the cumulative frequency distribution of the node strength in terms of the centrality values, both in logarithmic representation. The figure on the left displays the curves for real data (from $T=0.1$ to $T=1.2$), and the figure on the right represents the curves for randomized data (from $T=0.9$ to $T=1.2$). Again, the results are remarkably different from randomized data, but they do not seem to correspond to scale-free networks.

\begin{pspicture}(-1.5,-4.3)(5,1.7)
\psset{xunit=2,yunit=2}
\psline{->}(-0.5,0)(2.5,0) \psline{->}(0,-2)(0,0.5) \scriptsize \rput(2.9,0){$log(Nd)$} \rput(0.65,0.5){$\log(Cum.\ distr.)$} \psline(0.5,-0.05)(0.5,0.05) \rput(0.5,-0.15){$0.5$} \psline(1,-0.05)(1,0.05) \rput(1,-0.15){$1$} \psline(1.5,-0.05)(1.5,0.05) \rput(1.5,-0.15){$1.5$} \psline(2,-0.05)(2,0.05) \rput(2,-0.15){$2$} \psline(-0.05,-1.5)(0.05,-1.5) \rput(-0.2,-1.5){$-1.5$} \psline(-0.05,-1)(0.05,-1) \rput(-0.2,-1){$-1$} \psline(-0.05,-0.5)(0.05,-0.5) \rput(-0.2,-0.5){$-0.5$}
\psline[linecolor=red] (0.566,-0.845) (-0.030,-0.544) (-0.031,-0.368) (-0.032,-0.243) (-0.032,-0.146) (-0.039,-0.067) (-0.045,0.000)
\psline[linecolor=red] (0.900,-1.146) (0.887,-0.845) (0.838,-0.669) (0.776,-0.544) (0.710,-0.447) (0.625,-0.368) (0.625,-0.301) (0.522,-0.243) (0.516,-0.192) (0.388,-0.146) (-0.032,-0.105) (-0.032,-0.067) (-0.046,-0.032) (-0.046,0.000)
\psline[linecolor=red] (1.066,-1.380) (1.047,-1.079) (1.027,-0.903) (1.024,-0.778) (0.978,-0.681) (0.970,-0.602) (0.965,-0.535) (0.950,-0.477) (0.944,-0.426) (0.912,-0.380) (0.896,-0.339) (0.764,-0.301) (0.701,-0.266) (0.634,-0.234) (0.456,-0.204) (0.214,-0.176) (-0.032,-0.150) (-0.046,-0.125) (-0.046,-0.101) (-0.109,-0.079) (-0.109,-0.058) (-0.133,-0.038) (-0.133,-0.018) (-0.150,0.000)
\psline[linecolor=red] (1.151,-1.556) (1.143,-1.255) (1.118,-1.079) (1.108,-0.954) (1.106,-0.857) (1.105,-0.778) (1.094,-0.711) (1.071,-0.653) (1.057,-0.602) (1.051,-0.556) (1.048,-0.515) (1.039,-0.477) (1.018,-0.442) (0.981,-0.410) (0.946,-0.380) (0.873,-0.352) (0.774,-0.326) (0.643,-0.301) (0.583,-0.278) (0.557,-0.255) (0.529,-0.234) (0.470,-0.214) (0.429,-0.195) (0.424,-0.176) (0.291,-0.158) (0.276,-0.141) (0.115,-0.125) (0.104,-0.109) (0.103,-0.094) (0.093,-0.079) (-0.046,-0.065) (-0.109,-0.051) (-0.109,-0.038) (-0.133,-0.025) (-0.197,-0.012) (-0.219,0.000)
\psline[linecolor=red] (1.339,-1.681) (1.322,-1.380) (1.293,-1.204) (1.284,-1.079) (1.251,-0.982) (1.234,-0.903) (1.228,-0.836) (1.210,-0.778) (1.209,-0.727) (1.202,-0.681) (1.196,-0.640) (1.172,-0.602) (1.162,-0.567) (1.162,-0.535) (1.157,-0.505) (1.136,-0.477) (1.117,-0.451) (1.107,-0.426) (1.088,-0.402) (1.086,-0.380) (1.068,-0.359) (1.028,-0.339) (1.013,-0.320) (0.996,-0.301) (0.871,-0.283) (0.867,-0.266) (0.756,-0.250) (0.752,-0.234) (0.725,-0.219) (0.713,-0.204) (0.708,-0.190) (0.645,-0.176) (0.556,-0.163) (0.550,-0.150) (0.488,-0.137) (0.470,-0.125) (0.445,-0.113) (0.320,-0.101) (0.318,-0.090) (0.112,-0.079) (0.048,-0.068) (0.031,-0.058) (0.023,-0.048) (0.018,-0.038) (-0.259,-0.028) (-0.290,-0.018) (-0.290,-0.009) (-0.293,0.000)
\psline[linecolor=red] (1.389,-1.740) (1.383,-1.439) (1.376,-1.263) (1.368,-1.138) (1.368,-1.041) (1.356,-0.962) (1.350,-0.895) (1.342,-0.837) (1.340,-0.786) (1.337,-0.740) (1.337,-0.699) (1.337,-0.661) (1.325,-0.626) (1.290,-0.594) (1.276,-0.564) (1.248,-0.536) (1.247,-0.510) (1.241,-0.485) (1.206,-0.462) (1.187,-0.439) (1.183,-0.418) (1.175,-0.398) (1.159,-0.379) (1.143,-0.360) (1.130,-0.342) (1.100,-0.325) (1.092,-0.309) (1.072,-0.293) (1.067,-0.278) (1.053,-0.263) (1.040,-0.249) (1.038,-0.235) (1.033,-0.222) (1.029,-0.209) (1.008,-0.196) (1.000,-0.184) (0.979,-0.172) (0.898,-0.161) (0.864,-0.149) (0.784,-0.138) (0.686,-0.128) (0.630,-0.117) (0.625,-0.107) (0.622,-0.097) (0.615,-0.087) (0.564,-0.078) (0.556,-0.068) (0.144,-0.059) (0.097,-0.050) (0.095,-0.041) (-0.290,-0.033) (-0.290,-0.024) (-0.339,-0.016) (-0.339,-0.008) (-0.355,0.000)
\psline[linecolor=red] (1.435,-1.792) (1.428,-1.491) (1.426,-1.315) (1.421,-1.190) (1.418,-1.093) (1.417,-1.014) (1.415,-0.947) (1.407,-0.889) (1.403,-0.838) (1.402,-0.792) (1.392,-0.751) (1.387,-0.713) (1.386,-0.678) (1.382,-0.646) (1.376,-0.616) (1.374,-0.588) (1.373,-0.562) (1.352,-0.537) (1.342,-0.514) (1.336,-0.491) (1.334,-0.470) (1.322,-0.450) (1.297,-0.431) (1.290,-0.412) (1.271,-0.394) (1.254,-0.377) (1.224,-0.361) (1.216,-0.345) (1.214,-0.330) (1.208,-0.315) (1.204,-0.301) (1.203,-0.287) (1.195,-0.274) (1.182,-0.261) (1.177,-0.248) (1.176,-0.236) (1.162,-0.224) (1.161,-0.213) (1.156,-0.201) (1.155,-0.190) (1.145,-0.180) (1.142,-0.169) (1.141,-0.159) (1.119,-0.149) (1.058,-0.139) (1.024,-0.130) (1.020,-0.120) (0.992,-0.111) (0.828,-0.102) (0.691,-0.093) (0.688,-0.085) (0.673,-0.076) (0.603,-0.068) (0.499,-0.060) (0.271,-0.052) (0.207,-0.044) (0.160,-0.037) (0.074,-0.029) (0.026,-0.022) (-0.029,-0.014) (-0.113,-0.007) (-0.510,0.000)
\psline[linecolor=red] (1.462,-1.851) (1.461,-1.550) (1.457,-1.374) (1.453,-1.249) (1.449,-1.152) (1.448,-1.073) (1.448,-1.006) (1.442,-0.948) (1.437,-0.897) (1.429,-0.851) (1.427,-0.810) (1.419,-0.772) (1.418,-0.737) (1.417,-0.705) (1.412,-0.675) (1.407,-0.647) (1.399,-0.621) (1.399,-0.596) (1.392,-0.573) (1.390,-0.550) (1.382,-0.529) (1.370,-0.509) (1.369,-0.490) (1.349,-0.471) (1.347,-0.453) (1.332,-0.436) (1.316,-0.420) (1.311,-0.404) (1.310,-0.389) (1.297,-0.374) (1.288,-0.360) (1.283,-0.346) (1.281,-0.333) (1.276,-0.320) (1.271,-0.307) (1.266,-0.295) (1.255,-0.283) (1.255,-0.271) (1.253,-0.260) (1.249,-0.249) (1.249,-0.238) (1.244,-0.228) (1.236,-0.218) (1.228,-0.208) (1.224,-0.198) (1.216,-0.189) (1.198,-0.179) (1.196,-0.170) (1.160,-0.161) (1.158,-0.152) (1.130,-0.144) (1.113,-0.135) (1.063,-0.127) (1.014,-0.119) (0.951,-0.111) (0.900,-0.103) (0.877,-0.095) (0.866,-0.088) (0.824,-0.080) (0.724,-0.073) (0.515,-0.066) (0.494,-0.059) (0.406,-0.052) (0.269,-0.045) (0.027,-0.038) (-0.003,-0.032) (-0.314,-0.025) (-0.349,-0.019) (-0.369,-0.012) (-0.603,-0.006) (-0.658,0.000)
\psline[linecolor=red] (1.487,-1.940) (1.487,-1.638) (1.479,-1.462) (1.473,-1.337) (1.473,-1.241) (1.471,-1.161) (1.469,-1.094) (1.464,-1.036) (1.463,-0.985) (1.456,-0.940) (1.449,-0.898) (1.449,-0.860) (1.446,-0.826) (1.444,-0.793) (1.435,-0.763) (1.431,-0.735) (1.430,-0.709) (1.421,-0.684) (1.413,-0.661) (1.409,-0.638) (1.405,-0.617) (1.398,-0.597) (1.396,-0.578) (1.381,-0.559) (1.374,-0.542) (1.363,-0.525) (1.358,-0.508) (1.351,-0.492) (1.350,-0.477) (1.336,-0.462) (1.336,-0.448) (1.328,-0.434) (1.327,-0.421) (1.323,-0.408) (1.320,-0.395) (1.312,-0.383) (1.308,-0.371) (1.308,-0.360) (1.307,-0.348) (1.302,-0.337) (1.300,-0.327) (1.298,-0.316) (1.292,-0.306) (1.291,-0.296) (1.274,-0.286) (1.261,-0.277) (1.259,-0.267) (1.249,-0.258) (1.236,-0.249) (1.231,-0.241) (1.228,-0.232) (1.200,-0.224) (1.178,-0.215) (1.171,-0.207) (1.159,-0.199) (1.142,-0.191) (1.124,-0.184) (1.109,-0.176) (1.099,-0.169) (1.079,-0.161) (1.006,-0.154) (0.975,-0.147) (0.954,-0.140) (0.923,-0.133) (0.889,-0.127) (0.769,-0.120) (0.653,-0.113) (0.642,-0.107) (0.626,-0.101) (0.570,-0.094) (0.501,-0.088) (0.353,-0.082) (0.293,-0.076) (0.274,-0.070) (0.241,-0.064) (0.018,-0.059) (-0.211,-0.053) (-0.338,-0.047) (-0.420,-0.042) (-0.442,-0.036) (-0.484,-0.031) (-0.617,-0.026) (-0.633,-0.020) (-0.928,-0.015) (-0.940,-0.010) (-0.985,-0.005) (-0.999,0.000)
\psline[linecolor=red] (1.501,-1.964) (1.500,-1.663) (1.492,-1.487) (1.488,-1.362) (1.487,-1.265) (1.486,-1.186) (1.486,-1.119) (1.480,-1.061) (1.478,-1.010) (1.470,-0.964) (1.464,-0.922) (1.464,-0.885) (1.462,-0.850) (1.460,-0.818) (1.455,-0.788) (1.449,-0.760) (1.443,-0.733) (1.439,-0.709) (1.433,-0.685) (1.424,-0.663) (1.423,-0.642) (1.416,-0.621) (1.416,-0.602) (1.399,-0.584) (1.387,-0.566) (1.384,-0.549) (1.371,-0.532) (1.366,-0.517) (1.365,-0.501) (1.356,-0.487) (1.355,-0.472) (1.343,-0.459) (1.343,-0.445) (1.342,-0.432) (1.341,-0.420) (1.330,-0.407) (1.327,-0.396) (1.326,-0.384) (1.325,-0.373) (1.325,-0.362) (1.321,-0.351) (1.321,-0.341) (1.321,-0.330) (1.311,-0.320) (1.301,-0.311) (1.291,-0.301) (1.283,-0.292) (1.279,-0.283) (1.263,-0.274) (1.258,-0.265) (1.258,-0.256) (1.225,-0.248) (1.214,-0.240) (1.209,-0.231) (1.187,-0.223) (1.179,-0.216) (1.168,-0.208) (1.152,-0.200) (1.120,-0.193) (1.109,-0.186) (1.068,-0.178) (1.054,-0.171) (1.051,-0.164) (1.023,-0.158) (1.003,-0.151) (0.953,-0.144) (0.893,-0.138) (0.884,-0.131) (0.874,-0.125) (0.865,-0.119) (0.831,-0.113) (0.801,-0.106) (0.773,-0.100) (0.758,-0.095) (0.757,-0.089) (0.701,-0.083) (0.653,-0.077) (0.625,-0.072) (0.621,-0.066) (0.573,-0.061) (0.517,-0.055) (0.454,-0.050) (0.407,-0.045) (0.394,-0.040) (0.362,-0.034) (0.309,-0.029) (0.289,-0.024) (0.190,-0.019) (0.182,-0.014) (0.109,-0.010) (0.024,-0.005) (0.015,0.000)
\psline[linecolor=red] (1.500,-1.964) (1.499,-1.663) (1.490,-1.487) (1.486,-1.362) (1.486,-1.265) (1.485,-1.186) (1.485,-1.119) (1.479,-1.061) (1.478,-1.010) (1.469,-0.964) (1.462,-0.922) (1.462,-0.885) (1.460,-0.850) (1.458,-0.818) (1.453,-0.788) (1.448,-0.760) (1.441,-0.733) (1.437,-0.709) (1.431,-0.685) (1.423,-0.663) (1.422,-0.642) (1.415,-0.621) (1.415,-0.602) (1.397,-0.584) (1.386,-0.566) (1.383,-0.549) (1.371,-0.532) (1.365,-0.517) (1.364,-0.501) (1.355,-0.487) (1.354,-0.472) (1.342,-0.459) (1.342,-0.445) (1.341,-0.432) (1.339,-0.420) (1.329,-0.407) (1.326,-0.396) (1.325,-0.384) (1.324,-0.373) (1.322,-0.362) (1.321,-0.351) (1.320,-0.341) (1.319,-0.330) (1.310,-0.320) (1.298,-0.311) (1.286,-0.301) (1.279,-0.292) (1.274,-0.283) (1.263,-0.274) (1.257,-0.265) (1.257,-0.256) (1.224,-0.248) (1.211,-0.240) (1.207,-0.231) (1.185,-0.223) (1.178,-0.216) (1.166,-0.208) (1.151,-0.200) (1.118,-0.193) (1.106,-0.186) (1.063,-0.178) (1.053,-0.171) (1.050,-0.164) (1.018,-0.158) (1.003,-0.151) (0.951,-0.144) (0.887,-0.138) (0.881,-0.131) (0.873,-0.125) (0.861,-0.119) (0.827,-0.113) (0.799,-0.106) (0.771,-0.100) (0.750,-0.095) (0.748,-0.089) (0.684,-0.083) (0.639,-0.077) (0.613,-0.072) (0.609,-0.066) (0.555,-0.061) (0.484,-0.055) (0.402,-0.050) (0.371,-0.045) (0.333,-0.040) (0.263,-0.034) (0.230,-0.029) (0.191,-0.024) (0.020,-0.019) (-0.125,-0.014) (-0.580,-0.010)
\psline[linecolor=red] (1.500,-1.964) (1.499,-1.663) (1.490,-1.487) (1.486,-1.362) (1.486,-1.265) (1.485,-1.186) (1.485,-1.119) (1.479,-1.061) (1.478,-1.010) (1.469,-0.964) (1.462,-0.922) (1.462,-0.885) (1.460,-0.850) (1.458,-0.818) (1.453,-0.788) (1.448,-0.760) (1.441,-0.733) (1.437,-0.709) (1.431,-0.685) (1.423,-0.663) (1.422,-0.642) (1.415,-0.621) (1.415,-0.602) (1.397,-0.584) (1.386,-0.566) (1.383,-0.549) (1.371,-0.532) (1.365,-0.517) (1.364,-0.501) (1.355,-0.487) (1.354,-0.472) (1.342,-0.459) (1.342,-0.445) (1.341,-0.432) (1.339,-0.420) (1.329,-0.407) (1.326,-0.396) (1.325,-0.384) (1.324,-0.373) (1.322,-0.362) (1.321,-0.351) (1.320,-0.341) (1.319,-0.330) (1.310,-0.320) (1.298,-0.311) (1.286,-0.301) (1.279,-0.292) (1.274,-0.283) (1.263,-0.274) (1.257,-0.265) (1.257,-0.256) (1.224,-0.248) (1.211,-0.240) (1.207,-0.231) (1.185,-0.223) (1.178,-0.216) (1.166,-0.208) (1.151,-0.200) (1.118,-0.193) (1.106,-0.186) (1.063,-0.178) (1.053,-0.171) (1.050,-0.164) (1.018,-0.158) (1.003,-0.151) (0.951,-0.144) (0.887,-0.138) (0.881,-0.131) (0.873,-0.125) (0.861,-0.119) (0.827,-0.113) (0.799,-0.106) (0.771,-0.100) (0.750,-0.095) (0.748,-0.089) (0.684,-0.083) (0.628,-0.077) (0.613,-0.072) (0.609,-0.066) (0.555,-0.061) (0.484,-0.055) (0.402,-0.050) (0.371,-0.045) (0.333,-0.040) (0.263,-0.034) (0.230,-0.029) (0.191,-0.024) (0.020,-0.019) (-0.125,-0.014) (-0.580,-0.010)
\end{pspicture}
\begin{pspicture}(-8,-4.3)(1,1.7)
\psset{xunit=2,yunit=2}
\psline{->}(-2,0)(0.5,0) \psline{->}(0,-2)(0,0.5) \scriptsize \rput(0.85,0){$log(Nd)$} \rput(0.65,0.5){$\log(Cum.\ distr.)$} \psline(-0.5,-0.05)(-0.5,0.05) \rput(-0.5,-0.15){$-0.5$} \psline(-1,-0.05)(-1,0.05) \rput(-1,-0.15){$-1$} \psline(-1.5,-0.05)(-1.5,0.05) \rput(-1.5,-0.15){$-1.5$} \psline(-0.05,-1.5)(0.05,-1.5) \rput(-0.2,-1.5){$-1.5$} \psline(-0.05,-1)(0.05,-1) \rput(-0.2,-1){$-1$} \psline(-0.05,-0.5)(0.05,-0.5) \rput(-0.2,-0.5){$-0.5$}
\psline[linecolor=blue] (-1.042,-0.301) (-1.042,0.000)
\psline[linecolor=blue] (0.155,-1.964) (0.154,-1.663) (0.144,-1.487) (0.143,-1.362) (0.143,-1.265) (0.138,-1.186) (0.123,-1.119) (0.120,-1.061) (0.117,-1.010) (0.108,-0.964) (0.106,-0.922) (0.106,-0.885) (0.101,-0.850) (0.095,-0.818) (0.095,-0.788) (0.092,-0.760) (0.091,-0.733) (0.087,-0.709) (0.084,-0.685) (0.081,-0.663) (0.080,-0.642) (0.080,-0.621) (0.077,-0.602) (0.076,-0.584) (0.074,-0.566) (0.068,-0.549) (0.068,-0.532) (0.066,-0.517) (0.064,-0.501) (0.064,-0.487) (0.063,-0.472) (0.062,-0.459) (0.060,-0.445) (0.058,-0.432) (0.054,-0.420) (0.052,-0.407) (0.051,-0.396) (0.049,-0.384) (0.047,-0.373) (0.047,-0.362) (0.043,-0.351) (0.043,-0.341) (0.037,-0.330) (0.031,-0.320) (0.027,-0.311) (0.026,-0.301) (0.023,-0.292) (0.022,-0.283) (0.021,-0.274) (0.021,-0.265) (0.020,-0.256) (0.019,-0.248) (0.019,-0.240) (0.019,-0.231) (0.017,-0.223) (0.017,-0.216) (0.016,-0.208) (0.014,-0.200) (0.014,-0.193) (0.012,-0.186) (0.011,-0.178) (0.010,-0.171) (0.007,-0.164) (0.005,-0.158) (0.004,-0.151) (0.003,-0.144) (0.000,-0.138) (-0.002,-0.131) (-0.008,-0.125) (-0.010,-0.119) (-0.011,-0.113) (-0.016,-0.106) (-0.020,-0.100) (-0.024,-0.095) (-0.026,-0.089) (-0.027,-0.083) (-0.028,-0.077) (-0.029,-0.072) (-0.031,-0.066) (-0.034,-0.061) (-0.035,-0.055) (-0.038,-0.050) (-0.039,-0.045) (-0.040,-0.040) (-0.041,-0.034) (-0.052,-0.029) (-0.056,-0.024) (-0.056,-0.019) (-0.057,-0.014) (-0.063,-0.010) (-0.067,-0.005) (-0.087,0.000)
\psline[linecolor=blue] (-0.168,-1.964) (-0.190,-1.663) (-0.351,-1.487) (-0.369,-1.362) (-0.396,-1.265) (-0.398,-1.186) (-0.429,-1.119) (-0.486,-1.061) (-0.489,-1.010) (-0.504,-0.964) (-0.508,-0.922) (-0.520,-0.885) (-0.608,-0.850) (-0.642,-0.818) (-0.659,-0.788) (-0.664,-0.760) (-0.677,-0.733) (-0.759,-0.709) (-0.760,-0.685) (-0.784,-0.663) (-0.834,-0.642) (-0.847,-0.621) (-0.896,-0.602) (-0.941,-0.584) (-0.952,-0.566) (-0.967,-0.549) (-1.047,-0.532) (-1.101,-0.517) (-1.103,-0.501) (-1.208,-0.487) (-1.253,-0.472) (-1.307,-0.459) (-1.401,-0.445) (-1.517,-0.432) (-1.632,-0.420)
\psline[linecolor=blue] (-0.168,-1.964) (-0.190,-1.663) (-0.351,-1.487) (-0.369,-1.362) (-0.396,-1.265) (-0.398,-1.186) (-0.429,-1.119) (-0.486,-1.061) (-0.489,-1.010) (-0.504,-0.964) (-0.508,-0.922) (-0.608,-0.885) (-0.642,-0.850) (-0.659,-0.818) (-0.664,-0.788) (-0.677,-0.760) (-0.728,-0.733) (-0.759,-0.709) (-0.760,-0.685) (-0.784,-0.663) (-0.834,-0.642) (-0.847,-0.621) (-0.896,-0.602) (-0.941,-0.584) (-0.952,-0.566) (-0.967,-0.549) (-1.047,-0.532) (-1.101,-0.517) (-1.103,-0.501) (-1.208,-0.487) (-1.253,-0.472) (-1.307,-0.459) (-1.401,-0.445) (-1.517,-0.432) (-1.632,-0.420)
\end{pspicture}

\noindent Fig. 25. Cumulative frequency distribution of the node strength in terms of the centrality values, both in logarithmic representation. The figure on the left displays the curves for real data, and the figure on the right represents the curves for randomized data.

\vskip 0.3 cm

\subsection{Eigenvector centrality}

One node may have a low dregree, but it may be connected with other nodes with very high degrees, so it is, in some way, influent. A measure that takes into account the degree of neighbouring nodes when calculating the importance of a node is called eigenvector centrality. If one considers the eigenvectors of the adjacency matrix of the distance matrix for the asset graph, and choosing its largest value, one may then define the eigenvector with largest eigenvalue by the equation
\begin{equation}
\label{eigenvec}
AX=\lambda X\ ,
\end{equation}
where $X$ is the eigenvector with the largest eigenvector $\lambda $. The eigenvector centrality of a node $i$ is then defined as the $i$th element of eigenvalue $X$:
\begin{equation}
Ec^i=x_i\ ,
\end{equation}
where $x_i$ is the element of $X$ in row $i$.

Here, there is a clearer set of nodes which exhibit more eigenvector centrality values than the others in a consistent way for the thresholds closest to the limit of the noise region. The UK, France, the Netherlands, Germany, Austria, and Denmark appear as the top nodes from $T=0.1$ up to $T=0.6$, which is not surprising, since they inhabit a densely connected region of the asset graphs.

Figure 26 displays the cumulative frequency distribution of the eigenvector centrality in terms of the centrality values, both in logarithmic representation. The figure on the left displays the curves for real data, and the figure on the right represents the curves for randomized data. All curves are quite close to one anotherm and the curves for extreme values, $T=1.0$ and $T=1.1$, are very similar to the ones obtained for randomized data.

\begin{pspicture}(-6,-4)(1,1.2)
\psset{xunit=1.7,yunit=2}
\psline{->}(-3.5,0)(0.5,0) \psline{->}(0,-2)(0,0.5) \scriptsize \rput(0.9,0){$log(Ec)$} \rput(0.75,0.5){$\log(Cum.\ distr.)$} \psline(-3,-0.05)(-3,0.05) \rput(-3,-0.15){$-3$} \psline(-2.5,-0.05)(-2.5,0.05) \rput(-2.5,-0.15){$-2.5$} \psline(-2,-0.05)(-2,0.05) \rput(-2,-0.15){$-2$} \psline(-1.5,-0.05)(-1.5,0.05) \rput(-1.5,-0.15){$-1.5$} \psline(-1,-0.05)(-1,0.05) \rput(-1,-0.15){$-1$} \psline(-0.5,-0.05)(-0.5,0.05) \rput(-0.5,-0.15){$-0.5$} \psline(-0.05,-1.5)(0.05,-1.5) \rput(-0.2,-1.5){$-1.5$} \psline(-0.05,-1)(0.05,-1) \rput(-0.2,-1){$-1$} \psline(-0.05,-0.5)(0.05,-0.5) \rput(-0.2,-0.5){$-0.5$}
\psline[linecolor=red] (-0.350,-0.845) (-0.350,-0.544) (-0.350,-0.368) (-0.350,-0.243) (-0.350,-0.146)
\psline[linecolor=red] (-0.391,-1.146) (-0.391,-0.845) (-0.415,-0.669) (-0.447,-0.544) (-0.487,-0.447) (-0.541,-0.368) (-0.541,-0.301) (-0.623,-0.243) (-0.684,-0.192) (-0.807,-0.146)
\psline[linecolor=red] (-0.511,-1.380) (-0.511,-1.079) (-0.523,-0.903) (-0.529,-0.778) (-0.542,-0.681) (-0.542,-0.602) (-0.556,-0.535) (-0.564,-0.477) (-0.570,-0.426) (-0.570,-0.380) (-0.592,-0.339) (-0.701,-0.301) (-0.752,-0.266) (-0.812,-0.234) (-0.975,-0.204)
\psline[linecolor=red] (-0.577,-1.556) (-0.577,-1.255) (-0.583,-1.079) (-0.588,-0.954) (-0.590,-0.857) (-0.590,-0.778) (-0.599,-0.711) (-0.599,-0.653) (-0.606,-0.602) (-0.614,-0.556) (-0.614,-0.515) (-0.622,-0.477) (-0.635,-0.442) (-0.640,-0.410) (-0.662,-0.380) (-0.750,-0.352) (-0.866,-0.326) (-0.939,-0.301) (-0.987,-0.278) (-1.481,-0.255) (-1.921,-0.234) (-2.046,-0.214)
\psline[linecolor=red] (-0.650,-1.681) (-0.652,-1.380) (-0.658,-1.204) (-0.662,-1.079) (-0.670,-0.982) (-0.674,-0.903) (-0.674,-0.836) (-0.678,-0.778) (-0.684,-0.727) (-0.686,-0.681) (-0.686,-0.640) (-0.688,-0.602) (-0.688,-0.567) (-0.688,-0.535) (-0.688,-0.505) (-0.703,-0.477) (-0.721,-0.451) (-0.721,-0.426) (-0.721,-0.402) (-0.721,-0.380) (-0.738,-0.359) (-0.762,-0.339) (-0.780,-0.320) (-0.783,-0.301) (-1.097,-0.283) (-1.102,-0.266) (-1.222,-0.250) (-1.292,-0.234) (-1.292,-0.219) (-1.347,-0.204) (-1.409,-0.190) (-1.602,-0.176) (-1.602,-0.163) (-1.770,-0.150) (-2.046,-0.137) (-2.046,-0.125) (-3.000,-0.113) (-3.000,-0.101) (-3.000,-0.090) (-3.000,-0.079) (-3.000,-0.068)
\psline[linecolor=red] (-0.710,-1.740) (-0.712,-1.439) (-0.712,-1.263) (-0.712,-1.138) (-0.712,-1.041) (-0.717,-0.962) (-0.717,-0.895) (-0.717,-0.837) (-0.717,-0.786) (-0.724,-0.740) (-0.724,-0.699) (-0.726,-0.661) (-0.730,-0.626) (-0.733,-0.594) (-0.762,-0.564) (-0.764,-0.536) (-0.770,-0.510) (-0.772,-0.485) (-0.788,-0.462) (-0.804,-0.439) (-0.804,-0.418) (-0.812,-0.398) (-0.824,-0.379) (-0.824,-0.360) (-0.824,-0.342) (-0.851,-0.325) (-0.857,-0.309) (-0.857,-0.293) (-0.870,-0.278) (-0.876,-0.263) (-0.889,-0.249) (-0.910,-0.235) (-0.932,-0.222) (-0.932,-0.209) (-0.932,-0.196) (-0.975,-0.184) (-0.975,-0.172) (-1.056,-0.161) (-1.222,-0.149) (-1.409,-0.138) (-1.469,-0.128) (-1.658,-0.117) (-1.921,-0.107) (-1.921,-0.097) (-1.921,-0.087) (-2.046,-0.078) (-2.097,-0.068) (-2.097,-0.059) (-2.699,-0.050) (-2.699,-0.041) (-3.000,-0.033)
\psline[linecolor=red] (-0.772,-1.792) (-0.772,-1.491) (-0.775,-1.315) (-0.777,-1.190) (-0.783,-1.093) (-0.788,-1.014) (-0.788,-0.947) (-0.788,-0.889) (-0.788,-0.838) (-0.788,-0.792) (-0.788,-0.751) (-0.793,-0.713) (-0.793,-0.678) (-0.799,-0.646) (-0.799,-0.616) (-0.801,-0.588) (-0.801,-0.562) (-0.807,-0.537) (-0.807,-0.514) (-0.812,-0.491) (-0.815,-0.470) (-0.815,-0.450) (-0.815,-0.431) (-0.821,-0.412) (-0.839,-0.394) (-0.848,-0.377) (-0.851,-0.361) (-0.857,-0.345) (-0.857,-0.330) (-0.857,-0.315) (-0.866,-0.301) (-0.876,-0.287) (-0.879,-0.274) (-0.889,-0.261) (-0.903,-0.248) (-0.903,-0.236) (-0.903,-0.224) (-0.914,-0.213) (-0.914,-0.201) (-0.914,-0.190) (-0.924,-0.180) (-0.928,-0.169) (-0.936,-0.159) (-0.943,-0.149) (-0.967,-0.139) (-1.022,-0.130) (-1.076,-0.120) (-1.081,-0.111) (-1.149,-0.102) (-1.367,-0.093) (-1.377,-0.085) (-1.432,-0.076) (-1.509,-0.068) (-1.538,-0.060) (-1.770,-0.052) (-1.770,-0.044) (-2.000,-0.037) (-2.222,-0.029) (-3.000,-0.022) (-3.000,-0.014)
\psline[linecolor=red] (-0.830,-1.851) (-0.833,-1.550) (-0.836,-1.374) (-0.836,-1.249) (-0.839,-1.152) (-0.839,-1.073) (-0.839,-1.006) (-0.842,-0.948) (-0.842,-0.897) (-0.842,-0.851) (-0.842,-0.810) (-0.845,-0.772) (-0.845,-0.737) (-0.845,-0.705) (-0.848,-0.675) (-0.851,-0.647) (-0.854,-0.621) (-0.854,-0.596) (-0.854,-0.573) (-0.854,-0.550) (-0.854,-0.529) (-0.854,-0.509) (-0.857,-0.490) (-0.857,-0.471) (-0.857,-0.453) (-0.857,-0.436) (-0.857,-0.420) (-0.857,-0.404) (-0.860,-0.389) (-0.860,-0.374) (-0.860,-0.360) (-0.863,-0.346) (-0.866,-0.333) (-0.866,-0.320) (-0.870,-0.307) (-0.870,-0.295) (-0.873,-0.283) (-0.879,-0.271) (-0.886,-0.260) (-0.893,-0.249) (-0.893,-0.238) (-0.903,-0.228) (-0.907,-0.218) (-0.910,-0.208) (-0.914,-0.198) (-0.921,-0.189) (-0.947,-0.179) (-0.951,-0.170) (-0.951,-0.161) (-0.951,-0.152) (-0.959,-0.144) (-1.004,-0.135) (-1.004,-0.127) (-1.041,-0.119) (-1.060,-0.111) (-1.066,-0.103) (-1.125,-0.095) (-1.161,-0.088) (-1.229,-0.080) (-1.432,-0.073) (-1.495,-0.066) (-1.553,-0.059) (-1.620,-0.052) (-2.000,-0.045) (-2.301,-0.038) (-2.301,-0.032) (-2.699,-0.025) (-2.699,-0.019) (-3.000,-0.012)
\psline[linecolor=red] (-0.883,-1.940) (-0.883,-1.638) (-0.886,-1.462) (-0.886,-1.337) (-0.886,-1.241) (-0.889,-1.161) (-0.889,-1.094) (-0.889,-1.036) (-0.889,-0.985) (-0.889,-0.940) (-0.889,-0.898) (-0.889,-0.860) (-0.889,-0.826) (-0.893,-0.793) (-0.893,-0.763) (-0.893,-0.735) (-0.893,-0.709) (-0.893,-0.684) (-0.896,-0.661) (-0.896,-0.638) (-0.896,-0.617) (-0.900,-0.597) (-0.900,-0.578) (-0.900,-0.559) (-0.900,-0.542) (-0.900,-0.525) (-0.903,-0.508) (-0.903,-0.492) (-0.903,-0.477) (-0.903,-0.462) (-0.903,-0.448) (-0.903,-0.434) (-0.903,-0.421) (-0.903,-0.408) (-0.907,-0.395) (-0.907,-0.383) (-0.907,-0.371) (-0.907,-0.360) (-0.907,-0.348) (-0.907,-0.337) (-0.907,-0.327) (-0.910,-0.316) (-0.910,-0.306) (-0.910,-0.296) (-0.914,-0.286) (-0.914,-0.277) (-0.914,-0.267) (-0.917,-0.258) (-0.917,-0.249) (-0.921,-0.241) (-0.921,-0.232) (-0.921,-0.224) (-0.924,-0.215) (-0.928,-0.207) (-0.928,-0.199) (-0.936,-0.191) (-0.943,-0.184) (-0.959,-0.176) (-0.971,-0.169) (-0.971,-0.161) (-0.983,-0.154) (-1.004,-0.147) (-1.004,-0.140) (-1.009,-0.133) (-1.013,-0.127) (-1.161,-0.120) (-1.167,-0.113) (-1.208,-0.107) (-1.244,-0.101) (-1.310,-0.094) (-1.357,-0.088) (-1.523,-0.082) (-1.523,-0.076) (-1.602,-0.070) (-1.721,-0.064) (-1.886,-0.059) (-2.046,-0.053) (-2.222,-0.047) (-2.398,-0.042) (-2.398,-0.036) (-2.699,-0.031) (-2.699,-0.026) (-2.699,-0.020) (-2.699,-0.015)
\psline[linecolor=red] (-0.951,-1.964) (-0.951,-1.663) (-0.955,-1.487) (-0.959,-1.362) (-0.959,-1.265) (-0.959,-1.186) (-0.959,-1.119) (-0.959,-1.061) (-0.959,-1.010) (-0.959,-0.964) (-0.959,-0.922) (-0.959,-0.885) (-0.959,-0.850) (-0.959,-0.818) (-0.959,-0.788) (-0.959,-0.760) (-0.959,-0.733) (-0.959,-0.709) (-0.959,-0.685) (-0.959,-0.663) (-0.959,-0.642) (-0.959,-0.621) (-0.963,-0.602) (-0.963,-0.584) (-0.963,-0.566) (-0.963,-0.549) (-0.963,-0.532) (-0.963,-0.517) (-0.963,-0.501) (-0.963,-0.487) (-0.963,-0.472) (-0.963,-0.459) (-0.963,-0.445) (-0.963,-0.432) (-0.963,-0.420) (-0.967,-0.407) (-0.967,-0.396) (-0.967,-0.384) (-0.967,-0.373) (-0.967,-0.362) (-0.967,-0.351) (-0.967,-0.341) (-0.967,-0.330) (-0.967,-0.320) (-0.967,-0.311) (-0.967,-0.301) (-0.967,-0.292) (-0.967,-0.283) (-0.967,-0.274) (-0.967,-0.265) (-0.971,-0.256) (-0.971,-0.248) (-0.971,-0.240) (-0.971,-0.231) (-0.971,-0.223) (-0.971,-0.216) (-0.971,-0.208) (-0.971,-0.200) (-0.971,-0.193) (-0.975,-0.186) (-0.975,-0.178) (-0.975,-0.171) (-0.975,-0.164) (-0.975,-0.158) (-0.975,-0.151) (-0.975,-0.144) (-0.975,-0.138) (-0.979,-0.131) (-0.979,-0.125) (-0.979,-0.119) (-0.979,-0.113) (-0.983,-0.106) (-0.983,-0.100) (-0.987,-0.095) (-0.987,-0.089) (-0.987,-0.083) (-0.991,-0.077) (-0.996,-0.072) (-1.000,-0.066) (-1.009,-0.061) (-1.022,-0.055) (-1.032,-0.050) (-1.051,-0.045) (-1.056,-0.040) (-1.056,-0.034) (-1.086,-0.029) (-1.097,-0.024) (-1.119,-0.019) (-1.215,-0.014) (-1.222,-0.010) (-1.301,-0.005) (-1.301,0.000)
\psline[linecolor=red] (-0.983,-1.964) (-0.983,-0.010) (-0.987,-0.005) (-0.987,0.000)
\psline[linecolor=red] (-0.983,-1.964) (-0.983,0.000)
\end{pspicture}
\begin{pspicture}(-8,-4)(1,1.2)
\psset{xunit=1.7,yunit=2}
\psline{->}(-3.5,0)(0.5,0) \psline{->}(0,-2)(0,0.5) \scriptsize \rput(0.9,0){$log(Ec)$} \rput(0.75,0.5){$\log(Cum.\ distr.)$} \psline(-3,-0.05)(-3,0.05) \rput(-3,-0.15){$-3$} \psline(-2.5,-0.05)(-2.5,0.05) \rput(-2.5,-0.15){$-2.5$} \psline(-2,-0.05)(-2,0.05) \rput(-2,-0.15){$-2$} \psline(-1.5,-0.05)(-1.5,0.05) \rput(-1.5,-0.15){$-1.5$} \psline(-1,-0.05)(-1,0.05) \rput(-1,-0.15){$-1$} \psline(-0.5,-0.05)(-0.5,0.05) \rput(-0.5,-0.15){$-0.5$} \psline(-0.05,-1.5)(0.05,-1.5) \rput(-0.2,-1.5){$-1.5$} \psline(-0.05,-1)(0.05,-1) \rput(-0.2,-1){$-1$} \psline(-0.05,-0.5)(0.05,-0.5) \rput(-0.2,-0.5){$-0.5$}
\psline[linecolor=blue] (-0.151,-0.301) (-0.151,0.000)
\psline[linecolor=blue] (-0.873,-1.964) (-0.893,-1.663) (-0.903,-1.487) (-0.907,-1.362) (-0.907,-1.265) (-0.907,-1.186) (-0.910,-1.119) (-0.914,-1.061) (-0.921,-1.010) (-0.921,-0.964) (-0.928,-0.922) (-0.932,-0.885) (-0.932,-0.850) (-0.936,-0.818) (-0.936,-0.788) (-0.939,-0.760) (-0.939,-0.733) (-0.943,-0.709) (-0.943,-0.685) (-0.947,-0.663) (-0.947,-0.642) (-0.947,-0.621) (-0.951,-0.602) (-0.951,-0.584) (-0.951,-0.566) (-0.955,-0.549) (-0.955,-0.532) (-0.955,-0.517) (-0.959,-0.501) (-0.959,-0.487) (-0.959,-0.472) (-0.963,-0.459) (-0.963,-0.445) (-0.963,-0.432) (-0.967,-0.420) (-0.967,-0.407) (-0.967,-0.396) (-0.971,-0.384) (-0.975,-0.373) (-0.975,-0.362) (-0.975,-0.351) (-0.979,-0.341) (-0.979,-0.330) (-0.979,-0.320) (-0.983,-0.311) (-0.983,-0.301) (-0.987,-0.292) (-0.987,-0.283) (-0.991,-0.274) (-0.991,-0.265) (-0.991,-0.256) (-0.991,-0.248) (-0.991,-0.240) (-1.000,-0.231) (-1.000,-0.223) (-1.004,-0.216) (-1.009,-0.208) (-1.009,-0.200) (-1.009,-0.193) (-1.013,-0.186) (-1.013,-0.178) (-1.013,-0.171) (-1.013,-0.164) (-1.013,-0.158) (-1.018,-0.151) (-1.018,-0.144) (-1.018,-0.138) (-1.022,-0.131) (-1.022,-0.125) (-1.022,-0.119) (-1.027,-0.113) (-1.027,-0.106) (-1.032,-0.100) (-1.032,-0.095) (-1.036,-0.089) (-1.036,-0.083) (-1.041,-0.077) (-1.051,-0.072) (-1.051,-0.066) (-1.056,-0.061) (-1.056,-0.055) (-1.056,-0.050) (-1.056,-0.045) (-1.060,-0.040) (-1.060,-0.034) (-1.060,-0.029) (-1.060,-0.024) (-1.060,-0.019) (-1.071,-0.014) (-1.076,-0.010) (-1.097,-0.005) (-1.125,0.000)
\psline[linecolor=blue] (-0.983,-1.964) (-0.983,-0.010) (-0.987,-0.005) (-0.987,0.000)
\psline[linecolor=blue] (-0.983,-1.964) (-0.983,0.000)
\end{pspicture}

\noindent Fig. 26. Cumulative frequency distribution of the eigenvector centrality in terms of the centrality values, both in logarithmic representation. The figure on the left displays the curves for real data, and the figure on the right represents the curves for randomized data.

\vskip 0.3 cm

\subsection{Betweenness}

The betweenness centrality measures how often a node lies on the paths between other vertices. It is an important measure of how much a node is important as an intermediate between other nodes. It may be defined as
\begin{equation}
Bc^k=\sum_{i,j=1}^n\frac{n_{ij}^k}{m_{ij}}\ ,
\end{equation}
where $n_{ij}$ is the number of shortest paths (geodesic paths) between nodes $i$ and $j$ that pass through node $k$ and $m_{ij}$ is the total number of shortest paths between nodes $i$ and $j$. Our network is fully connected, so we need not worry about $m_{ij}$ being zero.

Here, one does not have a particular set of nodes that present high betweenness for most threshold values. One may highlight Singapore, which appears among the top four from $T=0.5$ to $T=0.8$, and the Czech Republic, which is among the top four if $T=0.5$ or $T=0.6$.

Figure 27 displays the cumulative frequency distribution of the betweenness centrality in terms of the centrality values, both in logarithmic representation. Like before, the figure on the left displays the curves for real data, and the figure on the right represents the curves for randomized data. The curves for real and randomized data are almost exactly the same for $T=1.0$ and $T=1.1$, which are well inside the random noise region. The curves for thresholds ranging from $T=0.4$ to $T=0.9$ get entangled and are quite distinct from the other curves.

\begin{pspicture}(-3,-5.2)(5,1.5)
\psset{xunit=1.4,yunit=2}
\psline{->}(-2.5,0)(3,0) \psline{->}(0,-2.5)(0,0.5) \scriptsize \rput(3.5,0){$log(Bc)$} \rput(0.93,0.5){$\log(Cum.\ distr.)$} \psline(-2,-0.05)(-2,0.05) \rput(-2,-0.15){$-2$} \psline(-1.5,-0.05)(-1.5,0.05) \rput(-1.5,-0.15){$-1.5$} \psline(-1,-0.05)(-1,0.05) \rput(-1,-0.15){$-1$} \psline(-0.5,-0.05)(-0.5,0.05) \rput(-0.5,-0.15){$-0.5$} \psline(0.5,-0.05)(0.5,0.05) \rput(0.5,-0.15){$0.5$} \psline(1,-0.05)(1,0.05) \rput(1,-0.15){$1$} \psline(1.5,-0.05)(1.5,0.05) \rput(1.5,-0.15){$1.5$} \psline(2,-0.05)(2,0.05) \rput(2,-0.15){$2$} \psline(2.5,-0.05)(2.5,0.05) \rput(2.5,-0.15){$2.5$} \psline(-0.05,-2)(0.05,-2) \rput(-0.2,-2){$-2$} \psline(-0.05,-1.5)(0.05,-1.5) \rput(-0.2,-1.5){$-1.5$} \psline(-0.05,-1)(0.05,-1) \rput(-0.2,-1){$-1$} \psline(-0.05,-0.5)(0.05,-0.5) \rput(-0.2,-0.5){$-0.5$}
\psline[linecolor=red] (0.778,-0.845)
\psline[linecolor=red] (0.736,-1.146) (0.736,-0.845) (0.418,-0.669) (-0.022,-0.544) (-0.478,-0.447) (-0.699,-0.368)
\psline[linecolor=red] (0.751,-1.380) (0.751,-1.079) (0.580,-0.903) (0.465,-0.778) (0.460,-0.681) (0.140,-0.602) (0.140,-0.535) (0.000,-0.477) (-0.079,-0.426) (-0.079,-0.380) (-0.243,-0.339) (-0.845,-0.301)
\psline[linecolor=red] (1.316,-1.556) (1.301,-1.255) (1.228,-1.079) (0.954,-0.954) (0.923,-0.857) (0.923,-0.778) (0.795,-0.711) (0.699,-0.653) (0.699,-0.602) (0.677,-0.556) (0.677,-0.515) (0.573,-0.477) (0.545,-0.442) (0.545,-0.410) (0.242,-0.380) (0.176,-0.352) (0.176,-0.326) (0.025,-0.301) (0.025,-0.278) (0.000,-0.255) (0.000,-0.234) (-0.740,-0.214) (-1.114,-0.195)
\psline[linecolor=red] (2.610,-1.681) (2.560,-1.380) (2.012,-1.204) (1.847,-1.079) (1.643,-0.982) (1.634,-0.903) (1.510,-0.836) (1.376,-0.778) (1.305,-0.727) (1.235,-0.681) (1.145,-0.640) (1.114,-0.602) (0.967,-0.567) (0.967,-0.535) (0.775,-0.505) (0.771,-0.477) (0.518,-0.451) (0.260,-0.426) (0.209,-0.402) (0.093,-0.380) (0.000,-0.359) (-0.130,-0.339) (-0.130,-0.320) (-0.130,-0.301) (-0.130,-0.283) (-0.227,-0.266) (-0.354,-0.250) (-0.472,-0.234) (-0.550,-0.219) (-0.664,-0.204) (-0.664,-0.190) (-0.785,-0.176) (-0.845,-0.163) (-1.276,-0.150)
\psline[linecolor=red] (2.230,-1.740) (2.012,-1.439) (1.832,-1.263) (1.807,-1.138) (1.734,-1.041) (1.617,-0.962) (1.571,-0.895) (1.480,-0.837) (1.464,-0.786) (1.337,-0.740) (1.253,-0.699) (1.253,-0.661) (1.253,-0.626) (1.253,-0.594) (1.173,-0.564) (1.159,-0.536) (1.155,-0.510) (1.155,-0.485) (1.155,-0.462) (1.123,-0.439) (1.080,-0.418) (1.012,-0.398) (0.875,-0.379) (0.870,-0.360) (0.779,-0.342) (0.658,-0.325) (0.579,-0.309) (0.372,-0.293) (0.341,-0.278) (0.339,-0.263) (0.203,-0.249) (0.117,-0.235) (0.101,-0.222) (0.101,-0.209) (0.013,-0.196) (-0.387,-0.184) (-0.425,-0.172) (-0.439,-0.161) (-0.469,-0.149) (-0.478,-0.138) (-0.740,-0.128) (-0.740,-0.117) (-0.740,-0.107) (-0.740,-0.097) (-0.801,-0.087) (-1.367,-0.078)
\psline[linecolor=red] (2.358,-1.792) (1.994,-1.491) (1.949,-1.315) (1.914,-1.190) (1.858,-1.093) (1.850,-1.014) (1.763,-0.947) (1.711,-0.889) (1.632,-0.838) (1.505,-0.792) (1.493,-0.751) (1.485,-0.713) (1.462,-0.678) (1.389,-0.646) (1.298,-0.616) (1.208,-0.588) (1.110,-0.562) (1.053,-0.537) (1.003,-0.514) (1.003,-0.491) (1.003,-0.470) (1.003,-0.450) (1.003,-0.431) (0.969,-0.412) (0.957,-0.394) (0.947,-0.377) (0.946,-0.361) (0.929,-0.345) (0.895,-0.330) (0.862,-0.315) (0.862,-0.301) (0.834,-0.287) (0.821,-0.274) (0.815,-0.261) (0.805,-0.248) (0.765,-0.236) (0.749,-0.224) (0.704,-0.213) (0.674,-0.201) (0.483,-0.190) (0.398,-0.180) (0.327,-0.169) (0.311,-0.159) (0.235,-0.149) (-0.179,-0.139) (-0.179,-0.130) (-0.614,-0.120) (-0.614,-0.111) (-0.614,-0.102) (-0.623,-0.093) (-0.860,-0.085) (-0.886,-0.076) (-1.081,-0.068) (-1.081,-0.060) (-1.161,-0.052)
\psline[linecolor=red] (2.186,-1.851) (2.034,-1.550) (1.919,-1.374) (1.843,-1.249) (1.827,-1.152) (1.764,-1.073) (1.759,-1.006) (1.736,-0.948) (1.723,-0.897) (1.577,-0.851) (1.571,-0.810) (1.541,-0.772) (1.533,-0.737) (1.420,-0.705) (1.390,-0.675) (1.314,-0.647) (1.301,-0.621) (1.290,-0.596) (1.210,-0.573) (1.204,-0.550) (1.131,-0.529) (1.098,-0.509) (1.097,-0.490) (1.085,-0.471) (1.066,-0.453) (1.066,-0.436) (1.047,-0.420) (1.000,-0.404) (1.000,-0.389) (0.999,-0.374) (0.935,-0.360) (0.911,-0.346) (0.871,-0.333) (0.752,-0.320) (0.744,-0.307) (0.689,-0.295) (0.686,-0.283) (0.686,-0.271) (0.686,-0.260) (0.682,-0.249) (0.674,-0.238) (0.674,-0.228) (0.669,-0.218) (0.584,-0.208) (0.584,-0.198) (0.584,-0.189) (0.584,-0.179) (0.584,-0.170) (0.487,-0.161) (0.408,-0.152) (0.408,-0.144) (0.314,-0.135) (0.239,-0.127) (0.160,-0.119) (0.089,-0.111) (0.040,-0.103) (-0.049,-0.095) (-0.284,-0.088) (-0.401,-0.080) (-0.425,-0.073) (-0.533,-0.066)
\psline[linecolor=red] (2.405,-1.940) (2.228,-1.638) (1.989,-1.462) (1.984,-1.337) (1.958,-1.241) (1.932,-1.161) (1.906,-1.094) (1.867,-1.036) (1.853,-0.985) (1.788,-0.940) (1.712,-0.898) (1.706,-0.860) (1.607,-0.826) (1.551,-0.793) (1.492,-0.763) (1.483,-0.735) (1.465,-0.709) (1.447,-0.684) (1.368,-0.661) (1.336,-0.638) (1.307,-0.617) (1.297,-0.597) (1.289,-0.578) (1.275,-0.559) (1.256,-0.542) (1.249,-0.525) (1.215,-0.508) (1.215,-0.492) (1.210,-0.477) (1.195,-0.462) (1.177,-0.448) (1.161,-0.434) (1.161,-0.421) (1.156,-0.408) (1.144,-0.395) (1.114,-0.383) (1.110,-0.371) (1.078,-0.360) (1.066,-0.348) (1.059,-0.337) (1.000,-0.327) (0.980,-0.316) (0.975,-0.306) (0.904,-0.296) (0.888,-0.286) (0.874,-0.277) (0.848,-0.267) (0.842,-0.258) (0.841,-0.249) (0.834,-0.241) (0.808,-0.232) (0.780,-0.224) (0.778,-0.215) (0.741,-0.207) (0.738,-0.199) (0.675,-0.191) (0.653,-0.184) (0.648,-0.176) (0.608,-0.169) (0.608,-0.161) (0.591,-0.154) (0.562,-0.147) (0.479,-0.140) (0.445,-0.133) (0.368,-0.127) (0.363,-0.120) (0.354,-0.113) (0.251,-0.107) (0.142,-0.101) (0.056,-0.094) (-0.046,-0.088) (-0.081,-0.082) (-0.524,-0.076) (-0.553,-0.070) (-0.583,-0.064) (-0.672,-0.059) (-0.712,-0.053) (-0.900,-0.047) (-0.987,-0.042) (-1.260,-0.036) (-1.469,-0.031) (-1.569,-0.026)
\psline[linecolor=red] (0.861,-1.964) (0.861,-1.663) (0.822,-1.487) (0.802,-1.362) (0.769,-1.265) (0.759,-1.186) (0.752,-1.119) (0.751,-1.061) (0.751,-1.010) (0.751,-0.964) (0.746,-0.922) (0.745,-0.885) (0.738,-0.850) (0.733,-0.818) (0.731,-0.788) (0.727,-0.760) (0.727,-0.733) (0.726,-0.709) (0.719,-0.685) (0.705,-0.663) (0.704,-0.642) (0.703,-0.621) (0.703,-0.602) (0.698,-0.584) (0.695,-0.566) (0.688,-0.549) (0.678,-0.532) (0.675,-0.517) (0.672,-0.501) (0.670,-0.487) (0.654,-0.472) (0.653,-0.459) (0.648,-0.445) (0.648,-0.432) (0.646,-0.420) (0.645,-0.407) (0.640,-0.396) (0.634,-0.384) (0.632,-0.373) (0.631,-0.362) (0.623,-0.351) (0.621,-0.341) (0.620,-0.330) (0.618,-0.320) (0.618,-0.311) (0.590,-0.301) (0.589,-0.292) (0.585,-0.283) (0.583,-0.274) (0.576,-0.265) (0.570,-0.256) (0.568,-0.248) (0.567,-0.240) (0.563,-0.231) (0.563,-0.223) (0.563,-0.216) (0.561,-0.208) (0.559,-0.200) (0.536,-0.193) (0.532,-0.186) (0.528,-0.178) (0.513,-0.171) (0.510,-0.164) (0.500,-0.158) (0.483,-0.151) (0.479,-0.144) (0.475,-0.138) (0.466,-0.131) (0.465,-0.125) (0.460,-0.119) (0.460,-0.113) (0.456,-0.106) (0.445,-0.100) (0.435,-0.095) (0.418,-0.089) (0.414,-0.083) (0.410,-0.077) (0.384,-0.072) (0.381,-0.066) (0.377,-0.061) (0.375,-0.055) (0.359,-0.050) (0.358,-0.045) (0.332,-0.040) (0.307,-0.034) (0.302,-0.029) (0.274,-0.024) (0.261,-0.019) (0.255,-0.014) (0.202,-0.010) (0.088,-0.005) (0.066,0.000)
\psline[linecolor=red] (-1.959,-1.964) (-1.959,-0.010)
\end{pspicture}
\begin{pspicture}(-4.3,-5.2)(5,1.5)
\psset{xunit=1.4,yunit=2}
\psline{->}(-2.5,0)(3,0) \psline{->}(0,-2.5)(0,0.5) \scriptsize \rput(3.5,0){$log(Bc)$} \rput(0.93,0.5){$\log(Cum.\ distr.)$} \psline(-2,-0.05)(-2,0.05) \rput(-2,-0.15){$-2$} \psline(-1.5,-0.05)(-1.5,0.05) \rput(-1.5,-0.15){$-1.5$} \psline(-1,-0.05)(-1,0.05) \rput(-1,-0.15){$-1$} \psline(-0.5,-0.05)(-0.5,0.05) \rput(-0.5,-0.15){$-0.5$} \psline(0.5,-0.05)(0.5,0.05) \rput(0.5,-0.15){$0.5$} \psline(1,-0.05)(1,0.05) \rput(1,-0.15){$1$} \psline(1.5,-0.05)(1.5,0.05) \rput(1.5,-0.15){$1.5$} \psline(2,-0.05)(2,0.05) \rput(2,-0.15){$2$} \psline(2.5,-0.05)(2.5,0.05) \rput(2.5,-0.15){$2.5$} \psline(-0.05,-2)(0.05,-2) \rput(-0.2,-2){$-2$} \psline(-0.05,-1.5)(0.05,-1.5) \rput(-0.2,-1.5){$-1.5$} \psline(-0.05,-1)(0.05,-1) \rput(-0.2,-1){$-1$} \psline(-0.05,-0.5)(0.05,-0.5) \rput(-0.2,-0.5){$-0.5$}
\psline[linecolor=blue] (1.573,-1.964) (1.543,-1.663) (1.529,-1.487) (1.522,-1.362) (1.514,-1.265) (1.510,-1.186) (1.488,-1.119) (1.487,-1.061) (1.482,-1.010) (1.474,-0.964) (1.465,-0.922) (1.459,-0.885) (1.458,-0.850) (1.456,-0.818) (1.449,-0.788) (1.445,-0.760) (1.437,-0.733) (1.436,-0.709) (1.435,-0.685) (1.429,-0.663) (1.422,-0.642) (1.419,-0.621) (1.415,-0.602) (1.413,-0.584) (1.413,-0.566) (1.409,-0.549) (1.407,-0.532) (1.402,-0.517) (1.396,-0.501) (1.391,-0.487) (1.388,-0.472) (1.386,-0.459) (1.379,-0.445) (1.374,-0.432) (1.372,-0.420) (1.369,-0.407) (1.365,-0.396) (1.364,-0.384) (1.359,-0.373) (1.358,-0.362) (1.355,-0.351) (1.353,-0.341) (1.352,-0.330) (1.348,-0.320) (1.346,-0.311) (1.345,-0.301) (1.343,-0.292) (1.343,-0.283) (1.342,-0.274) (1.339,-0.265) (1.338,-0.256) (1.334,-0.248) (1.334,-0.240) (1.333,-0.231) (1.328,-0.223) (1.327,-0.216) (1.323,-0.208) (1.322,-0.200) (1.321,-0.193) (1.319,-0.186) (1.314,-0.178) (1.309,-0.171) (1.305,-0.164) (1.302,-0.158) (1.300,-0.151) (1.300,-0.144) (1.297,-0.138) (1.289,-0.131) (1.282,-0.125) (1.276,-0.119) (1.273,-0.113) (1.263,-0.106) (1.253,-0.100) (1.251,-0.095) (1.242,-0.089) (1.235,-0.083) (1.232,-0.077) (1.228,-0.072) (1.226,-0.066) (1.224,-0.061) (1.218,-0.055) (1.204,-0.050) (1.197,-0.045) (1.191,-0.040) (1.190,-0.034) (1.179,-0.029) (1.166,-0.024) (1.162,-0.019) (1.161,-0.014) (1.160,-0.010) (1.140,-0.005) (1.133,0.000)
\psline[linecolor=blue] (-1.959,-1.964) (-1.959,-0.010)
\end{pspicture}

\noindent Fig. 27. Cumulative frequency distribution of the betweenness centrality in terms of the centrality values, both in logarithmic representation. The figure on the left displays the curves for real data, and the figure on the right represents the curves for randomized data.

\vskip 0.3 cm

\subsection{Harmonic closeness}

Another measure of centrality is the {\sl closeness centrality}, which measures the average distance between one node and all the others. It is defined as the measure of the mean geodesic distance for a given node $i$, which is given by
\begin{equation}
\ell _i=\frac{1}{n}\sum_{j=1}^nd_{ij}\ ,
\end{equation}
where $n$ is the number of vertices and $d_{ij}$ is a geodesic (minimum path) distance from node $i$ to node $j$. This measure is small for highly connected vertices and large for distant or poorly connected ones. In order to obtain a measure that is large for highly connected nodes and small for poorly connected ones, one then defines the {\sl harmonic closeness centrality} of node $i$ as
\begin{equation}
Hc^i=\frac{1}{n-1}\sum_{i=1,i\neq j}^n\frac{1}{d_{ij}}\ .
\end{equation}

There is again a set of nodes which consistently have the highest harmonic closeness centralities for assert graphs close to the noise threshold. These are Central European indices, like the UK, France, Germany, the Netherlands, Italy, and with an Eastern European index, from the Czech Republic.

Figure 28 displays the cumulative frequency distribution of the harmonic closeness centrality in terms of the centrality values, both in logarithmic representation. Once more, the figure on the left displays the curves for real data, and the figure on the right represents the curves for randomized data. Again, only the curves for thresholds $T=1.0$ and $T=1.1$ are similar to the ones obtained based on randomized data.

\begin{pspicture}(-1,-4.4)(5,1.5)
\psset{xunit=2,yunit=2}
\psline{->}(-0.5,0)(2.5,0) \psline{->}(0,-2)(0,0.5) \scriptsize \rput(2.85,0){$log(Hc)$} \rput(0.65,0.5){$\log(Cum.\ distr.)$} \psline(0.5,-0.05)(0.5,0.05) \rput(0.5,-0.15){$0.5$} \psline(1,-0.05)(1,0.05) \rput(1,-0.15){$1$} \psline(1.5,-0.05)(1.5,0.05) \rput(1.5,-0.15){$1.5$} \psline(2,-0.05)(2,0.05) \rput(2,-0.15){$2$} \psline(-0.05,-1.5)(0.05,-1.5) \rput(-0.25,-1.5){$-1.5$} \psline(-0.05,-1)(0.05,-1) \rput(-0.2,-1){$-1$} \psline(-0.05,-0.5)(0.05,-0.5) \rput(-0.25,-0.5){$-0.5$}
\psline[linecolor=red] (0.602,-0.845) (0.398,-0.544) (0.398,-0.368) (0.398,-0.243) (0.398,-0.146) (0.000,-0.067) (0.000,0.000)
\psline[linecolor=red] (0.954,-1.146) (0.954,-0.845) (0.929,-0.669) (0.903,-0.544) (0.875,-0.447) (0.845,-0.368) (0.845,-0.301) (0.813,-0.243) (0.813,-0.192) (0.778,-0.146) (0.000,-0.105) (0.000,-0.067) (0.000,-0.032) (0.000,0.000)
\psline[linecolor=red] (1.146,-1.380) (1.146,-1.079) (1.130,-0.903) (1.130,-0.778) (1.114,-0.681) (1.114,-0.602) (1.114,-0.535) (1.097,-0.477) (1.097,-0.426) (1.097,-0.380) (1.079,-0.339) (1.041,-0.301) (1.021,-0.266) (1.000,-0.234) (0.954,-0.204) (0.301,-0.176) (0.176,-0.150) (0.176,-0.125) (0.000,-0.101) (0.000,-0.079) (0.000,-0.058) (0.000,-0.038) (0.000,-0.018) (0.000,0.000)
\psline[linecolor=red] (1.290,-1.556) (1.290,-1.255) (1.290,-1.079) (1.279,-0.954) (1.275,-0.857) (1.275,-0.778) (1.267,-0.711) (1.267,-0.653) (1.263,-0.602) (1.251,-0.556) (1.251,-0.515) (1.251,-0.477) (1.247,-0.442) (1.235,-0.410) (1.222,-0.380) (1.213,-0.352) (1.171,-0.326) (1.141,-0.301) (1.114,-0.278) (1.041,-0.255) (0.974,-0.234) (0.950,-0.214) (0.699,-0.195) (0.699,-0.176) (0.653,-0.158) (0.653,-0.141) (0.653,-0.125) (0.602,-0.109) (0.602,-0.094) (0.544,-0.079) (0.523,-0.065) (0.523,-0.051) (0.426,-0.038) (0.426,-0.025) (0.000,-0.012) (0.000,0.000)
\psline[linecolor=red] (1.565,-1.681) (1.559,-1.380) (1.547,-1.204) (1.542,-1.079) (1.541,-0.982) (1.528,-0.903) (1.522,-0.836) (1.522,-0.778) (1.522,-0.727) (1.515,-0.681) (1.515,-0.640) (1.509,-0.602) (1.509,-0.567) (1.509,-0.535) (1.509,-0.505) (1.502,-0.477) (1.495,-0.451) (1.495,-0.426) (1.495,-0.402) (1.495,-0.380) (1.495,-0.359) (1.481,-0.339) (1.473,-0.320) (1.473,-0.301) (1.414,-0.283) (1.412,-0.266) (1.406,-0.250) (1.379,-0.234) (1.379,-0.219) (1.379,-0.204) (1.361,-0.190) (1.361,-0.176) (1.331,-0.163) (1.331,-0.150) (1.312,-0.137) (1.292,-0.125) (1.281,-0.113) (1.279,-0.101) (1.267,-0.090) (1.267,-0.079) (1.267,-0.068) (1.226,-0.058) (1.226,-0.048) (1.226,-0.038) (1.226,-0.028) (1.119,-0.018) (0.000,-0.009) (0.000,0.000)
\psline[linecolor=red] (1.647,-1.740) (1.640,-1.439) (1.638,-1.263) (1.638,-1.138) (1.638,-1.041) (1.638,-0.962) (1.633,-0.895) (1.633,-0.837) (1.633,-0.786) (1.633,-0.740) (1.628,-0.699) (1.628,-0.661) (1.628,-0.626) (1.628,-0.594) (1.623,-0.564) (1.613,-0.536) (1.613,-0.510) (1.607,-0.485) (1.600,-0.462) (1.597,-0.439) (1.593,-0.418) (1.585,-0.398) (1.580,-0.379) (1.570,-0.360) (1.564,-0.342) (1.564,-0.325) (1.558,-0.309) (1.556,-0.293) (1.556,-0.278) (1.552,-0.263) (1.550,-0.249) (1.546,-0.235) (1.538,-0.222) (1.531,-0.209) (1.531,-0.196) (1.531,-0.184) (1.519,-0.172) (1.515,-0.161) (1.503,-0.149) (1.484,-0.138) (1.460,-0.128) (1.457,-0.117) (1.431,-0.107) (1.431,-0.097) (1.431,-0.087) (1.423,-0.078) (1.415,-0.068) (1.401,-0.059) (1.322,-0.050) (1.319,-0.041) (1.288,-0.033) (0.000,-0.024) (0.000,-0.016) (0.000,-0.008) (0.000,0.000)
\psline[linecolor=red] (1.753,-1.792) (1.749,-1.491) (1.746,-1.315) (1.742,-1.190) (1.742,-1.093) (1.730,-1.014) (1.726,-0.947) (1.726,-0.889) (1.726,-0.838) (1.726,-0.792) (1.726,-0.751) (1.726,-0.713) (1.726,-0.678) (1.722,-0.646) (1.722,-0.616) (1.717,-0.588) (1.717,-0.562) (1.717,-0.537) (1.717,-0.514) (1.713,-0.491) (1.713,-0.470) (1.709,-0.450) (1.709,-0.431) (1.709,-0.412) (1.708,-0.394) (1.705,-0.377) (1.700,-0.361) (1.700,-0.345) (1.696,-0.330) (1.692,-0.315) (1.690,-0.301) (1.687,-0.287) (1.686,-0.274) (1.686,-0.261) (1.683,-0.248) (1.683,-0.236) (1.667,-0.224) (1.667,-0.213) (1.663,-0.201) (1.658,-0.190) (1.658,-0.180) (1.658,-0.169) (1.653,-0.159) (1.648,-0.149) (1.645,-0.139) (1.638,-0.130) (1.635,-0.120) (1.625,-0.111) (1.602,-0.102) (1.580,-0.093) (1.564,-0.085) (1.558,-0.076) (1.546,-0.068) (1.534,-0.060) (1.501,-0.052) (1.494,-0.044) (1.477,-0.037) (1.457,-0.029) (1.401,-0.022) (1.392,-0.014) (1.263,-0.007) (1.251,0.000)
\psline[linecolor=red] (1.820,-1.851) (1.811,-1.550) (1.806,-1.374) (1.805,-1.249) (1.803,-1.152) (1.803,-1.073) (1.799,-1.006) (1.799,-0.948) (1.799,-0.897) (1.799,-0.851) (1.798,-0.810) (1.798,-0.772) (1.797,-0.737) (1.797,-0.705) (1.797,-0.675) (1.797,-0.647) (1.796,-0.621) (1.796,-0.596) (1.796,-0.573) (1.795,-0.550) (1.791,-0.529) (1.789,-0.509) (1.787,-0.490) (1.787,-0.471) (1.787,-0.453) (1.787,-0.436) (1.787,-0.420) (1.783,-0.404) (1.783,-0.389) (1.783,-0.374) (1.783,-0.360) (1.783,-0.346) (1.783,-0.333) (1.783,-0.320) (1.781,-0.307) (1.779,-0.295) (1.779,-0.283) (1.779,-0.271) (1.777,-0.260) (1.776,-0.249) (1.776,-0.238) (1.771,-0.228) (1.767,-0.218) (1.766,-0.208) (1.756,-0.198) (1.756,-0.189) (1.752,-0.179) (1.739,-0.170) (1.739,-0.161) (1.736,-0.152) (1.735,-0.144) (1.722,-0.135) (1.719,-0.127) (1.717,-0.119) (1.717,-0.111) (1.703,-0.103) (1.702,-0.095) (1.692,-0.088) (1.681,-0.080) (1.653,-0.073) (1.623,-0.066) (1.602,-0.059) (1.598,-0.052) (1.574,-0.045) (1.536,-0.038) (1.531,-0.032) (1.484,-0.025) (1.472,-0.019) (1.442,-0.012) (1.418,-0.006) (1.418,0.000)
\psline[linecolor=red] (1.899,-1.940) (1.899,-1.638) (1.896,-1.462) (1.893,-1.337) (1.892,-1.241) (1.892,-1.161) (1.890,-1.094) (1.890,-1.036) (1.889,-0.985) (1.889,-0.940) (1.889,-0.898) (1.889,-0.860) (1.887,-0.826) (1.886,-0.793) (1.886,-0.763) (1.886,-0.735) (1.886,-0.709) (1.885,-0.684) (1.884,-0.661) (1.884,-0.638) (1.884,-0.617) (1.881,-0.597) (1.881,-0.578) (1.879,-0.559) (1.878,-0.542) (1.878,-0.525) (1.878,-0.508) (1.878,-0.492) (1.877,-0.477) (1.877,-0.462) (1.875,-0.448) (1.875,-0.434) (1.875,-0.421) (1.875,-0.408) (1.875,-0.395) (1.875,-0.383) (1.875,-0.371) (1.875,-0.360) (1.875,-0.348) (1.875,-0.337) (1.875,-0.327) (1.874,-0.316) (1.872,-0.306) (1.872,-0.296) (1.871,-0.286) (1.871,-0.277) (1.871,-0.267) (1.870,-0.258) (1.867,-0.249) (1.866,-0.241) (1.865,-0.232) (1.865,-0.224) (1.864,-0.215) (1.864,-0.207) (1.860,-0.199) (1.854,-0.191) (1.851,-0.184) (1.848,-0.176) (1.847,-0.169) (1.846,-0.161) (1.838,-0.154) (1.838,-0.147) (1.838,-0.140) (1.824,-0.133) (1.821,-0.127) (1.789,-0.120) (1.789,-0.113) (1.762,-0.107) (1.761,-0.101) (1.740,-0.094) (1.740,-0.088) (1.702,-0.082) (1.698,-0.076) (1.693,-0.070) (1.674,-0.064) (1.651,-0.059) (1.643,-0.053) (1.633,-0.047) (1.625,-0.042) (1.618,-0.036) (1.613,-0.031) (1.603,-0.026) (1.598,-0.020) (1.592,-0.015) (1.492,-0.010) (1.435,-0.005) (1.435,0.000)
\psline[linecolor=red] (1.959,-1.964) (1.959,-1.663) (1.957,-1.487) (1.954,-1.362) (1.954,-1.265) (1.954,-1.186) (1.954,-1.119) (1.954,-1.061) (1.954,-1.010) (1.954,-0.964) (1.954,-0.922) (1.954,-0.885) (1.954,-0.850) (1.954,-0.818) (1.954,-0.788) (1.952,-0.760) (1.952,-0.733) (1.952,-0.709) (1.952,-0.685) (1.952,-0.663) (1.952,-0.642) (1.952,-0.621) (1.952,-0.602) (1.952,-0.584) (1.952,-0.566) (1.952,-0.549) (1.952,-0.532) (1.952,-0.517) (1.952,-0.501) (1.952,-0.487) (1.952,-0.472) (1.949,-0.459) (1.949,-0.445) (1.949,-0.432) (1.949,-0.420) (1.949,-0.407) (1.949,-0.396) (1.949,-0.384) (1.949,-0.373) (1.949,-0.362) (1.949,-0.351) (1.949,-0.341) (1.949,-0.330) (1.947,-0.320) (1.947,-0.311) (1.947,-0.301) (1.947,-0.292) (1.947,-0.283) (1.947,-0.274) (1.947,-0.265) (1.947,-0.256) (1.947,-0.248) (1.947,-0.240) (1.947,-0.231) (1.947,-0.223) (1.944,-0.216) (1.944,-0.208) (1.944,-0.200) (1.944,-0.193) (1.944,-0.186) (1.944,-0.178) (1.942,-0.171) (1.942,-0.164) (1.942,-0.158) (1.942,-0.151) (1.942,-0.144) (1.942,-0.138) (1.942,-0.131) (1.940,-0.125) (1.940,-0.119) (1.940,-0.113) (1.940,-0.106) (1.940,-0.100) (1.940,-0.095) (1.940,-0.089) (1.937,-0.083) (1.937,-0.077) (1.932,-0.072) (1.932,-0.066) (1.927,-0.061) (1.924,-0.055) (1.919,-0.050) (1.911,-0.045) (1.911,-0.040) (1.908,-0.034) (1.892,-0.029) (1.889,-0.024) (1.881,-0.019) (1.848,-0.014) (1.848,-0.010) (1.823,-0.005) (1.816,0.000)
\psline[linecolor=red] (1.959,-1.964) (1.959,-0.010) (1.957,-0.005) (1.957,0.000)
\psline[linecolor=red] (1.959,-1.964) (1.959,0.000)
\end{pspicture}
\begin{pspicture}(-4,-4.4)(5,1.5)
\psset{xunit=2,yunit=2}
\psline{->}(-0.5,0)(2.5,0) \psline{->}(0,-2)(0,0.5) \scriptsize \rput(2.85,0){$log(Hc)$} \rput(0.65,0.5){$\log(Cum.\ distr.)$} \psline(0.5,-0.05)(0.5,0.05) \rput(0.5,-0.15){$0.5$} \psline(1,-0.05)(1,0.05) \rput(1,-0.15){$1$} \psline(1.5,-0.05)(1.5,0.05) \rput(1.5,-0.15){$1.5$} \psline(2,-0.05)(2,0.05) \rput(2,-0.15){$2$} \psline(-0.05,-1.5)(0.05,-1.5) \rput(-0.25,-1.5){$-1.5$} \psline(-0.05,-1)(0.05,-1) \rput(-0.2,-1){$-1$} \psline(-0.05,-0.5)(0.05,-0.5) \rput(-0.25,-0.5){$-0.5$}
\psline[linecolor=blue] (0.000,-0.301) (0.000,0.000)
\psline[linecolor=blue] (1.875,-1.964) (1.866,-1.663) (1.863,-1.487) (1.863,-1.362) (1.863,-1.265) (1.863,-1.186) (1.860,-1.119) (1.857,-1.061) (1.857,-1.010) (1.854,-0.964) (1.854,-0.922) (1.854,-0.885) (1.851,-0.850) (1.851,-0.818) (1.851,-0.788) (1.851,-0.760) (1.851,-0.733) (1.851,-0.709) (1.848,-0.685) (1.848,-0.663) (1.848,-0.642) (1.845,-0.621) (1.845,-0.602) (1.845,-0.584) (1.845,-0.566) (1.845,-0.549) (1.842,-0.532) (1.842,-0.517) (1.842,-0.501) (1.842,-0.487) (1.842,-0.472) (1.842,-0.459) (1.842,-0.445) (1.839,-0.432) (1.839,-0.420) (1.839,-0.407) (1.839,-0.396) (1.839,-0.384) (1.839,-0.373) (1.839,-0.362) (1.839,-0.351) (1.836,-0.341) (1.836,-0.330) (1.836,-0.320) (1.836,-0.311) (1.836,-0.301) (1.836,-0.292) (1.833,-0.283) (1.833,-0.274) (1.833,-0.265) (1.833,-0.256) (1.833,-0.248) (1.833,-0.240) (1.829,-0.231) (1.829,-0.223) (1.829,-0.216) (1.829,-0.208) (1.829,-0.200) (1.829,-0.193) (1.826,-0.186) (1.826,-0.178) (1.826,-0.171) (1.826,-0.164) (1.826,-0.158) (1.826,-0.151) (1.826,-0.144) (1.823,-0.138) (1.823,-0.131) (1.823,-0.125) (1.823,-0.119) (1.823,-0.113) (1.820,-0.106) (1.820,-0.100) (1.820,-0.095) (1.820,-0.089) (1.820,-0.083) (1.816,-0.077) (1.813,-0.072) (1.813,-0.066) (1.813,-0.061) (1.813,-0.055) (1.813,-0.050) (1.813,-0.045) (1.813,-0.040) (1.810,-0.034) (1.810,-0.029) (1.810,-0.024) (1.810,-0.019) (1.810,-0.014) (1.806,-0.010) (1.803,-0.005) (1.796,0.000)
\psline[linecolor=blue] (1.959,-1.964) (1.959,-0.010) (1.957,-0.005) (1.957,0.000)
\psline[linecolor=blue] (1.959,-1.964) (1.959,0.000)
\end{pspicture}

\noindent Fig. 28. Cumulative frequency distribution of the harmonic closeness centrality in terms of the centrality values, both in logarithmic representation. The figure on the left displays the curves for real data, and the figure on the right represents the curves for randomized data.

\vskip 0.3 cm

\subsection{Overview}

Our analysis of the centrality measures showed that there is a strong dependence of centralities on the threshold chosen to define a particular asset graph, and that, although the probability frequency distributions and the cummulative frequency distributions of the four centralities we have researched do not resemble those of scale-free networks, they are remarkably different from their counterparts based on randomized data, except for high values of the threshold. The values of the threshold between $T=0.5$ and $T=0.8$, close to the noise threshold, are believed to be the ones that generate asset graphs with more information also in terms of centralities. Central European indices, like the UK, France, the Netherlands, Germany, Austria, and Italy, carry high centrality values, in general, and Singapore and the Czech Republic presente high betweenness centrality, Singapore because it connects most Pacifi Asian indices with some European ones, and the Czech Republic because it somehow connects Eastern and Central European indices. The final figure of this article (Figure 29) shows the five centrality measures discussed in this section as functions of indices. 

\begin{pspicture}(-10,-6)(1,1)
\psset{xunit=1,yunit=1,Alpha=30,Beta=20} \scriptsize
\pstThreeDLine[linecolor=black](0,0,0)(11,0,0) \pstThreeDLine[linecolor=black](0,1.7,0)(11,1.7,0) \pstThreeDLine[linecolor=black](0,2.7,0)(11,2.7,0) \pstThreeDLine[linecolor=black](0,6.1,0)(11,6.1,0) \pstThreeDLine[linecolor=black](0,8.5,0)(11,8.5,0) \pstThreeDLine[linecolor=black](0,10.3,0)(11,10.3,0) \pstThreeDLine[linecolor=black](0,12.3,0)(11,12.3,0) \pstThreeDLine[linecolor=black](0,12.7,0)(11,12.7,0) \pstThreeDLine[linecolor=black](0,13.2,0)(11,13.2,0)
\pstThreeDLine[linecolor=black]	(10,1,0.597) (10,1.2,0.597) (10,1.4,0.661) (10,1.6,0.726)
(10,1.8,0.677) (10,2,0.677) (10,2.2,0.742) (10,2.4,0.742) (10,2.6,0.823) (10,2.8,0.855) (10,3,0.903) (10,3.2,0.855) (10,3.4,0.855) (10,3.6,0.839) (10,3.8,0.935) (10,4,0.887) (10,4.2,0.903) (10,4.4,0.871) (10,4.6,0.919) (10,4.8,0.823) (10,5,0.919) (10,5.2,0.871) (10,5.4,0.855) (10,5.6,0.661) (10,5.8,0.855) (10,6,0.871) (10,6.2,0.919) (10,6.4,0.952) (10,6.6,0.855) (10,6.8,0.226) (10,7,0.919) (10,7.2,0.468) (10,7.4,0.016) (10,7.6,0.016) (10,7.8,0.032) (10,8,0.871) (10,8.2,0.919) (10,8.4,0.565) (10,8.6,0.774) (10,8.8,0.177) (10,9,0.823) (10,9.2,0.903) (10,9.4,0.855) (10,9.6,0.500) (10,9.8,0.871) (10,10,0.806) (10,10.2,0.823) (10,10.4,0.065) (10,10.6,0.516) (10,10.8,0.065) (10,11,0.032) (10,11.2,0.323) (10,11.4,0.435) (10,11.6,0.113) (10,11.8,0.952) (10,12,0.855) (10,12.2,0.968) (10,12.4,0.194) (10,12.6,0.823) (10,12.8,0.903) (10,13,0.903) (10,13.2,0.032)
\pstThreeDLine[linecolor=black]	(8,1,0.567) (8,1.2,0.545) (8,1.4,0.583) (8,1.6,0.644) (8,1.8,0.612) (8,2,0.595) (8,2.2,0.612) (8,2.4,0.542) (8,2.6,0.685) (8,2.8,0.970) (8,3,0.865) (8,3.2,0.998) (8,3.4,0.968) (8,3.6,0.906) (8,3.8,0.989) (8,4,0.969) (8,4.2,0.980) (8,4.4,1.000) (8,4.6,0.865) (8,4.8,0.903) (8,5,0.955) (8,5.2,0.943) (8,5.4,0.901) (8,5.6,0.399) (8,5.8,0.928) (8,6,0.881) (8,6.2,0.809) (8,6.4,0.923) (8,6.6,0.741) (8,6.8,0.108) (8,7,0.662) (8,7.2,0.260) (8,7.4,0.009) (8,7.6,0.008) (8,7.8,0.015) (8,8,0.848) (8,8.2,0.704) (8,8.4,0.309) (8,8.6,0.466) (8,8.8,0.088) (8,9,0.499) (8,9.2,0.618) (8,9.4,0.770) (8,9.6,0.275) (8,9.8,0.807) (8,10,0.651) (8,10.2,0.671) (8,10.4,0.037) (8,10.6,0.254) (8,10.8,0.034) (8,11,0.017) (8,11.2,0.183) (8,11.4,0.230) (8,11.6,0.064) (8,11.8,0.714) (8,12,0.605) (8,12.2,0.767) (8,12.4,0.113) (8,12.6,0.578) (8,12.8,0.659) (8,13,0.636) (8,13.2,0.015)
\pstThreeDLine[linecolor=black]	(6,1,0.669) (6,1.2,0.669) (6,1.4,0.743) (6,1.6,0.811) (6,1.8,0.757) (6,2,0.757) (6,2.2,0.831) (6,2.4,0.824) (6,2.6,0.912) (6,2.8,0.939) (6,3,0.966) (6,3.2,0.939) (6,3.4,0.939) (6,3.6,0.926) (6,3.8,0.986) (6,4,0.959) (6,4.2,0.966) (6,4.4,0.946) (6,4.6,0.973) (6,4.8,0.912) (6,5,0.980) (6,5.2,0.946) (6,5.4,0.939) (6,5.6,0.757) (6,5.8,0.939) (6,6,0.946) (6,6.2,0.980) (6,6.4,0.993) (6,6.6,0.932) (6,6.8,0.162) (6,7,0.980) (6,7.2,0.466) (6,7.4,0.000) (6,7.6,0.000) (6,7.8,0.014) (6,8,0.946) (6,8.2,0.966) (6,8.4,0.581) (6,8.6,0.838) (6,8.8,0.189) (6,9,0.865) (6,9.2,0.939) (6,9.4,0.932) (6,9.6,0.507) (6,9.8,0.946) (6,10,0.892) (6,10.2,0.905) (6,10.4,0.034) (6,10.6,0.588) (6,10.8,0.014) (6,11,0.007) (6,11.2,0.250) (6,11.4,0.399) (6,11.6,0.068) (6,11.8,0.986) (6,12,0.878) (6,12.2,0.973) (6,12.4,0.216) (6,12.6,0.865) (6,12.8,0.932) (6,13,0.953) (6,13.2,0.034)
\pstThreeDLine[linecolor=black]	(4,1,0.000) (4,1.2,0.000) (4,1.4,0.002) (4,1.6,0.006) (4,1.8,0.002) (4,2,0.003) (4,2.2,0.007) (4,2.4,0.009) (4,2.6,0.017) (4,2.8,0.025) (4,3,0.073) (4,3.2,0.025) (4,3.4,0.025) (4,3.6,0.020) (4,3.8,0.088) (4,4,0.053) (4,4.2,0.056) (4,4.4,0.032) (4,4.6,0.082) (4,4.8,0.017) (4,5,0.065) (4,5.2,0.032) (4,5.4,0.025) (4,5.6,0.003) (4,5.8,0.025) (4,6,0.032) (4,6.2,0.079) (4,6.4,0.127) (4,6.6,0.031) (4,6.8,1.000) (4,7,0.065) (4,7.2,0.374) (4,7.4,0.000) (4,7.6,0.000) (4,7.8,0.000) (4,8,0.032) (4,8.2,0.106) (4,8.4,0.082) (4,8.6,0.048) (4,8.8,0.000) (4,9,0.172) (4,9.2,0.226) (4,9.4,0.030) (4,9.6,0.065) (4,9.8,0.031) (4,10,0.013) (4,10.2,0.037) (4,10.4,0.000) (4,10.6,0.008) (4,10.8,0.011) (4,11,0.000) (4,11.2,0.705) (4,11.4,0.379) (4,11.6,0.031) (4,11.8,0.134) (4,12,0.222) (4,12.2,0.344) (4,12.4,0.000) (4,12.6,0.104) (4,12.8,0.160) (4,13,0.246) (4,13.2,0.000)
\pstThreeDLine[linecolor=black]	(2,1,0.790) (2,1.2,0.790) (2,1.4,0.823) (2,1.6,0.856) (2,1.8,0.831) (2,2,0.831) (2,2.2,0.864) (2,2.4,0.864) (2,2.6,0.904) (2,2.8,0.919) (2,3,0.947) (2,3.2,0.919) (2,3.4,0.919) (2,3.6,0.912) (2,3.8,0.962) (2,4,0.937) (2,4.2,0.944) (2,4.4,0.927) (2,4.6,0.955) (2,4.8,0.904) (2,5,0.952) (2,5.2,0.927) (2,5.4,0.919) (2,5.6,0.826) (2,5.8,0.919) (2,6,0.927) (2,6.2,0.955) (2,6.4,0.970) (2,6.6,0.919) (2,6.8,0.636) (2,7,0.952) (2,7.2,0.745) (2,7.4,0.396) (2,7.6,0.396) (2,7.8,0.449) (2,8,0.927) (2,8.2,0.955) (2,8.4,0.793) (2,8.6,0.884) (2,8.8,0.601) (2,9,0.914) (2,9.2,0.949) (2,9.4,0.919) (2,9.6,0.765) (2,9.8,0.927) (2,10,0.894) (2,10.2,0.907) (2,10.4,0.520) (2,10.6,0.763) (2,10.8,0.462) (2,11,0.419) (2,11.2,0.682) (2,11.4,0.727) (2,11.6,0.568) (2,11.8,0.967) (2,12,0.932) (2,12.2,0.980) (2,12.4,0.606) (2,12.6,0.912) (2,12.8,0.949) (2,13,0.955) (2,13.2,0.515)
\pstThreeDPut(12,0.85,0){NoAm} \pstThreeDPut(12,2.2,0){SoAm} \pstThreeDPut(12,4.4,0){CeEu} \pstThreeDPut(12,7.3,0){EaEu} \pstThreeDPut(12,9.4,0){CeAs} \pstThreeDPut(12,11.3,0){PaAs} \pstThreeDPut(12,12.5,0){Oc} \pstThreeDPut(12,12.95,0){Af} \pstThreeDPut(8,-1,0.6){\large $T=0.7$} \normalsize \pstThreeDPut(10,14,0){$Nd$} \pstThreeDPut(8,14,0){$Ns$} \pstThreeDPut(6,14,0){$Ec$} \pstThreeDPut(4,14,0){$Bc$} \pstThreeDPut(2,14,0){$Hc$}
\end{pspicture}

\begin{pspicture}(-10,-6)(1,1)
\psset{xunit=1,yunit=1,Alpha=30,Beta=20} \scriptsize
\pstThreeDLine[linecolor=black](0,0,0)(11,0,0) \pstThreeDLine[linecolor=black](0,1.7,0)(11,1.7,0) \pstThreeDLine[linecolor=black](0,2.7,0)(11,2.7,0) \pstThreeDLine[linecolor=black](0,5.9,0)(11,5.9,0) \pstThreeDLine[linecolor=black](0,7.9,0)(11,7.9,0) \pstThreeDLine[linecolor=black](0,9.1,0)(11,9.1,0) \pstThreeDLine[linecolor=black](0,11.1,0)(11,11.1,0) \pstThreeDLine[linecolor=black](0,11.5,0)(11,11.5,0) \pstThreeDLine[linecolor=black](0,11.8,0)(11,11.8,0)
\pstThreeDLine[linecolor=black]	(10,1,0.564) (10,1.2,0.487) (10,1.4,0.538) (10,1.6,0.615) (10,1.8,0.538) (10,2,0.538) (10,2.2,0.641) (10,2.4,0.359) (10,2.6,0.692) (10,2.8,0.949) (10,3,0.795) (10,3.2,0.974) (10,3.4,0.974) (10,3.6,0.949) (10,3.8,0.974) (10,4,0.974) (10,4.2,0.949) (10,4.4,0.974) (10,4.6,0.769) (10,4.8,0.923) (10,5,1.000) (10,5.2,0.923) (10,5.4,0.949) (10,5.6,0.923) (10,5.8,0.897) (10,6,0.692) (10,6.2,0.897) (10,6.4,0.667) (10,6.6,0.256) (10,6.8,0.821) (10,7,0.590) (10,7.2,0.026) (10,7.4,0.026) (10,7.6,0.077) (10,7.8,0.667) (10,8,0.718) (10,8.2,0.590) (10,8.4,0.615) (10,8.6,0.026) (10,8.8,0.026) (10,9,0.564) (10,9.2,0.205) (10,9.4,0.359) (10,9.6,0.026) (10,9.8,0.179) (10,10,0.231) (10,10.2,0.205) (10,10.4,0.231) (10,10.6,0.821) (10,10.8,0.231) (10,11,0.077) (10,11.2,0.410) (10,11.4,0.077) (10,11.6,0.744) (10,11.8,0.821)
\pstThreeDLine[linecolor=black]	(8,1,0.482) (8,1.2,0.416) (8,1.4,0.441) (8,1.6,0.504) (8,1.8,0.446) (8,2,0.437) (8,2.2,0.477) (8,2.4,0.248) (8,2.6,0.514) (8,2.8,0.953) (8,3,0.711) (8,3.2,1.000) (8,3.4,0.971) (8,3.6,0.887) (8,3.8,0.914) (8,4,0.953) (8,4.2,0.928) (8,4.4,0.987) (8,4.6,0.656) (8,4.8,0.886) (8,5,0.893) (8,5.2,0.887) (8,5.4,0.864) (8,5.6,0.897) (8,5.8,0.797) (8,6,0.589) (8,6.2,0.771) (8,6.4,0.551) (8,6.6,0.172) (8,6.8,0.720) (8,7,0.408) (8,7.2,0.021) (8,7.4,0.021) (8,7.6,0.051) (8,7.8,0.567) (8,8,0.611) (8,8.2,0.447) (8,8.4,0.461) (8,8.6,0.019) (8,8.8,0.019) (8,9,0.389) (8,9.2,0.174) (8,9.4,0.299) (8,9.6,0.018) (8,9.8,0.150) (8,10,0.198) (8,10.2,0.147) (8,10.4,0.171) (8,10.6,0.628) (8,10.8,0.168) (8,11,0.051) (8,11.2,0.323) (8,11.4,0.057) (8,11.6,0.623) (8,11.8,0.722)
\pstThreeDLine[linecolor=black]	(6,1,0.631) (6,1.2,0.544) (6,1.4,0.600) (6,1.6,0.682) (6,1.8,0.600) (6,2,0.600) (6,2.2,0.713) (6,2.4,0.451) (6,2.6,0.769) (6,2.8,0.985) (6,3,0.872) (6,3.2,0.995) (6,3.4,0.995) (6,3.6,0.985) (6,3.8,0.985) (6,4,0.995) (6,4.2,0.985) (6,4.4,0.995) (6,4.6,0.836) (6,4.8,0.964) (6,5,1.000) (6,5.2,0.954) (6,5.4,0.969) (6,5.6,0.969) (6,5.8,0.949) (6,6,0.790) (6,6.2,0.882) (6,6.4,0.769) (6,6.6,0.308) (6,6.8,0.887) (6,7,0.662) (6,7.2,0.000) (6,7.4,0.000) (6,7.6,0.062) (6,7.8,0.769) (6,8,0.805) (6,8.2,0.692) (6,8.4,0.723) (6,8.6,0.000) (6,8.8,0.000) (6,9,0.544) (6,9.2,0.041) (6,9.4,0.174) (6,9.6,0.005) (6,9.8,0.041) (6,10,0.062) (6,10.2,0.113) (6,10.4,0.046) (6,10.6,0.713) (6,10.8,0.062) (6,11,0.010) (6,11.2,0.200) (6,11.4,0.010) (6,11.6,0.805) (6,11.8,0.867)
\pstThreeDLine[linecolor=black]	(4,1,0.002) (4,1.2,0.001) (4,1.4,0.001) (4,1.6,0.006) (4,1.8,0.001) (4,2,0.001) (4,2.2,0.009) (4,2.4,0.000) (4,2.6,0.013) (4,2.8,0.084) (4,3,0.022) (4,3.2,0.105) (4,3.4,0.105) (4,3.6,0.084) (4,3.8,0.244) (4,4,0.105) (4,4.2,0.084) (4,4.4,0.105) (4,4.6,0.088) (4,4.8,0.078) (4,5,0.319) (4,5.2,0.085) (4,5.4,0.178) (4,5.6,0.071) (4,5.8,0.061) (4,6,0.008) (4,6.2,0.399) (4,6.4,0.002) (4,6.6,0.000) (4,6.8,0.044) (4,7,0.014) (4,7.2,0.000) (4,7.4,0.000) (4,7.6,0.000) (4,7.8,0.002) (4,8,0.013) (4,8.2,0.001) (4,8.4,0.000) (4,8.6,0.000) (4,8.8,0.000) (4,9,0.219) (4,9.2,0.035) (4,9.4,0.377) (4,9.6,0.000) (4,9.8,0.000) (4,10,0.007) (4,10.2,0.027) (4,10.4,0.044) (4,10.6,1.000) (4,10.8,0.007) (4,11,0.002) (4,11.2,0.605) (4,11.4,0.002) (4,11.6,0.128) (4,11.8,0.171)
\pstThreeDLine[linecolor=black]	(2,1,0.778) (2,1.2,0.739) (2,1.4,0.767) (2,1.6,0.801) (2,1.8,0.767) (2,2,0.767) (2,2.2,0.816) (2,2.4,0.688) (2,2.6,0.838) (2,2.8,0.970) (2,3,0.883) (2,3.2,0.981) (2,3.4,0.981) (2,3.6,0.970) (2,3.8,0.985) (2,4,0.981) (2,4.2,0.970) (2,4.4,0.981) (2,4.6,0.898) (2,4.8,0.959) (2,5,1.000) (2,5.2,0.959) (2,5.4,0.970) (2,5.6,0.959) (2,5.8,0.947) (2,6,0.857) (2,6.2,0.959) (2,6.4,0.827) (2,6.6,0.647) (2,6.8,0.914) (2,7,0.812) (2,7.2,0.023) (2,7.4,0.023) (2,7.6,0.568) (2,7.8,0.827) (2,8,0.868) (2,8.2,0.793) (2,8.4,0.805) (2,8.6,0.023) (2,8.8,0.023) (2,9,0.812) (2,9.2,0.598) (2,9.4,0.718) (2,9.6,0.438) (2,9.8,0.586) (2,10,0.609) (2,10.2,0.650) (2,10.4,0.609) (2,10.6,0.925) (2,10.8,0.609) (2,11,0.474) (2,11.2,0.744) (2,11.4,0.470) (2,11.6,0.891) (2,11.8,0.925)
\pstThreeDPut(12,0.85,0){NoAm} \pstThreeDPut(12,2.2,0){SoAm} \pstThreeDPut(12,4.3,0){CeEu} \pstThreeDPut(12,6.9,0){EaEu} \pstThreeDPut(12,8.5,0){CeAs} \pstThreeDPut(12,10.1,0){PaAs} \pstThreeDPut(12,11.3,0){Oc} \pstThreeDPut(12,11.65,0){Af} \pstThreeDPut(8,-1,0.6){\large $T=0.6$} \normalsize \pstThreeDPut(10,12.6,0){$Nd$} \pstThreeDPut(8,12.6,0){$Ns$} \pstThreeDPut(6,12.6,0){$Ec$} \pstThreeDPut(4,12.6,0){$Bc$} \pstThreeDPut(2,12.6,0){$Hc$}
\end{pspicture}

\begin{pspicture}(-10,-6)(1,0.3)
\psset{xunit=1,yunit=1,Alpha=30,Beta=20} \scriptsize
\pstThreeDLine[linecolor=black](0,0,0)(11,0,0) \pstThreeDLine[linecolor=black](0,1.7,0)(11,1.7,0) \pstThreeDLine[linecolor=black](0,2.5,0)(11,2.5,0) \pstThreeDLine[linecolor=black](0,5.7,0)(11,5.7,0) \pstThreeDLine[linecolor=black](0,7.3,0)(11,7.3,0) \pstThreeDLine[linecolor=black](0,8.1,0)(11,8.1,0) \pstThreeDLine[linecolor=black](0,9.7,0)(11,9.7,0) \pstThreeDLine[linecolor=black](0,9.9,0)(11,9.9,0) \pstThreeDLine[linecolor=black](0,10.4,0)(11,10.4,0)
\pstThreeDLine[linecolor=black]	(10,1,0.375) (10,1.2,0.281) (10,1.4,0.281) (10,1.6,0.406) (10,1.8,0.219) (10,2,0.281) (10,2.2,0.125) (10,2.4,0.219) (10,2.6,0.906) (10,2.8,0.656) (10,3,1.000) (10,3.2,0.875) (10,3.4,0.656) (10,3.6,0.750) (10,3.8,0.813) (10,4,0.781) (10,4.2,0.969) (10,4.4,0.656) (10,4.6,0.719) (10,4.8,0.719) (10,5,0.750) (10,5.2,0.781) (10,5.4,0.781) (10,5.6,0.719) (10,5.8,0.656) (10,6,0.781) (10,6.2,0.563) (10,6.4,0.719) (10,6.6,0.031) (10,6.8,0.031) (10,7,0.031) (10,7.2,0.594) (10,7.4,0.688) (10,7.6,0.063) (10,7.8,0.125) (10,8,0.063) (10,8.2,0.156) (10,8.4,0.281) (10,8.6,0.156) (10,8.8,0.156) (10,9,0.063) (10,9.2,0.063) (10,9.4,0.313) (10,9.6,0.063) (10,9.8,0.188) (10,10,0.031) (10,10.2,0.563) (10,10.4,0.656)
\pstThreeDLine[linecolor=black]	(8,1,0.337) (8,1.2,0.258) (8,1.4,0.243) (8,1.6,0.340) (8,1.8,0.202) (8,2,0.234) (8,2.2,0.096) (8,2.4,0.164) (8,2.6,0.898) (8,2.8,0.600) (8,3,1.000) (8,3.2,0.880) (8,3.4,0.665) (8,3.6,0.740) (8,3.8,0.815) (8,4,0.784) (8,4.2,0.961) (8,4.4,0.558) (8,4.6,0.719) (8,4.8,0.680) (8,5,0.742) (8,5.2,0.728) (8,5.4,0.773) (8,5.6,0.657) (8,5.8,0.535) (8,6,0.665) (8,6.2,0.454) (8,6.4,0.626) (8,6.6,0.023) (8,6.8,0.023) (8,7,0.023) (8,7.2,0.489) (8,7.4,0.561) (8,7.6,0.059) (8,7.8,0.095) (8,8,0.049) (8,8.2,0.135) (8,8.4,0.236) (8,8.6,0.128) (8,8.8,0.141) (8,9,0.048) (8,9.2,0.048) (8,9.4,0.261) (8,9.6,0.051) (8,9.8,0.162) (8,10,0.025) (8,10.2,0.471) (8,10.4,0.585)
\pstThreeDLine[linecolor=black]	(6,1,0.353) (6,1.2,0.228) (6,1.4,0.201) (6,1.6,0.357) (6,1.8,0.112) (6,2,0.228) (6,2.2,0.112) (6,2.4,0.268) (6,2.6,0.982) (6,2.8,0.848) (6,3,1.000) (6,3.2,0.973) (6,3.4,0.848) (6,3.6,0.920) (6,3.8,0.955) (6,4,0.946) (6,4.2,0.996) (6,4.4,0.848) (6,4.6,0.915) (6,4.8,0.915) (6,5,0.924) (6,5.2,0.937) (6,5.4,0.946) (6,5.6,0.915) (6,5.8,0.817) (6,6,0.920) (6,6.2,0.741) (6,6.4,0.915) (6,6.6,0.040) (6,6.8,0.000) (6,7,0.000) (6,7.2,0.772) (6,7.4,0.884) (6,7.6,0.076) (6,7.8,0.174) (6,8,0.000) (6,8.2,0.004) (6,8.4,0.004) (6,8.6,0.004) (6,8.8,0.004) (6,9,0.000) (6,9.2,0.000) (6,9.4,0.040) (6,9.6,0.000) (6,9.8,0.004) (6,10,0.000) (6,10.2,0.737) (6,10.4,0.848)
\pstThreeDLine[linecolor=black]	(4,1,0.008) (4,1.2,0.003) (4,1.4,0.004) (4,1.6,0.015) (4,1.8,0.002) (4,2,0.004) (4,2.2,0.000) (4,2.4,0.001) (4,2.6,0.106) (4,2.8,0.001) (4,3,0.252) (4,3.2,0.079) (4,3.4,0.001) (4,3.6,0.058) (4,3.8,0.042) (4,4,0.023) (4,4.2,0.172) (4,4.4,0.001) (4,4.6,0.002) (4,4.8,0.002) (4,5,0.014) (4,5.2,0.034) (4,5.4,0.023) (4,5.6,0.002) (4,5.8,0.049) (4,6,1.000) (4,6.2,0.000) (4,6.4,0.002) (4,6.6,0.000) (4,6.8,0.000) (4,7,0.000) (4,7.2,0.000) (4,7.4,0.001) (4,7.6,0.000) (4,7.8,0.000) (4,8,0.000) (4,8.2,0.000) (4,8.4,0.032) (4,8.6,0.000) (4,8.8,0.000) (4,9,0.000) (4,9.2,0.000) (4,9.4,0.890) (4,9.6,0.000) (4,9.8,0.108) (4,10,0.000) (4,10.2,0.000) (4,10.4,0.001)
\pstThreeDLine[linecolor=black]	(2,1,0.693) (2,1.2,0.652) (2,1.4,0.652) (2,1.6,0.706) (2,1.8,0.624) (2,2,0.652) (2,2.2,0.584) (2,2.4,0.624) (2,2.6,0.959) (2,2.8,0.850) (2,3,1.000) (2,3.2,0.946) (2,3.4,0.850) (2,3.6,0.891) (2,3.8,0.918) (2,4,0.905) (2,4.2,0.986) (2,4.4,0.850) (2,4.6,0.878) (2,4.8,0.878) (2,5,0.891) (2,5.2,0.905) (2,5.4,0.905) (2,5.6,0.878) (2,5.8,0.850) (2,6,0.948) (2,6.2,0.810) (2,6.4,0.878) (2,6.6,0.533) (2,6.8,0.027) (2,7,0.027) (2,7.2,0.823) (2,7.4,0.864) (2,7.6,0.520) (2,7.8,0.584) (2,8,0.458) (2,8.2,0.503) (2,8.4,0.558) (2,8.6,0.503) (2,8.8,0.503) (2,9,0.458) (2,9.2,0.458) (2,9.4,0.703) (2,9.6,0.458) (2,9.8,0.517) (2,10,0.358) (2,10.2,0.810) (2,10.4,0.850)
\pstThreeDPut(12,0.85,0){NoAm} \pstThreeDPut(12,2.1,0){SoAm} \pstThreeDPut(12,4.1,0){CeEu} \pstThreeDPut(12,6.5,0){EaEu} \pstThreeDPut(12,7.7,0){CeAs} \pstThreeDPut(12,8.9,0){PaAs} \pstThreeDPut(12,9.8,0){Oc} \pstThreeDPut(12,10.15,0){Af} \pstThreeDPut(8,-1,0.6){\large $T=0.5$} \normalsize \pstThreeDPut(10,11.3,0){$Nd$} \pstThreeDPut(8,11.3,0){$Ns$} \pstThreeDPut(6,11.3,0){$Ec$} \pstThreeDPut(4,11.3,0){$Bc$} \pstThreeDPut(2,11.3,0){$Hc$}
\end{pspicture}

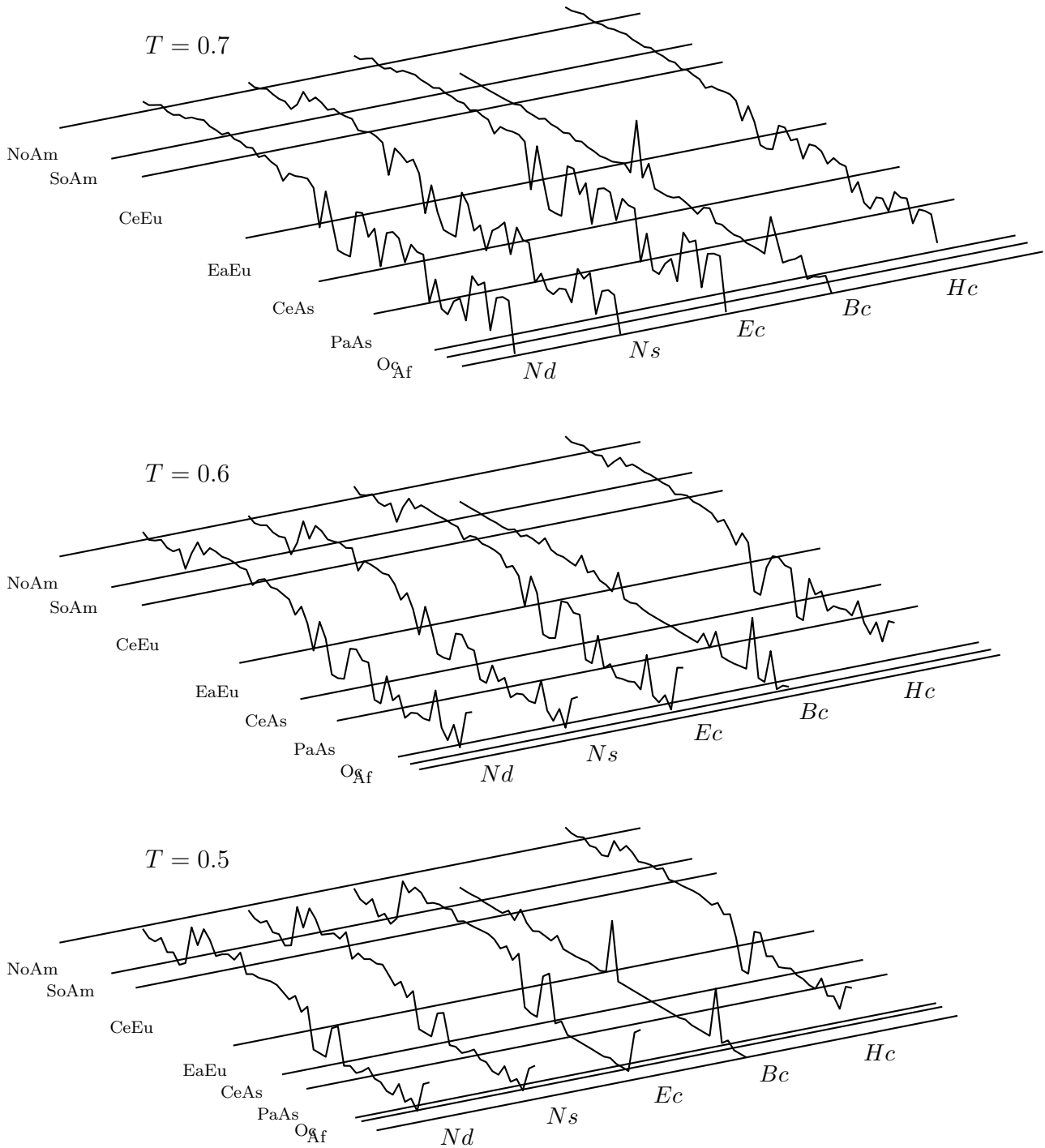
\noindent Fig. 29. Centrality measures for asset graphs defined by $T=0.7$, $T=0,6$, and $T=0.5$ in terms of node. The horizontal lines indicate divisions between regions: NoAm for North America, SoAm for South America, CeEu for Central Europe, EaEu for Eastern Europe, CeAs for Central Asia, Pa As for Pacific Asia, Oc for Oceania, and Af for Africa.

\vskip 0.3 cm 

From the figure, one can see that Central Europe and part of Eastern Europe present a consistent large number of centralities, expect for betweenness. North America and part of the Pacific Asia also behave, in a lesse degree, in the same way. For betweenness centrality, we have peaks in the Czech Republic and Singapore, and their centralities were explained in the previous paragraph. The eigenvector centrality and betweenness centrality are particularly useful in determining high centrality, in its two main aspects, nodes that are well connected in regions of densely connected nodes, and nodes that link smaller clusters of nodes.

\vskip 0.3 cm

\section{Conclusion}

The aims of this article may be divided into three parts: the first one was to use asset graphs based on thresholds for a distance measure obtained from the correlation matrices of time series of world stock market indices in order to study their behavior due to changes in the threshold value and also their evolution in time; the second was to analyze the survivability of the connections between indices in asset graphs and, as a consequence, of clusters based on regional, economical, and cultural ties, and their dependence on the threshold values; the third was to analyze some centrality measures of the nodes in the previous asset graphs, and their dependence on threshold values.

What we have shown is that financial markets tend to group themselves according to geographical region, and also according to economic and cultural ties. Two main clusters were detected, an American one, and an Eurpean one, both existing already at low threshold values (small distances) and very stable in time (high survivability). Other clusters form at higher levels of the threshold, like a Pacific Asian cluster and, at higher threshold values, an Arab one. All clusters tend to integrate at higher levels of the threshold, beginning by an absorption of some Eastern European nodes, plus Israel and South Africa, by the European cluster, and an enlargement of the North American cluster by the addition of South American nodes. An unification of the American, European, and Pacific Asian clusters occurs at intermediate values of the threshold. We could also see that, in times of crisis, like 2008, those asset graphs tend to shrink in size and augment in number of nodes, both consequences of the higher correlations between indices. These results are robust with respect to both the choice of correlation coefficient (Spearman or Pearson, specifically), and the lagging of some of the indices in order to take into account the difference in operating hours. We also used simulations with randomized data in order to establish threshold values above which random noise has a strong influence in the results.

Another result was that the second highest eigenvalue of the correlation matrix for the time series of stock exchange indices seems to be connected with the difference in operating hours, a result already obtained in \cite{clusters1}, separating the indices in two groups, Western and Eastern ones. By lagging the second group of indices in one day, we then obtain a new structure for the second eigenvalue, separating the European cluster from the others.

We could also see that Central Europe exhibits high values of centrality according to four centrality measures, two of them (node degree and node strenght) based on the number and strenght of connections of a node, one of them (eigenvector centrality) which also takes into account how connected each nodes' neighbors are, and one (harmonic closeness centrality) measuring the average (inverse) distance of a node to all the others. We also saw that the Czech Republic and Singapore appear as highly central with respect to a fifth measure, betweeness centrality, which measures how often a node is in between shortest paths connecting other nodes. We also made a study of the evolution of the average node degree in time and showed it is highly correlated with the average correlation of nodes in time.

All results showed there is a lot of information that can be obtained from asset graphs based on distance thresholds, and that the most significant ones seem to be those around the noise threshold. Most results have been compared with the ones obtained by simulations with randomized data, and they are remarkably different from those, except for high values of the threshold, which leads to the belief that the connections between international stock exchange indices are not formed at random, but have a complex structure underlying their behavior. The stability of some of the main connections also reinforce the idea that those ties are not formed at random. Finally, asset graphs do not exhibit characteristics of scale-free networks when analyzed in their centralities, although they are still quite different from random ones in terms of these same centrality measures.

\vskip 0.6 cm

\noindent{\bf \large Acknowledgements}

\vskip 0.4 cm

The author acknowledges the support of this work by a grant from Insper, Instituto de Ensino e Pesquisa. This article was written using \LaTeX, all figures were made using PSTricks, and the calculations were made using Matlab and Excel. All data are freely available upon request on leonidassj@insper.edu.br.

\appendix

\section{Stock Market Indices}

The next table (table 2) shows the stock market indices we used, their original countries, the symbols we used for them in the main text, and their codes in Bloomberg. In the tables, we use ``SX'' as short for ``Stock Exchange''. Some of the indices changed names and/or method of calculation and are designated by the two names, prior to and after the changing date.

\[ \begin{array}{|l|l|c|c|} \hline \text{Index} & \text{Country} & \text{Symbol} & \text{Code in Bloomberg} \\ \hline \text{\bf North America} \\ \hline \text{S\&P 500} & \text{United States of America} & \text{S\&P} & \text{SPX} \\ \text{Nasdaq Composite} & \text{United States of America} & \text{Nasd} & \text{CCMP} \\ \text{S\&P/TSX Composite} & \text{Canada} & \text{Cana} & \text{SPTSX} \\ \text{IPC} & \text{Mexico} & \text{Mexi} & \text{MEXBOL} \\ \hline \text{\bf Central America} \\ \hline \text{Bolsa de Panama General} & \text{Panama} & \text{Pana} & \text{BVPSBVPS} \\ \text{BCT Corp Costa Rica} & \text{Costa Rica} & \text{CoRi} & \text{CRSMBCT} \\ \hline \text{\bf Caribbean} \\ \hline \text{Jamaica SX Market Index} & \text{Jamaica} & \text{Jama} & \text{JMSMX} \\ \hline \text{\bf British overseas territories} \\ \hline \text{Bermuda SX Index} & \text{Bermuda} & \text{Berm} & \text{BSX} \\ \hline \text{\bf South America} \\ \hline \text{Ibovespa} & \text{Brazil} & \text{Braz} & \text{IBOV} \\ \text{Merval} & \text{Argentina} & \text{Arge} & \text{MERVAL} \\ \text{IPSA} & \text{Chile} & \text{Chil} & \text{IPSA} \\ \text{IGBC} & \text{Colombia} & \text{Colo} & \text{IGBC} \\ \text{IBC} & \text{Venezuela} & \text{Vene} & \text{IBVC} \\ \text{IGBVL} & \text{Peru} & \text{Peru} & \text{IGBVL} \\ \hline \text{\bf Western and Central Europe} \\ \hline \text{FTSE 100} & \text{United Kingdom} & \text{UK} & \text{UKX} \\ \text{ISEQ} & \text{Ireland} & \text{Irel} & \text{ISEQ} \\ \text{CAC 40} & \text{France} & \text{Fran} & \text{CAC} \\ \text{DAX} & \text{Germany} & \text{Germ} & \text{DAX} \\ \text{SMI} & \text{Switzerland} & \text{Swit} & \text{SMI} \\ \text{ATX} & \text{Austria} & \text{Autr} & \text{ATX} \\ \text{FTSE MIB or MIB-30} & \text{Italy} & \text{Ital} & \text{FTSEMIB} \\ \text{Malta SX Index} & \text{Malta} & \text{Malt} & \text{MALTEX} \\ \text{BEL 20} & \text{Belgium} & \text{Belg} & \text{BEL20} \\ \text{AEX} & \text{Netherlands} & \text{Neth} & \text{AEX} \\ \text{Luxembourg LuxX} & \text{Luxembourg} & \text{Luxe} & \text{LUXXX} \\ \text{OMX Stockholm 30} & \text{Sweden} & \text{Swed} & \text{OMX} \\ \text{OMX Copenhagen 20} & \text{Denmark} & \text{Denm} & \text{KFX} \\ \text{OMX Helsinki} & \text{Finland} & \text{Finl} & \text{HEX} \\ \text{OBX} & \text{Norway} & \text{Norw} & \text{OBX} \\ \text{OMX Iceland All-Share Index} & \text{Iceland} & \text{Icel} & \text{ICEXI} \\ \text{IBEX 35} & \text{Spain} & \text{Spai} & \text{IBEX} \\ \text{PSI 20} & \text{Portugal} & \text{Port} & \text{PSI20} \\ \text{Athens SX General Index} & \text{Greece} & \text{Gree} & \text{ASE} \\ \hline \end{array} \]

\[ \begin{array}{|l|l|c|c|} \hline \text{Index} & \text{Country} & \text{Symbol} & \text{Code in Bloomberg} \\ \hline \text{\bf Eastern Europe} \\ \hline \text{PX or PX50} & \text{Czech Republic} & \text{CzRe} & \text{PX} \\ \text{SAX} & \text{Slovakia} & \text{Slok} & \text{SKSM} \\ \text{Budapest SX Index} & \text{Hungary} & \text{Hung} & \text{BUX} \\ \text{BELEX 15} & \text{Serbia} & \text{Serb} & \text{BELEX15} \\ \text{CROBEX} & \text{Croatia} & \text{Croa} & \text{CRO} \\ \text{SBI TOP} & \text{Slovenia} & \text{Slov} & \text{SBITOP} \\ \text{SASE 10} & \text{Bosnia and Herzegovina} & \text{BoHe} & \text{SASX10} \\ \text{MOSTE} & \text{Montenegro} & \text{Mont} & \text{MOSTE} \\ \text{MBI 10} & \text{Macedonia} & \text{Mace} & \text{MBI} \\ \text{WIG} & \text{Poland} & \text{Pola} & \text{WIG} \\ \text{BET} & \text{Romania} & \text{Roma} & \text{BET} \\ \text{SOFIX} & \text{Bulgaria} & \text{Bulg} & \text{SOFIX} \\ \text{OMXT} & \text{Estonia} & \text{Esto} & \text{TALSE} \\ \text{OMXR} & \text{Latvia} & \text{Latv} & \text{RIGSE} \\ \text{OMXV} & \text{Lithuania} & \text{Lith} & \text{VILSE} \\ \text{PFTS} & \text{Ukraine} & \text{Ukra} & \text{PFTS} \\ \hline \text{\bf Eurasia} \\ \hline \text{MICEX} & \text{Russia} & \text{Russ} & \text{INDEXCF} \\ \text{ISE National 100} & \text{Turkey} & \text{Turk} & \text{XU100} \\ \hline \text{\bf Western and Central Asia} \\ \hline \text{KASE} & \text{Kazakhstan} & \text{Kaza} & \text{KZKAK} \\ \text{CSE} & \text{Cyprus} & \text{Cypr} & \text{CYSMMAPA} \\ \text{Tel Aviv 25} & \text{Israel} & \text{Isra} & \text{TA-25} \\ \text{Al Quds} & \text{Palestine} & \text{Pale} & \text{PASISI} \\ \text{BLOM} & \text{Lebanon} & \text{Leba} & \text{BLOM} \\ \text{ASE General Index} & \text{Jordan} & \text{Jord} & \text{JOSMGNFF} \\ \text{TASI} & \text{Saudi Arabia} & \text{SaAr} & \text{SASEIDX} \\ \text{Kuwait SE Weighted Index} & \text{Kuwait} & \text{Kuwa} & \text{SECTMIND} \\ \text{Bahrain All Share Index} & \text{Bahrain} & \text{Bahr} & \text{BHSEASI} \\ \text{QE or DSM 20} & \text{Qatar} & \text{Qata} & \text{DSM} \\ \text{ADX General Index} & \text{United Arab Emirates} & \text{UAE} & \text{ADSMI} \\ \text{MSM 30} & \text{Ohman} & \text{Ohma} & \text{MSM30} \\ \hline \text{\bf South Asia} \\ \hline \text{Karachi 100} & \text{Pakistan} & \text{Paki} & \text{KSE100} \\ \text{SENSEX 30} & \text{India} & \text{Indi} & \text{SENSEX} \\ \text{Colombo All-Share Index} & \text{Sri Lanka} & \text{SrLa} & \text{CSEALL} \\ \text{DSE General Index} & \text{Bangladesh} & \text{Bang} & \text{DHAKA} \\ \hline \end{array} \]

\[ \begin{array}{|l|l|c|c|} \hline \text{Index} & \text{Country} & \text{Symbol} & \text{Code in Bloomberg} \\ \hline \text{\bf Asia-Pacific} \\ \hline \text{Nikkei 25} & \text{Japan} & \text{Japa} & \text{NKY} \\ \text{Hang Seng} & \text{Hong Kong} & \text{HoKo} & \text{HSI} \\ \text{Shangai SE Composite} & \text{China} & \text{Chin} & \text{SHCOMP} \\ \text{MSE TOP 20} & \text{Mongolia} & \text{Mong} & \text{MSETOP} \\ \text{TAIEX} & \text{Taiwan} & \text{Taiw} & \text{TWSE} \\ \text{KOSPI} & \text{South Korea} & \text{SoKo} & \text{KOSPI} \\ \text{SET} & \text{Thailand} & \text{Thai} & \text{SET} \\ \text{VN-Index} & \text{Vietnam} & \text{Viet} & \text{VNINDEX} \\ \text{KLCI} & \text{Malaysia} & \text{Mala} & \text{FBMKLCI} \\ \text{Straits Times} & \text{Singapore} & \text{Sing} & \text{FSSTI} \\ \text{Jakarta Composite Index} & \text{Indonesia} & \text{Indo} & \text{JCI} \\ \text{PSEi} & \text{Philippines} & \text{Phil} & \text{PCOMP} \\ \hline \text{\bf Oceania} \\ \hline \text{S\&P/ASX 200} & \text{Australia} & \text{Aust} & \text{AS51} \\ \text{NZX 50} & \text{New Zealand} & \text{NeZe} & \text{NZSE50FG} \\ \hline \text{\bf Northern Africa} \\ \hline \text{CFG 25} & \text{Morocco} & \text{Moro} & \text{MCSINDEX} \\ \text{TUNINDEX} & \text{Tunisia} & \text{Tuni} & \text{TUSISE} \\ \text{EGX 30} & \text{Egypt} & \text{Egyp} & \text{CASE} \\ \hline \text{\bf Central and Southern Africa} \\ \hline \text{Ghana All Share Index} & \text{Ghana} & \text{Ghan} & \text{GGSEGSE} \\ \text{Nigeria SX All Share Index} & \text{Nigeria} & \text{Nige} & \text{NGSEINDX} \\ \text{NSE 20} & \text{Kenya} & \text{Keny} & \text{KNSMIDX} \\ \text{DSEI} & \text{Tanzania} & \text{Tanz} & \text{DARSDSEI} \\ \text{FTSE/Namibia Overall} & \text{Namibia} & \text{Nami} & \text{FTN098} \\ \text{Gaborone} & \text{Botswana} & \text{Bots} & \text{BGSMDC} \\ \text{FTSE/JSE Africa All Share} & \text{South Africa} & \text{SoAf} & \text{JALSH} \\ \text{SEMDEX} & \text{Mauritius} & \text{Maur} & \text{SEMDEX} \\ \hline \end{array} \]
\hskip 2.4 cm Table 2: names, codes, and abbreviations of the stock market indices used in this article.

\end{document}